\documentclass[acmsmall,nonacm]{acmart}
\newif\ifreviewing
\reviewingtrue %

\usepackage[all]{nowidow}
\usepackage{amsmath,amsfonts}
\usepackage{proof}
\usepackage{array}
\usepackage{verbatimbox}
\usepackage[noend]{algpseudocode}
\usepackage{graphicx}
\usepackage{textcomp}
\usepackage{xspace}
\usepackage{xcolor}
\usepackage{xstring} %

\usepackage{listings}
\usepackage{booktabs}
\usepackage{algorithm}
\usepackage{tabularx}
\usepackage{makecell}
\usepackage{multicol}
\usepackage{multirow}
\usepackage{soul}
\usepackage{tikz}
\usetikzlibrary{calc}
\usetikzlibrary{positioning,fit}
\usetikzlibrary{shapes.multipart}
\usetikzlibrary{backgrounds}
\usepackage{comment}
\usepackage{colortbl} %
\usepackage[breakable]{tcolorbox}
\usepackage{hyperref}
\usepackage{subcaption}
\usepackage{pgfkeys}

\usepackage{enumitem}

\definecolor{darkblueA}{rgb}{0.0, 0.0, 0.4}
\definecolor{darkgreen}{RGB}{0, 130, 0}
\definecolor{mygray}{HTML}{e3e6e8}
\definecolor{darkblue}{RGB}{0, 0, 150}
\definecolor{cyanHighlight}{HTML}{5CFFFF}
\definecolor{yellowHighlight}{HTML}{FBFF00}
\definecolor{lightgreen}{RGB}{144,238,144}
\definecolor{markrepgray}{RGB}{170, 170, 170}

\ifreviewing

  \newcommand{\todel}[1]{} %
  \newcommand{\toadd}[1]{\ignorespaces#1\unskip}

  \newcommand\bo[1]{} %
  \newcommand\brandon[1]{} %
  \newcommand\prateek[1]{} %
  \newcommand\umang[1]{} %
  \newcommand\joey[1]{} %
  \newcommand\daniel[1]{} %

\else

  \newcommand{\todel}[1]{\textcolor{red}{\ignorespaces#1\unskip}}
  \newcommand{\toadd}[1]{\textcolor{darkgreen}{\ignorespaces#1\unskip}}

  \newcommand\bo[1]{\textcolor{brown}{Bo: #1}}
  \newcommand\brandon[1]{\textcolor{orange}{Brandon: #1}}
  \newcommand\prateek[1]{\textcolor{red}{Prateek: #1}}
  \newcommand\umang[1]{\textcolor{purple}{Umang: #1}}
  \newcommand\joey[1]{\textcolor{blue}{Joey: #1}}
  \newcommand\daniel[1]{\textcolor{cyan}{Daniel: #1}}

\fi

\newcommand{\code}[1]{{\setlength\fboxsep{1pt}\colorbox{mygray}{\texttt{\lstinline|#1|}}}}

\newcommand{\tool}{\textsc{Reboot}\xspace}

\newcommand{\nobgcode}[1]{{\setlength\fboxsep{1pt}{\lstinline|#1|}}}

\newcommand\ignore[1]{}

\newenvironment{myparagraph}[1]{\vspace{0.05in}\noindent{\bf #1}}{}

\definecolor{lstgray}{HTML}{666666}		%
\definecolor{lstlightblue}{HTML}{006699}		%
\definecolor{lstlightgreen}{HTML}{669900}		%
\definecolor{lstbluegreen}{HTML}{33997e}		%
\definecolor{lstmagenta}{HTML}{d94a7a}		%
\definecolor{lstorange}{HTML}{e2661a}		%
\definecolor{lstpurple}{HTML}{7d4793}		%
\definecolor{lstgreen}{HTML}{718a62}		%
\lstdefinelanguage{CRS}{
  morekeywords=[1]{void, int, float, double, char, short, long, unsigned, signed},
  morekeywords=[2]{if, else, for, while, do, switch, case, default, return, break, continue, goto, true, false},
  morekeywords=[3]{sizeof, typeof, enum, struct, union, typedef, static, extern, volatile, const, volatile, restrict, inline, __restrict, __restrict__},
  morekeywords=[4]{define, include},
  keywordstyle=[1]{\color{lstlightblue}},
  keywordstyle=[2]{\color{lstlightgreen}},
  keywordstyle=[3]{\color{lstbluegreen}},
  keywordstyle=[4]{\color{lstpurple}},
  morecomment=[l]{//},
  morecomment=[s]{/*}{*/},
  commentstyle={\color{lstorange}\bfseries},
  morestring=[b]',
  morestring=[b]",
  stringstyle=\color{lstmagenta},
}

\lstdefinelanguage{Rust}{
  morekeywords=[1]{i8, i16, i32, i64, i128, isize, u8, u16, u32, u64, u128, usize, f32, f64, bool, char, str, String},
  morekeywords=[2]{if, else, for, while, loop, match, break, continue, return, true, false},
  morekeywords=[3]{fn, struct, enum, trait, impl, pub, mod, use, let, mut, const, static, type, where, move, ref, unsafe},
  morekeywords=[4]{Vec, Option, Result, Box, Arc, Rc, Cell, RefCell, Mutex, RwLock},
  keywordstyle=[1]{\color{lstlightblue}},
  keywordstyle=[2]{\color{lstlightgreen}},
  keywordstyle=[3]{\color{lstbluegreen}},
  keywordstyle=[4]{\color{lstpurple}},
  morecomment=[l]{//},
  morecomment=[s]{/*}{*/},
  commentstyle={\color{lstorange}\bfseries},
  morestring=[b]',
  morestring=[b]",
  stringstyle=\color{lstmagenta},
}

\lstdefinestyle{CRSstyle}{
  language=CRS,
  basicstyle=\ttfamily\scriptsize,
  frame=single,
  lineskip=-0.4ex,
  basewidth=0.53em,
}

\lstdefinestyle{Ruststyle}{
  language=Rust,
  basicstyle=\ttfamily\scriptsize,
  frame=single,
  lineskip=-0.4ex,
  basewidth=0.53em,
}

\usepackage{pifont}

\newcommand{\awkCdebugTotalSec}{1.02}
\newcommand{\awkCdebugTotalPct}{100.0\%}
\newcommand{\awkCdebugMedianMs}{1.66}
\newcommand{\awkCdebugMedianPct}{100.0\%}
\newcommand{\awkREBOOTdebugTotalSec}{7.03}
\newcommand{\awkREBOOTdebugTotalPct}{703.4\%}
\newcommand{\awkREBOOTdebugMedianMs}{2.57}
\newcommand{\awkREBOOTdebugMedianPct}{257.4\%}
\newcommand{\awkREBOOTNFdebugTotalSec}{127.96}
\newcommand{\awkREBOOTNFdebugTotalPct}{12795.8\%}
\newcommand{\awkREBOOTNFdebugMedianMs}{6.12}
\newcommand{\awkREBOOTNFdebugMedianPct}{612.3\%}

\newcommand{\gnubcCdebugTotalSec}{14.48}
\newcommand{\gnubcCdebugTotalPct}{100.0\%}
\newcommand{\gnubcCdebugMedianMs}{0.99}
\newcommand{\gnubcCdebugMedianPct}{100.0\%}
\newcommand{\gnubcREBOOTdebugTotalSec}{11.35}
\newcommand{\gnubcREBOOTdebugTotalPct}{1135.3\%}
\newcommand{\gnubcREBOOTdebugMedianMs}{1.61}
\newcommand{\gnubcREBOOTdebugMedianPct}{161.5\%}
\newcommand{\gnubcREBOOTNFdebugTotalSec}{13.65}
\newcommand{\gnubcREBOOTNFdebugTotalPct}{1365.3\%}
\newcommand{\gnubcREBOOTNFdebugMedianMs}{1.86}
\newcommand{\gnubcREBOOTNFdebugMedianPct}{186.2\%}

\newcommand{\mujsCdebugTotalSec}{0.52}
\newcommand{\mujsCdebugTotalPct}{100.0\%}
\newcommand{\mujsCdebugMedianMs}{2.02}
\newcommand{\mujsCdebugMedianPct}{100.0\%}
\newcommand{\mujsREBOOTdebugTotalSec}{2.15}
\newcommand{\mujsREBOOTdebugTotalPct}{214.5\%}
\newcommand{\mujsREBOOTdebugMedianMs}{2.08}
\newcommand{\mujsREBOOTdebugMedianPct}{207.8\%}
\newcommand{\mujsREBOOTNFdebugTotalSec}{2.77}
\newcommand{\mujsREBOOTNFdebugTotalPct}{276.7\%}
\newcommand{\mujsREBOOTNFdebugMedianMs}{2.25}
\newcommand{\mujsREBOOTNFdebugMedianPct}{224.6\%}

\newcommand{\picocCdebugTotalSec}{0.38}
\newcommand{\picocCdebugTotalPct}{100.0\%}
\newcommand{\picocCdebugMedianMs}{1.45}
\newcommand{\picocCdebugMedianPct}{100.0\%}
\newcommand{\picocREBOOTdebugTotalSec}{2.01}
\newcommand{\picocREBOOTdebugTotalPct}{201.0\%}
\newcommand{\picocREBOOTdebugMedianMs}{1.35}
\newcommand{\picocREBOOTdebugMedianPct}{134.7\%}
\newcommand{\picocREBOOTNFdebugTotalSec}{3.37}
\newcommand{\picocREBOOTNFdebugTotalPct}{336.8\%}
\newcommand{\picocREBOOTNFdebugMedianMs}{1.78}
\newcommand{\picocREBOOTNFdebugMedianPct}{178.5\%}

\newcommand{\pocketpyCdebugTotalSec}{2.29}
\newcommand{\pocketpyCdebugTotalPct}{100.0\%}
\newcommand{\pocketpyCdebugMedianMs}{2.90}
\newcommand{\pocketpyCdebugMedianPct}{100.0\%}
\newcommand{\pocketpyREBOOTdebugTotalSec}{7.70}
\newcommand{\pocketpyREBOOTdebugTotalPct}{769.9\%}
\newcommand{\pocketpyREBOOTdebugMedianMs}{1.41}
\newcommand{\pocketpyREBOOTdebugMedianPct}{140.5\%}

\newcommand{\wrenCdebugTotalSec}{3.32}
\newcommand{\wrenCdebugTotalPct}{100.0\%}
\newcommand{\wrenCdebugMedianMs}{2.29}
\newcommand{\wrenCdebugMedianPct}{100.0\%}
\newcommand{\wrenREBOOTdebugTotalSec}{1.57}
\newcommand{\wrenREBOOTdebugTotalPct}{157.4\%}
\newcommand{\wrenREBOOTdebugMedianMs}{1.00}
\newcommand{\wrenREBOOTdebugMedianPct}{100.1\%}
\newcommand{\wrenREBOOTNFdebugTotalSec}{57.85}
\newcommand{\wrenREBOOTNFdebugTotalPct}{5784.7\%}
\newcommand{\wrenREBOOTNFdebugMedianMs}{2.02}
\newcommand{\wrenREBOOTNFdebugMedianPct}{201.8\%}

\newcommand{\awkCreleaseTotalSec}{0.89}
\newcommand{\awkCreleaseTotalPct}{100.0\%}
\newcommand{\awkCreleaseMedianMs}{1.57}
\newcommand{\awkCreleaseMedianPct}{100.0\%}
\newcommand{\awkREBOOTreleaseTotalSec}{2.36}
\newcommand{\awkREBOOTreleaseTotalPct}{235.6\%}
\newcommand{\awkREBOOTreleaseMedianMs}{1.51}
\newcommand{\awkREBOOTreleaseMedianPct}{150.7\%}
\newcommand{\awkREBOOTNFreleaseTotalSec}{71.35}
\newcommand{\awkREBOOTNFreleaseTotalPct}{7135.5\%}
\newcommand{\awkREBOOTNFreleaseMedianMs}{4.27}
\newcommand{\awkREBOOTNFreleaseMedianPct}{426.5\%}

\newcommand{\gnubcCreleaseTotalSec}{7.75}
\newcommand{\gnubcCreleaseTotalPct}{100.0\%}
\newcommand{\gnubcCreleaseMedianMs}{1.01}
\newcommand{\gnubcCreleaseMedianPct}{100.0\%}
\newcommand{\gnubcREBOOTreleaseTotalSec}{1.27}
\newcommand{\gnubcREBOOTreleaseTotalPct}{126.8\%}
\newcommand{\gnubcREBOOTreleaseMedianMs}{1.32}
\newcommand{\gnubcREBOOTreleaseMedianPct}{131.7\%}
\newcommand{\gnubcREBOOTNFreleaseTotalSec}{2.58}
\newcommand{\gnubcREBOOTNFreleaseTotalPct}{257.6\%}
\newcommand{\gnubcREBOOTNFreleaseMedianMs}{1.57}
\newcommand{\gnubcREBOOTNFreleaseMedianPct}{157.1\%}

\newcommand{\mujsCreleaseTotalSec}{0.46}
\newcommand{\mujsCreleaseTotalPct}{100.0\%}
\newcommand{\mujsCreleaseMedianMs}{1.76}
\newcommand{\mujsCreleaseMedianPct}{100.0\%}
\newcommand{\mujsREBOOTreleaseTotalSec}{1.43}
\newcommand{\mujsREBOOTreleaseTotalPct}{143.0\%}
\newcommand{\mujsREBOOTreleaseMedianMs}{1.39}
\newcommand{\mujsREBOOTreleaseMedianPct}{139.4\%}
\newcommand{\mujsREBOOTNFreleaseTotalSec}{1.14}
\newcommand{\mujsREBOOTNFreleaseTotalPct}{113.8\%}
\newcommand{\mujsREBOOTNFreleaseMedianMs}{1.11}
\newcommand{\mujsREBOOTNFreleaseMedianPct}{111.5\%}

\newcommand{\picocCreleaseTotalSec}{0.27}
\newcommand{\picocCreleaseTotalPct}{100.0\%}
\newcommand{\picocCreleaseMedianMs}{1.39}
\newcommand{\picocCreleaseMedianPct}{100.0\%}
\newcommand{\picocREBOOTreleaseTotalSec}{2.62}
\newcommand{\picocREBOOTreleaseTotalPct}{261.7\%}
\newcommand{\picocREBOOTreleaseMedianMs}{1.34}
\newcommand{\picocREBOOTreleaseMedianPct}{133.7\%}
\newcommand{\picocREBOOTNFreleaseTotalSec}{1.35}
\newcommand{\picocREBOOTNFreleaseTotalPct}{134.5\%}
\newcommand{\picocREBOOTNFreleaseMedianMs}{1.19}
\newcommand{\picocREBOOTNFreleaseMedianPct}{119.4\%}

\newcommand{\pocketpyCreleaseTotalSec}{1.62}
\newcommand{\pocketpyCreleaseTotalPct}{100.0\%}
\newcommand{\pocketpyCreleaseMedianMs}{2.55}
\newcommand{\pocketpyCreleaseMedianPct}{100.0\%}
\newcommand{\pocketpyREBOOTreleaseTotalSec}{1.16}
\newcommand{\pocketpyREBOOTreleaseTotalPct}{116.1\%}
\newcommand{\pocketpyREBOOTreleaseMedianMs}{0.79}
\newcommand{\pocketpyREBOOTreleaseMedianPct}{79.5\%}

\newcommand{\wrenCreleaseTotalSec}{2.24}
\newcommand{\wrenCreleaseTotalPct}{100.0\%}
\newcommand{\wrenCreleaseMedianMs}{1.77}
\newcommand{\wrenCreleaseMedianPct}{100.0\%}
\newcommand{\wrenREBOOTreleaseTotalSec}{2.22}
\newcommand{\wrenREBOOTreleaseTotalPct}{222.5\%}
\newcommand{\wrenREBOOTreleaseMedianMs}{1.28}
\newcommand{\wrenREBOOTreleaseMedianPct}{127.8\%}
\newcommand{\wrenREBOOTNFreleaseTotalSec}{6.11}
\newcommand{\wrenREBOOTNFreleaseTotalPct}{610.9\%}
\newcommand{\wrenREBOOTNFreleaseMedianMs}{1.44}
\newcommand{\wrenREBOOTNFreleaseMedianPct}{144.2\%}

\newcommand{\perfReleaseTotalMinSec}{1.16}
\newcommand{\perfReleaseTotalMinPct}{116.1\%}
\newcommand{\perfReleaseTotalMaxSec}{2.62}
\newcommand{\perfReleaseTotalMaxPct}{261.7\%}

\newcommand{\perfReleaseSlowdownMinMs}{1.28}
\newcommand{\perfReleaseSlowdownMaxMs}{1.51}
\newcommand{\perfReleaseSlowdownCount}{5}

\newcommand{\bmawkLoc}{6,332}
\newcommand{\bmawkLocRounded}{6k}
\newcommand{\bmawkTests}{279}
\newcommand{\bmawkCovPct}{74\%}

\newcommand{\bmgnubcLoc}{7,525}
\newcommand{\bmgnubcLocRounded}{8k}
\newcommand{\bmgnubcTests}{134}
\newcommand{\bmgnubcCovPct}{85\%}

\newcommand{\bmpicocLoc}{8,486}
\newcommand{\bmpicocLocRounded}{8k}
\newcommand{\bmpicocTests}{154}
\newcommand{\bmpicocCovPct}{81\%}

\newcommand{\bmwrenLoc}{8,325}
\newcommand{\bmwrenLocRounded}{8k}
\newcommand{\bmwrenTests}{851}
\newcommand{\bmwrenCovPct}{88\%}

\newcommand{\bmmujsLoc}{17,090}
\newcommand{\bmmujsLocRounded}{17k}
\newcommand{\bmmujsTests}{187}
\newcommand{\bmmujsCovPct}{88\%}

\newcommand{\bmpocketpyLoc}{23,271}
\newcommand{\bmpocketpyLocRounded}{23k}
\newcommand{\bmpocketpyTests}{83}
\newcommand{\bmpocketpyCovPct}{81\%}

\newcommand{\bmMinLocRounded}{6k}
\newcommand{\bmMaxLocRounded}{23k}
\newcommand{\bmMinCovPct}{74\%}
\newcommand{\bmMaxCovPct}{88\%}

\newcommand{\resawkRsLoc}{9,514}
\newcommand{\resawkTestPass}{279/279}
\newcommand{\resawkProvPct}{100\%}
\newcommand{\resawkTime}{36.0h}
\newcommand{\resawkCost}{\$713.62}

\newcommand{\resgnubcRsLoc}{6,784}
\newcommand{\resgnubcTestPass}{134/134}
\newcommand{\resgnubcProvPct}{100\%}
\newcommand{\resgnubcTime}{27.7h}
\newcommand{\resgnubcCost}{\$463.46}

\newcommand{\respicocRsLoc}{14,259}
\newcommand{\respicocTestPass}{154/154}
\newcommand{\respicocProvPct}{100\%}
\newcommand{\respicocTime}{46.2h}
\newcommand{\respicocCost}{\$971.04}

\newcommand{\reswrenRsLoc}{12,191}
\newcommand{\reswrenTestPass}{851/851}
\newcommand{\reswrenProvPct}{100\%}
\newcommand{\reswrenTime}{46.2h}
\newcommand{\reswrenCost}{\$949.83}

\newcommand{\resmujsRsLoc}{16,235}
\newcommand{\resmujsTestPass}{187/187}
\newcommand{\resmujsProvPct}{100\%}
\newcommand{\resmujsTime}{44.8h}
\newcommand{\resmujsCost}{\$957.75}

\newcommand{\respocketpyRsLoc}{24,348}
\newcommand{\respocketpyTestPass}{83/83}
\newcommand{\respocketpyProvPct}{100\%}
\newcommand{\respocketpyTime}{90.5h}
\newcommand{\respocketpyCost}{\$1781.79}

\newcommand{\resMinTimeRounded}{28}
\newcommand{\resMaxTimeRounded}{90}

\newcommand{\resMinCostRounded}{460}
\newcommand{\resMaxCostRounded}{1,780}

\newcommand{\valawkPct}{78.82\%}
\newcommand{\valawkFrac}{160/203}

\newcommand{\valgnubcPct}{78.57\%}
\newcommand{\valgnubcFrac}{99/126}

\newcommand{\valpicocPct}{69.50\%}
\newcommand{\valpicocFrac}{139/200}

\newcommand{\valwrenPct}{91.78\%}
\newcommand{\valwrenFrac}{134/146}

\newcommand{\valmujsPct}{74.77\%}
\newcommand{\valmujsFrac}{163/218}

\newcommand{\valpocketpyPct}{61.58\%}
\newcommand{\valpocketpyFrac}{125/203}

\newcommand{\valMinPct}{62\%}
\newcommand{\valMaxPct}{92\%}

\newcommand{\valawkCov}{79.63\%}
\newcommand{\valgnubcCov}{80.31\%}
\newcommand{\valmujsCov}{84.85\%}
\newcommand{\valpicocCov}{82.24\%}
\newcommand{\valpocketpyCov}{62.98\%}
\newcommand{\valwrenCov}{82.73\%}

\newcommand{\valCovMin}{62.98\%}
\newcommand{\valCovMax}{84.85\%}
\newcommand{\valCovAvg}{78.79\%}

\newcommand{\ablNFawkTime}{34.6h}
\newcommand{\ablNFawkUserInt}{6}
\newcommand{\ablNFawkProvFrac}{279/279}
\newcommand{\ablNFawkProvPct}{100\%}
\newcommand{\ablNFawkValFrac}{146/203}
\newcommand{\ablNFawkValPct}{71.9\%}

\newcommand{\ablNFgnubcTime}{35.5h}
\newcommand{\ablNFgnubcUserInt}{6}
\newcommand{\ablNFgnubcProvFrac}{133/134}
\newcommand{\ablNFgnubcProvPct}{99\%}
\newcommand{\ablNFgnubcValFrac}{85/126}
\newcommand{\ablNFgnubcValPct}{67.5\%}

\newcommand{\ablNFpicocTime}{19.2h}
\newcommand{\ablNFpicocUserInt}{2}
\newcommand{\ablNFpicocProvFrac}{152/154}
\newcommand{\ablNFpicocProvPct}{99\%}
\newcommand{\ablNFpicocValFrac}{98/200}
\newcommand{\ablNFpicocValPct}{49.0\%}

\newcommand{\ablNFwrenTime}{80.7h}
\newcommand{\ablNFwrenUserInt}{13}
\newcommand{\ablNFwrenProvFrac}{851/851}
\newcommand{\ablNFwrenProvPct}{100\%}
\newcommand{\ablNFwrenValFrac}{125/146}
\newcommand{\ablNFwrenValPct}{85.6\%}

\newcommand{\ablNFmujsTime}{52.6h}
\newcommand{\ablNFmujsUserInt}{12}
\newcommand{\ablNFmujsProvFrac}{187/187}
\newcommand{\ablNFmujsProvPct}{100\%}
\newcommand{\ablNFmujsValFrac}{143/218}
\newcommand{\ablNFmujsValPct}{65.6\%}

\newcommand{\ablNFpocketpyTime}{5.4h}
\newcommand{\ablNFpocketpyUserInt}{0}
\newcommand{\ablNFpocketpyProvFrac}{11/83}
\newcommand{\ablNFpocketpyProvPct}{13\%}
\newcommand{\ablNFpocketpyValFrac}{30/203}
\newcommand{\ablNFpocketpyValPct}{14.8\%}

\newcommand{\ablNFiipocketpyTime}{19.8h}
\newcommand{\ablNFiipocketpyUserInt}{2}
\newcommand{\ablNFiipocketpyProvFrac}{50/83}
\newcommand{\ablNFiipocketpyProvPct}{60\%}
\newcommand{\ablNFiipocketpyValFrac}{99/203}
\newcommand{\ablNFiipocketpyValPct}{48.8\%}

\newcommand{\ablTFullTotal}{11,725}
\newcommand{\ablTmujsPass}{8,268}
\newcommand{\ablTmujsConform}{9,853}
\newcommand{\ablTmujsFrac}{8,268/9,853}
\newcommand{\ablTmujsPct}{83.91\%}

\newcommand{\ablTNFmujsFrac}{3,538/9,853}
\newcommand{\ablTNFmujsPct}{35.91\%}

\newcommand{\ablCVETime}{58.0h}
\newcommand{\ablCVEUserInt}{4}
\newcommand{\ablCVEProvFrac}{187/187}
\newcommand{\ablCVEProvPct}{100\%}
\newcommand{\ablCVEValFrac}{170/218}
\newcommand{\ablCVEValPct}{78.0\%}
\newcommand{\ablCVETFrac}{7,554/9,853}
\newcommand{\ablCVETPct}{76.67\%}

\newcommand{\ablCVENFTime}{28.0h}
\newcommand{\ablCVENFUserInt}{1}
\newcommand{\ablCVENFProvFrac}{187/187}
\newcommand{\ablCVENFProvPct}{100\%}
\newcommand{\ablCVENFValFrac}{144/218}
\newcommand{\ablCVENFValPct}{66.1\%}
\newcommand{\ablCVENFTFrac}{2,835/9,853}
\newcommand{\ablCVENFTPct}{28.77\%}

\newcommand{\ablImprovMin}{6\%}
\newcommand{\ablImprovMax}{20\%}

\newcommand{\escawkUserInt}{1}
\newcommand{\escawkMinor}{2}

\newcommand{\escgnubcUserInt}{1}

\newcommand{\escpicocUserInt}{4}
\newcommand{\escpicocMinor}{1}

\newcommand{\escwrenUserInt}{8}

\newcommand{\escmujsUserInt}{4}

\newcommand{\escpocketpyUserInt}{11}
\newcommand{\escpocketpyMinor}{1}

\newcommand{\escCatTask}{10}
\newcommand{\escCatWorkflow}{7}
\newcommand{\escCatMisbehavior}{6}
\newcommand{\escCatDesign}{3}
\newcommand{\escCatSystem}{3}

\newcommand{\escTotalUserInt}{29}
\newcommand{\escTotalMinor}{4}
\newcommand{\escTotalEsc}{125}
\newcommand{\escTotalAuto}{96}
\newcommand{\escAutoPct}{77\%}

\newcommand{\cveAll}{30}
\newcommand{\cveAppl}{20}
\newcommand{\cveExcl}{10}

\newcommand{\cveElim}{14}
\newcommand{\cveMitig}{4}
\newcommand{\cveSurv}{2}
\newcommand{\cveElimOrMitig}{18}

\newcommand{\cveHeapBOF}{6}
\newcommand{\cveHeapUAF}{3}
\newcommand{\cveStackExhaust}{3}
\newcommand{\cveIntOverflow}{2}
\newcommand{\cveBytecodeLogic}{2}
\newcommand{\cveNULLDeref}{1}
\newcommand{\cveOOBRead}{1}
\newcommand{\cveStackBOF}{1}
\newcommand{\cveGlobalBOF}{1}

\newcommand{\cveFactArch}{6}
\newcommand{\cveFactArchPri}{5}
\newcommand{\cveFactArchCon}{1}
\newcommand{\cveFactTypeSafety}{5}
\newcommand{\cveFactTypeSafetyPri}{2}
\newcommand{\cveFactTypeSafetyCon}{3}
\newcommand{\cveFactBounds}{4}
\newcommand{\cveFactBoundsPri}{1}
\newcommand{\cveFactBoundsCon}{3}
\newcommand{\cveFactAPI}{3}
\newcommand{\cveFactAPIPri}{3}
\newcommand{\cveFactAPICon}{0}
\newcommand{\cveFactOwnership}{2}
\newcommand{\cveFactOwnershipPri}{2}
\newcommand{\cveFactOwnershipCon}{0}
\newcommand{\cveFactOther}{2}
\newcommand{\cveFactOtherPri}{2}
\newcommand{\cveFactOtherCon}{0}

\begin{document}
\title{Mostly Automatic Translation of Language Interpreters\\ from C to Safe Rust}

\author{Bo Wang}
\authornote{Work done during an internship at Amazon.}
\affiliation{%
  \institution{National University of Singapore}
  \country{Singapore}
}
\email{bo_wang@u.nus.edu}

\author{Brandon Paulsen}
\affiliation{%
  \institution{Amazon}
  \country{USA}
}
\email{bpaulse@amazon.com}

\author{Joey Dodds}
\affiliation{%
  \institution{Amazon}
  \country{USA}
}
\email{jldodds@amazon.com}

\author{Daniel Kroening}
\affiliation{%
  \institution{Amazon}
  \country{USA}
}
\email{dkr@amazon.com}

\author{Umang Mathur}
\affiliation{%
  \institution{National University of Singapore}
  \country{Singapore}
}
\email{umathur@nus.edu.sg}

\author{Prateek Saxena}
\affiliation{%
  \institution{National University of Singapore}
  \country{Singapore}
}
\email{dcsprs@nus.edu.sg}

\begin{abstract}
Translating C programs to safe Rust is challenging owing to significant differences in typing constraints, ownership, and borrowing rules.
Interpreter programs are particularly important targets for such translation, as they often handle untrusted inputs and suffer from memory-related vulnerabilities.
We present \tool, a mostly-automatic technique that translates real-world interpreter programs from C to safe Rust.
Using \tool, we have translated six interpreters ranging from \bmMinLocRounded{} to \bmMaxLocRounded{} lines of C code to safe Rust, with each translation requiring only \escawkUserInt{} to \escpocketpyUserInt{} brief user interventions.
All translations pass 100\% of the provided test suites, and achieve \valMinPct--\valMaxPct{} pass rates on separately created validation tests that were never exposed to the system.
A security case study on \code{mujs} shows that memory vulnerabilities such as heap buffer overflows and use-after-free present in C are eliminated in the safe Rust translation.
Two ideas underpin \tool.
First, \emph{feature reduction} decomposes the translation by program features, creating a sequence of milestones where each is a complete, testable program; the translation starts from the simplest version and incrementally restores features, with each milestone validated before proceeding.
Second, a \emph{multi-agent architecture} orchestrates inherently unreliable coding agents through automated validation and feedback, keeping long-running translation workflows on track with minimal human involvement.
An ablation study confirms that feature reduction improves translation correctness compared to using multi-agent translation alone, \toadd{with \ablImprovMin--\ablImprovMax{} improvements in pass rates on validation test suites}.

\end{abstract}

\maketitle %

\section{Introduction}

C has been widely used for implementing language interpreters, where low-level memory control is essential for performance.
However, C's lack of memory safety often leads to vulnerabilities such as spatial and temporal memory errors~\cite{sokwar,ossfuzz}, and Rust is gaining popularity as an alternative that provides strong memory safety guarantees while retaining low-level control~\cite{microsoftrust,linuxrust}.
Within the body of software written in C, interpreter programs are particularly important targets for migration to Rust: they appear in a wide range of contexts—as standalone tools, as scripting engines embedded in databases and browsers, and in many other applications—and are especially security-critical because they often handle untrusted inputs.
Many widely-deployed interpreters written in C suffer from memory-related security vulnerabilities~\cite{chen2013security,park2018bytecode,park2020nojitsu,jang2016integer,jiang2021pyguard}, and translating them to safe Rust would significantly improve their security posture.

Automatic translation from C to safe Rust is challenging.
The significant differences in typing constraints, ownership, and borrowing rules mean that producing safe, idiomatic Rust often requires substantial restructuring of the original C code.
Existing rule-based translators~\cite{c2rust,zhang2023ownership, emre2023aliasing, ling2022rust} typically preserve the low-level structure of the C source, producing non-idiomatic Rust with limited safety guarantees.
\toadd{LLM-based approaches~\cite{eniser2024towards,yang2024vert,syzygy,shiraishi2024smartc2rust,cai2025rustmap,ou2025enhancing,zhou2025llm,farrukh2025safetrans,dehghan2025translating,wang2026his2trans,2025rustassure} are more promising for synthesizing safe Rust, but most decompose programs by functions. Since individually translated functions cannot be tested as part of a running program, validating partial translations requires nontrivial additional techniques, and success has been demonstrated primarily on libraries. A recent agent-based approach~\cite{actor} avoids function-level decomposition, achieving reasonable correctness on standalone programs of hundreds to thousands of lines of code.}
However, the existing approaches have not yet been shown to scale to the size and complexity of real-world interpreter programs, which are often large, have complex internal data types, and exhibit cross-cutting data flow that often needs to be restructured to satisfy Rust's language rules.

\myparagraph{Our Results.}
In this work, we present \tool, an agent-based technique for translating interpreter programs from C to safe Rust.
\tool takes a C program and a test suite as input, and produces a safe Rust translation that passes all provided tests; a human user is only occasionally needed to make design decisions or provide clarifications that the system cannot resolve on its own.
Using \tool, we have translated six interpreter programs—\code{awk} (\bmawkLocRounded{} LoC), \code{picoc} (\bmpicocLocRounded{} LoC), \code{gnu-bc} (\bmgnubcLocRounded{} LoC), \code{wren} (\bmwrenLocRounded{} LoC), \code{mujs} (\bmmujsLocRounded{} LoC), and \code{pocketpy} (\bmpocketpyLocRounded{} LoC)—from C to safe Rust.
Each program is paired with a test suite (sourced from the project or supplemented by us 
), achieving \bmMinCovPct--\bmMaxCovPct{} line coverage on the C source, and all translations pass all of the provided test suites.
To evaluate correctness beyond the provided tests, we created validation test suites independently of and separately from the provided tests, never exposed to the system during translation; our translations achieve \valMinPct--\valMaxPct{} pass rates on these unseen tests.
Each translation takes \resMinTimeRounded{} to \resMaxTimeRounded{} hours of wall-clock time, costs \$\resMinCostRounded{} to \$\resMaxCostRounded{}, and requires only \escawkUserInt{} to \escpocketpyUserInt{} user interventions, each taking roughly 5 minutes.
To evaluate security improvements, we prepared a version of \code{mujs} with \cveAppl{} of its historical CVEs re-introduced into the latest codebase, and translated it using \tool; the results show that memory-related vulnerabilities such as heap buffer overflows, use-after-free, and stack buffer overflows are indeed eliminated in the safe Rust translation.
Performance measurements on release builds show that most translated programs have median slowdowns of $\sim$\perfReleaseSlowdownMinMs{}x--\perfReleaseSlowdownMaxMs{}x compared to the original C programs.

\myparagraph{Our Approach.}
Two ideas underpin \tool.
First, to handle the unreliability of coding agents, \tool uses a \emph{multi-agent architecture} that orchestrates long-running translation workflows.
Agents are inherently unreliable: they may fail to complete a task, claim completion when the output is incorrect, get stuck in unproductive loops, or crash unexpectedly.
\tool addresses this by employing additional agents that validate outputs and provide automated feedback, detecting and recovering from failures without human involvement.
A finite-state-machine guard enforces the workflow protocol, ensuring that agents follow the expected sequence of operations.
Together, these mechanisms allow the system to stay on track for days with most issues resolved automatically and only a small number of cases escalated to the user.

Second, to handle the complexity of translating real-world interpreters, \tool uses a new approach that we call \emph{feature reduction}. 
\toadd{Feature reduction decomposes the translation into a sequence of validated milestones, each defined by program features rather than by syntactic structure.}
\toadd{Prior work typically decomposes by syntactic units such as functions; however, interpreter features like exception handling or closures are cross-cutting---they span multiple functions and files across the codebase, which motivates decomposing by features instead.}
Feature reduction progressively simplifies the full interpreter into a sequence of feature levels, where each level is a complete, runnable program that can be independently tested and validated.
The translation starts from the simplest version and incrementally restores features, with each milestone validated before proceeding.
This creates manageable steps without imposing structural constraints on how the code is organized in Rust, giving agents the freedom to restructure code as needed for Rust's ownership and borrowing rules.
An ablation study confirms that feature reduction improves translation correctness compared to using the multi-agent architecture alone without feature reduction, \toadd{with \ablImprovMin--\ablImprovMax{} improvements in pass rates on validation test suites}.

The rest of this paper is organized as follows.
Section~\ref{sec:overview} gives an overview of feature reduction and multi-agent orchestration.
Section~\ref{sec:approach} describes the system in detail.
Section~\ref{sec:impl} covers the implementation.
Section~\ref{sec:eval} presents the evaluation, and Section~\ref{sec:related} discusses related work.

\section{Overview}
\label{sec:overview}

Translating complex interpreter programs from C to safe Rust using LLM agents presents two high-level design challenges: decomposing the translation task into manageable pieces, and orchestrating a self-correcting workflow of  agents that fail probabilistically.

\subsection{Decomposition Strategy}
\label{sec:overview-decomposition}

Translating a large program all at once is difficult, so decomposing the task is necessary and standard.
Most prior work decomposes by functions and user-defined data types (e.g. structs) both for C to Rust translation~\cite{syzygy,shiraishi2024smartc2rust,cai2025rustmap,ou2025enhancing,zhou2025llm,dehghan2025translating,wang2026his2trans,2025rustassure} and for other language pairs~\cite{zhang2025scalable,ibrahimzada2025alphatrans,skel,zhang2026validatedcodetranslationprojects}.
However, interpreter features are cross-cutting: a single feature like exception handling is implemented by code scattered across the lexer, parser, runtime, and bytecode executor.
We propose a new approach called \emph{feature reduction} that decomposes by program features instead.
A feature in an interpreter is a supported language capability, such as for-loops, closures, or exception handling.
Removing a feature means removing all related code segments across these components, yielding a simpler yet complete, runnable interpreter.

Feature reduction applies this process progressively, removing one or a few features at a time.
Each step produces a \emph{feature level} (FL)—a version of the program that supports a specific subset of features.
The removal order respects feature dependencies: features that other features depend on are removed later, so each feature level remains a coherent, runnable program.
Each feature level has its own test suite, adapted from the original by removing or simplifying tests that depend on unsupported features.
The result is a sequence of feature levels, from the full-featured program (FL$_N$) down to a minimal version (FL$_0$) that implements only basic functionality.

\begin{figure*}[t]
\centering
\includegraphics[width=0.9\textwidth,trim=0.3cm 1.5cm 1.8cm 0cm]{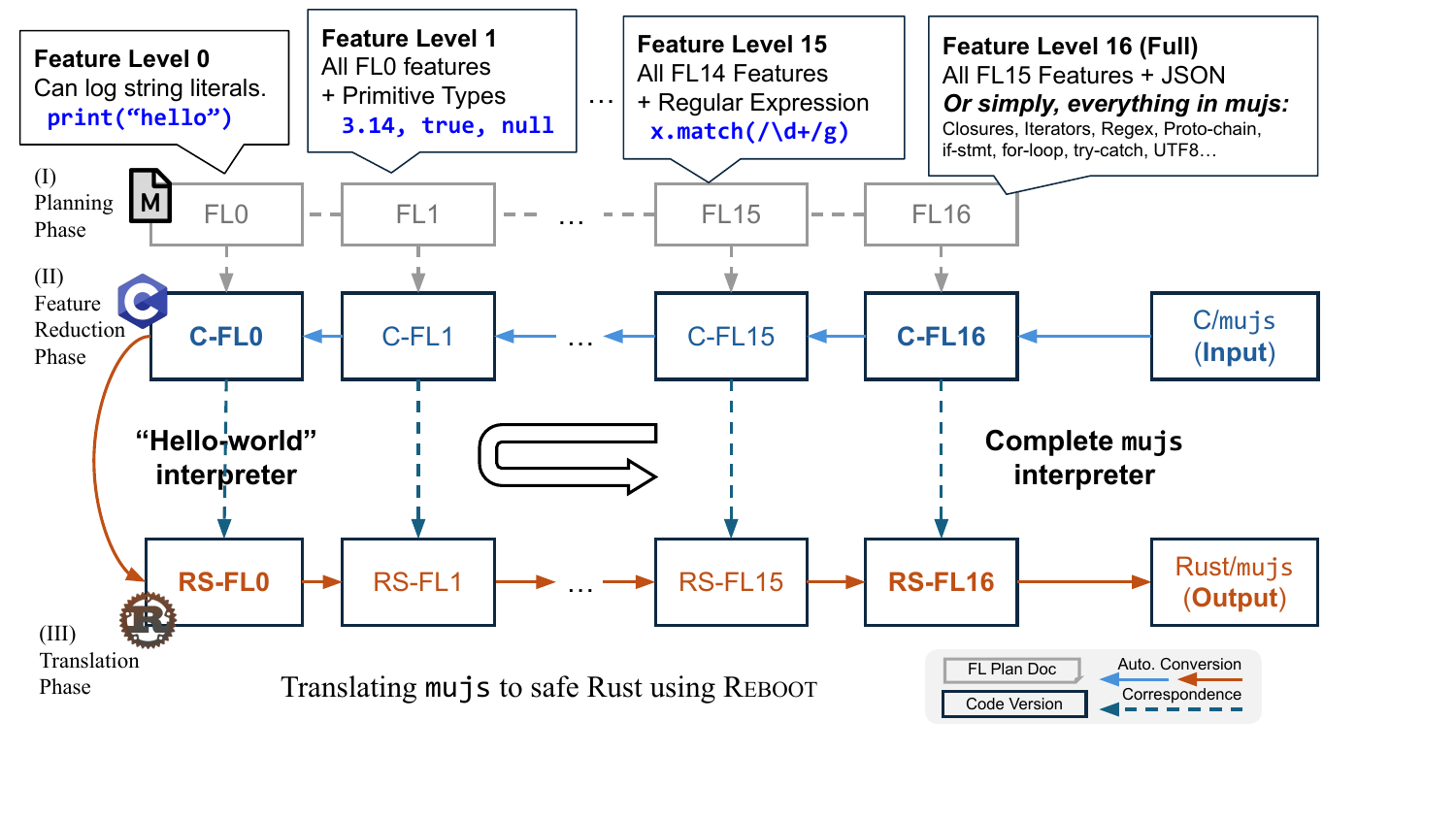} 
\caption{
The workflow of the translation process using \tool. 
}
\label{fig:overview}
\end{figure*} %
\begin{figure*}[t]
\centering
\includegraphics[width=0.98\textwidth,trim=0.3cm 5.4cm 0cm 0cm]{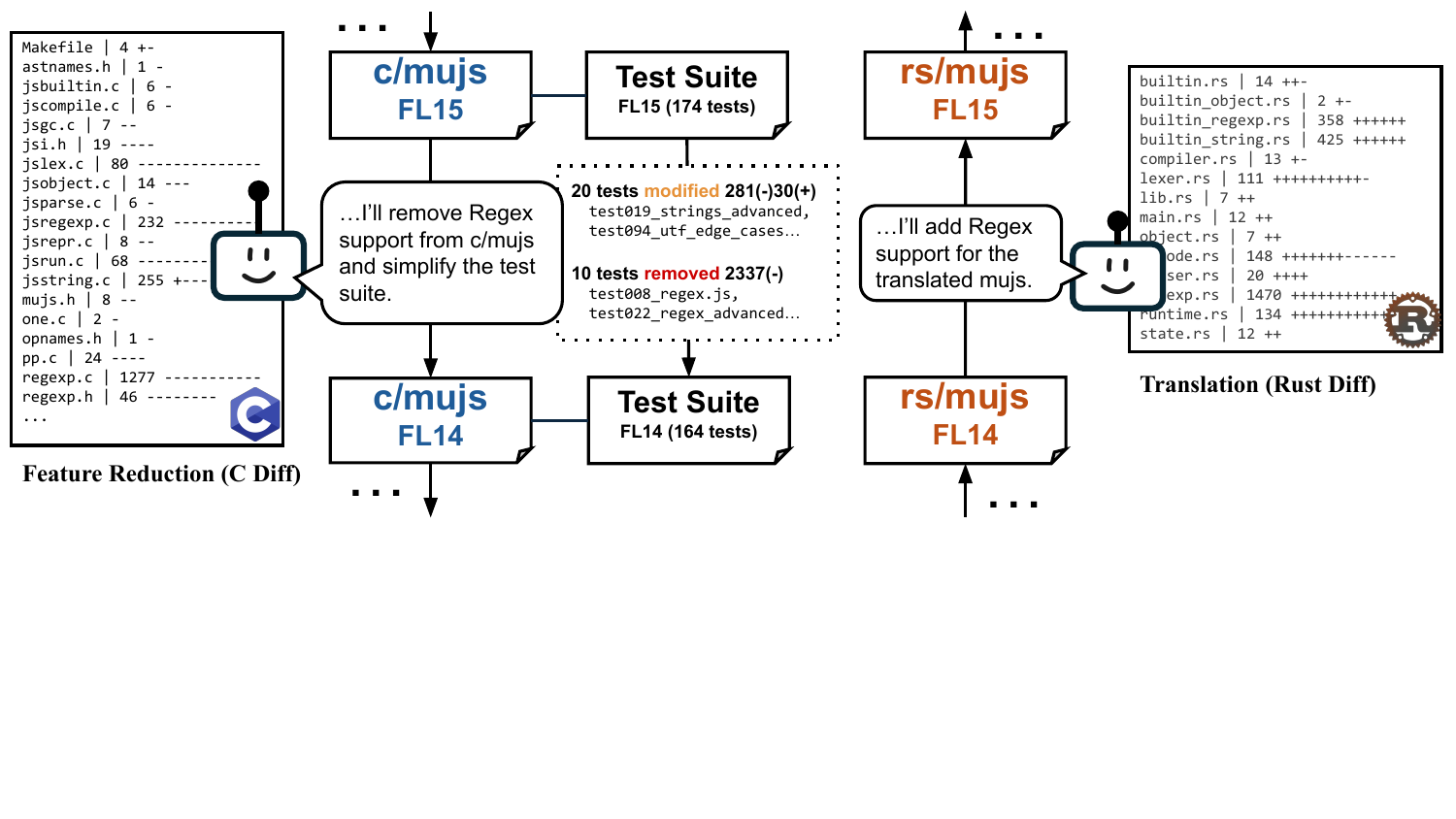} 
\caption{
An example of changes in source code as well as the test suite across feature levels during translation.
}
\label{fig:overview-diff}
\end{figure*} %

Figure~\ref{fig:overview} illustrates the overall process using mujs, a JavaScript interpreter, as an example.
The process consists of three main phases.
In the \emph{Planning Phase}, we analyze the input C program to identify its features and create a reduction plan that defines the sequence of feature levels.
In the \emph{Feature Reduction Phase}, we progressively simplify the C program from the full version (C-FL$_N$) down to a minimal version (C-FL$_0$), producing a validated C program at each level.
In the \emph{Translation Phase}, we start by translating the simplest C program (C-FL$_0$) to Rust (RS-FL$_0$).
We then incrementally restore features, translating RS-FL$_0$ to RS-FL$_1$, then to RS-FL$_2$, and so on, until we reach the full translation (RS-FL$_N$).
Each step in the feature reduction and translation phases is validated before proceeding to the next.

In the mujs example, FL$_{16}$ is the complete JavaScript interpreter with all features: closures, iterators, prototype chains, regular expressions, exception handling, and more.
By removing features from the full program, we obtain a sequence of simpler versions.
FL$_{15}$, obtained by removing only CLI options, retains nearly all language features including regular expressions.
Intermediate levels support progressively fewer capabilities—FL$_1$ supports only binary operators and numbers (e.g., \code{3.14*(R**2)}), and FL$_0$ is a minimal ``hello-world'' interpreter limited to logging string literals (e.g., \code{print("hello")}).
Each feature level is a working JavaScript interpreter for a subset of the language, obtained by feature reduction.

Figure~\ref{fig:overview-diff} shows a single feature-level transition in each phase.
On the left, the transition from C-FL$_{15}$ to C-FL$_{14}$ removes the support for regular expressions from the C codebase.
The diff spans many files, including \code{regexp.c} ($\sim$1.3k lines), \code{jsstring.c} ($\sim$250 lines), \code{jsregexp.c} ($\sim$230 lines), and others. Many of these changes are related to the same feature, demonstrating that a single feature can be cross-cutting.
On the right, the translation from RS-FL$_{14}$ to RS-FL$_{15}$ adds regular expression support to the Rust translation.
The Rust changes similarly span multiple files, with \code{regexp.rs} ($\sim$1.5k lines) and \code{builtin\_string.rs} ($\sim$400 lines) among the largest.

\begin{figure}[t]
\centering
\includegraphics[width=0.65\textwidth,trim=0cm 3cm 10cm 0cm]{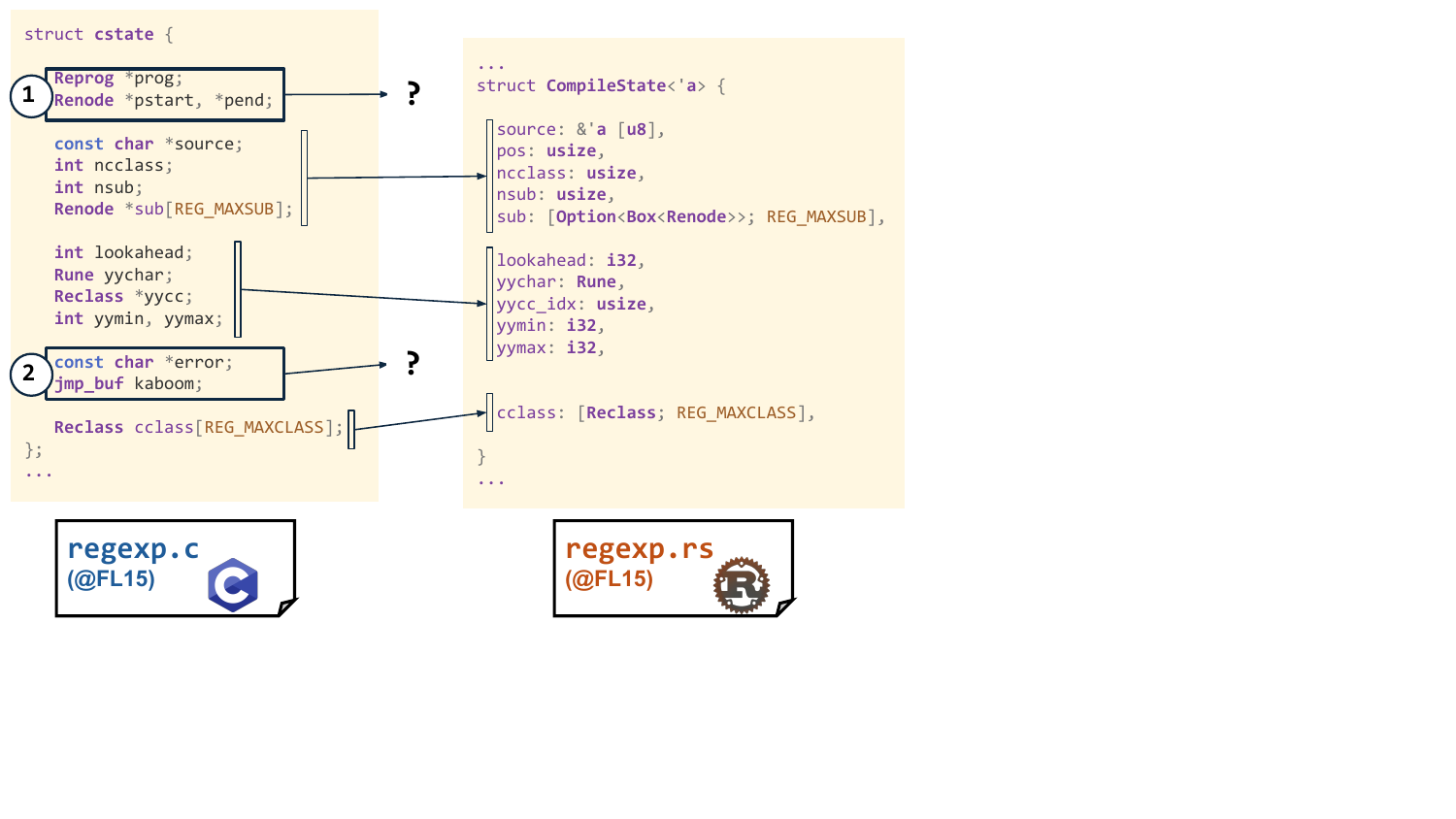}
\caption{
Comparison of the regex compilation state in C (\code{struct cstate}) and Rust (\code{CompileState}).
Corresponding fields are connected by lines.
\ding{192}~Fields for the output program and arena allocator are absent in Rust, replaced by local variables and individual heap allocations.
\ding{193}~Fields for \code{setjmp}/\code{longjmp} error handling are absent in Rust, replaced by \code{Result}-based error propagation.
}
\label{fig:overview-cstate}
\end{figure}

Beyond spanning multiple files, translating a feature may also require redesigning the program's data structures to utilize Rust language features and satisfy safe Rust constraints.
Figure~\ref{fig:overview-cstate} compares the central compilation state for the regex feature in C (\code{struct cstate}) and in the Rust translation (\code{CompileState}).
While most fields have direct correspondences, two groups of fields are absent in Rust (\ding{192} and \ding{193} in the figure).
In \ding{192}, the C struct stores a pointer to the output program (\code{prog}) and an arena allocator (\code{pstart}, \code{pend}) for parse tree nodes; these are kept in the shared state so that the error handling logic can free them manually on failure. In Rust, ownership-based resource management makes this unnecessary: the output program is a local variable, and nodes are individually heap-allocated with \code{Box}.
In \ding{193}, the C struct uses \code{setjmp}/\code{longjmp} for non-local error handling (\code{error}, \code{kaboom}); the Rust translation replaces this with \code{Result}-based error propagation~\cite{rustresulttype,setjmp}.
These data structure changes are not isolated—they propagate to every function that uses the state, affecting function signatures and error handling throughout the feature's implementation.
Feature reduction gives agents the freedom to make such restructuring decisions, since it imposes no constraints on how the Rust code is organized; the resulting program at each feature level is validated as a whole before proceeding to the next.

\subsection{Orchestration of Faulty Agents}
\label{sec:overview-orchestration}

\toadd{
\begin{table*}[t]
\centering
\caption{Agent fault types and handling mechanisms. The system observes only the worker's status, not the underlying reality. $H$ (history-feedback) and $V$ (validity) handle each status; the outcome depends on whether the worker's claim matches reality. \textbf{Escalate} means pausing and requesting external assistance (from a higher-level system or user).}
\label{tab:failure-mechanisms}
\resizebox{\textwidth}{!}{%
\begin{tabular}{@{}lllll@{}}
\toprule
\textbf{Agent Says} & \textbf{Reality} & \textbf{Type} & \textbf{Mechanism} & \textbf{Examples} \\
\midrule
\multirow{3}{*}{\texttt{BLOCKED}}
  & Actually stuck & A.1 & $H$: suggest next & Bug too complex, needs major refactoring \\
  & Not stuck & B.1 & $H$: suggest next & Agent unwilling to work \\
  & Repeated (loop) & C.1 & $H$: loop $\rightarrow$ suggest next / \textbf{Escalate} & Repeated \texttt{BLOCKED} without progress \\
\midrule
\multirow{3}{*}{\texttt{MORE\_WORK}}
  & Making progress & A.2 & $H$: progress $\rightarrow$ continue & Tests improving across attempts \\
  & No progress & B.2 & $H$: no-progress $\rightarrow$ suggest next & Agent retries without meaningful change \\
  & Repeated (loop) & C.2 & $H$: loop $\rightarrow$ suggest next / \textbf{Escalate} & Repeated \texttt{MORE\_WORK} without progress \\
\midrule
\multirow{3}{*}{\texttt{DONE}}
  & Actually done & -- & $H$$\rightarrow$$V$: valid $\rightarrow$ proceed & All tests pass, code review clean \\
  & Not done & B.3 & $H$$\rightarrow$$V$: invalid $\rightarrow$ $H$ suggest next & Tests fail, or missing test cases \\
  & Repeated (loop) & C.3 & $H$$\rightarrow$$V$: invalid $\rightarrow$ \textbf{Escalate} & Worker ignores $V$'s complaints, keeps claiming \texttt{DONE} \\
\midrule
\multirow{2}{*}{\texttt{INFEASIBLE}}
  & Actually infeasible & A.4 & $H$ $\rightarrow$ \textbf{Escalate} & No safe Rust equivalent; wrong tests \\
  & Not infeasible & B.4 & $H$ $\rightarrow$ \textbf{Escalate} & Agent claims infeasible; alternative exists \\
\midrule
\texttt{ERROR}
  & Agent failed & D & $H$: resume / restart / \textbf{Escalate} & Crash, Hang, malformed output \\
\bottomrule
\end{tabular}%
}
\end{table*}
}

An independent challenge arises in orchestration of unreliable agents.
\toadd{At each feature-level transition, a multi-agent system works toward an \emph{objective}---such as reducing the program by one feature level, or translating one feature level to Rust. The primary worker agent makes multiple attempts, reporting a \emph{status} after completion (e.g., \texttt{DONE}, \texttt{BLOCKED}). The reported status may not match \emph{reality}---what the worker actually achieved, usually due to hallucination. Any outcome other than successfully meeting the objective is a \emph{fault}. We model this problem abstractly below.}

\paragraph{Faults.}
\toadd{After each attempt, the worker reports a status indicating what it believes it has achieved toward the objective, but this status may not reflect reality.}
\toadd{Table~\ref{tab:failure-mechanisms} categorizes agent faults based on two dimensions: what the agent reports and what is actually true in reality.}
\toadd{The worker reports one of five statuses: \texttt{BLOCKED} (cannot proceed), \texttt{MORE\_WORK} (objective not yet met), \texttt{DONE} (objective met), \texttt{INFEASIBLE} (objective is fundamentally infeasible), or \texttt{ERROR} (agent crashed or produced malformed output).}
\toadd{When the status matches reality but the objective remains unmet, we have type A faults (A.1 for blocked, A.2 for needs more work, A.4 for genuinely infeasible).}
\toadd{When the status mismatches reality, we have type B faults: false blocking (B.1), misassessment of progress (B.2), false completion (B.3), or false infeasibility (B.4).}
\toadd{When the worker repeatedly reports the same status without actual progress, we have type C faults (unproductive loops): repeated blocking (C.1), repeated incomplete work with no visible progress (C.2), or repeated false completion where the worker ignores validation complaints (C.3).}
\toadd{Finally, type D faults cover agent crashes and malformed outputs.}
\toadd{Since the system only observes the worker's status, not the underlying reality, handling these faults requires additional mechanisms.}

\begin{figure}[t]
\centering
\begin{minipage}[t]{0.47\columnwidth}
\centering
{\small\sffamily\bfseries (a)~mujs RS-FL$_{15}$: Regex}\\[0.15em]
{\color{gray}\rule{0.9\linewidth}{0.4pt}}
\vspace{0.15em}

\begin{tikzpicture}
\node[text width=0.93\linewidth, font=\ttfamily\tiny, align=left, inner sep=0.2em] {
\textbf{W:} Translates core regex engine ($\sim$1.5k lines)\\[-0.1em]
\hspace{2em}Tests: 122/168. Reports \texttt{MORE\_WORK}\\[-0.1em]
\colorbox{green!15}{\textbf{$H$:} Progress detected $\rightarrow$ continue} \hfill {\color{gray}\sffamily [A.2]}\\[0.25em]
\textbf{W:} Adds RegExp prototype methods\\[-0.1em]
\hspace{2em}Tests: 129/168. Reports \texttt{MORE\_WORK}\\[-0.1em]
\colorbox{green!15}{\textbf{$H$:} Progress detected $\rightarrow$ continue} \hfill {\color{gray}\sffamily [A.2]}\\[0.2em]
\hfil\textcolor{gray}{$\cdots$ {\rmfamily\scriptsize 2 more attempts (143, 150/168)} $\cdots$}\hfil\\[0.2em]
\textbf{W:} Fixes regex error handling (try-catch)\\[-0.1em]
\hspace{2em}Tests: 168/168. Reports \texttt{DONE}\\[0.25em]
\textbf{$V$:} Tests: 168/168 passed $\checkmark$\\[-0.1em]
\hspace{2em}No cheating detected $\checkmark$\\[-0.1em]
\textbf{$V$:} Code review: 11 compiler warnings\\[-0.1em]
\colorbox{green!15}{\textbf{$H$:} Assigns investigation + fix tasks}\\[0.25em]
\textbf{W:} Fixes all 11 warnings\\[-0.1em]
\hspace{2em}(unused vars, dead code, unreachable)\\[-0.1em]
\hspace{2em}Tests: 168/168, warnings: 0\\[0.25em]
\textbf{$V$:} Re-validates: tests pass $\checkmark$\\[-0.1em]
\textbf{$V$:} Code review: 0 warnings $\checkmark$\\[0.3em]
\textcolor{gray}{\rule{0.7\linewidth}{0.3pt}}\\[0.1em]
\hfil{\rmfamily\scriptsize 14 tasks $\cdot$ 0 escalations $\cdot$ fully automatic}\hfil
};
\end{tikzpicture}

\vspace{0.8em}
{\scriptsize
\colorbox{green!15}{$H$ auto-feedback} \quad
\colorbox{yellow!30}{Escalation} \quad
\colorbox{blue!15}{User guidance}\\[0.3em]
{\sffamily\color{gray}[X.n]} fault type (Table~\ref{tab:failure-mechanisms})
}
\end{minipage}%
\hfill%
\begin{minipage}[t]{0.47\columnwidth}
\centering
{\small\sffamily\bfseries (b)~picoc RS-FL$_{15}$ + RS-FL$_{6.4}$}\\[0.15em]
{\color{gray}\rule{0.9\linewidth}{0.4pt}}
\vspace{0.15em}

\begin{tikzpicture}
\node[text width=0.93\linewidth, font=\ttfamily\tiny, align=left, inner sep=0.2em] {
{\sffamily\scriptsize\bfseries RS-FL$_{15}$: Eliminate Unsafe Blocks}\\[0.15em]
\textbf{$V$:} Code review: 87 \texttt{unsafe} blocks\\[-0.1em]
\textbf{W:} Reports \texttt{INFEASIBLE}: ``cannot eliminate'' \hfill {\color{gray}\sffamily [B.4]}\\[-0.1em]
\colorbox{yellow!30}{\textbf{$H$:} Conflicts with requirement $\rightarrow$ \textbf{ESCALATE}}\\[0.2em]
\colorbox{blue!15}{\textbf{User:} ``Use nix/chrono. Zero unsafe non-negotiable.''}\\[0.2em]
\textbf{W:} Correct approach. Eliminates unsafe blocks.\\[-0.1em]
\hspace{2em}Unsafe: 87 $\rightarrow$ 25. Reports \texttt{MORE\_WORK}\\[-0.1em]
\colorbox{green!15}{\textbf{$H$:} Progress $\rightarrow$ continue} \hfill {\color{gray}\sffamily [A.2]}\\[0.1em]
\hfil\textcolor{gray}{$\cdots$ {\rmfamily\scriptsize several iterations} $\cdots$}\hfil\\[0.1em]
\textbf{W:} Encapsulates remaining 4 in safe wrappers\\[-0.1em]
\hspace{2em}Tests: 154/154. Reports \texttt{DONE} \hfill {\color{gray}\sffamily [C.3]}\\[0.15em]
\colorbox{yellow!30}{\textbf{$H$:} $V$ rejected, worker ignores $\rightarrow$ \textbf{ESCALATE}}\\[-0.1em]
\hspace{2em}{\rmfamily\tiny Worker: ``achieves spirit of 100\% safe''}\\[-0.1em]
\hspace{2em}{\rmfamily\tiny Requirement: zero \texttt{unsafe} blocks}\\[0.2em]
\colorbox{blue!15}{\textbf{User:} ``Drop fork() support. Safe enum, not transmute.''}\\[0.2em]
\textbf{W:} Eliminates final 4 blocks\\[-0.1em]
\hspace{2em}Tests: 154/154, unsafe: 0\\[-0.1em]
\textbf{$V$:} Re-validates: all tests pass $\checkmark$\\[0.15em]
\hfil{\rmfamily\scriptsize 26 tasks $\cdot$ 2 escalations $\cdot$ $\sim$5.5h}\hfil\\[0.35em]
{\sffamily\scriptsize\bfseries RS-FL$_{6.4}$: Struct Pointer Members}\\[0.15em]
\textbf{W:} Translates struct support\\[-0.1em]
\hspace{2em}Tests: 118/118. Reports \texttt{DONE} \hfill {\color{gray}\sffamily [B.3]}\\[0.15em]
\textbf{$V$:} Checks coverage---1 test not in runner!\\[-0.1em]
\hspace{2em}Runs missing test $\rightarrow$ \textbf{FAILS}\\[-0.1em]
\hspace{2em}$\rightarrow$ Rejects \texttt{DONE}, sends back to W\\[0.15em]
\textbf{W:} Fixes struct pointer support + runner\\[-0.1em]
\hspace{2em}Tests: 119/119\\[-0.1em]
\textbf{$V$:} Re-validates: all tests pass $\checkmark$\\[0.15em]
\hfil{\rmfamily\scriptsize 14 tasks $\cdot$ 0 escalations $\cdot$ $\sim$1.5h}\hfil
};
\end{tikzpicture}
\end{minipage}

\caption{Condensed system logs (simplified from real trajectories) showing $V$ and $H$ mechanisms during translation using \tool.
\textbf{(a)}~mujs FL$_{15}$: iterative progress tracked by $H$ (A.2), multi-level validation by $V$ (tests + code review), fully automatic---no user intervention needed.
\textbf{(b)}~picoc FL$_{15}$+FL$_{6.4}$: worker claims objective infeasible (B.4, overridden by user), worker repeatedly claims completion ignoring $V$'s complaints (C.3, escalated), and $V$ detecting a missing test in the runner (B.3).
}
\label{fig:system-log}
\end{figure}
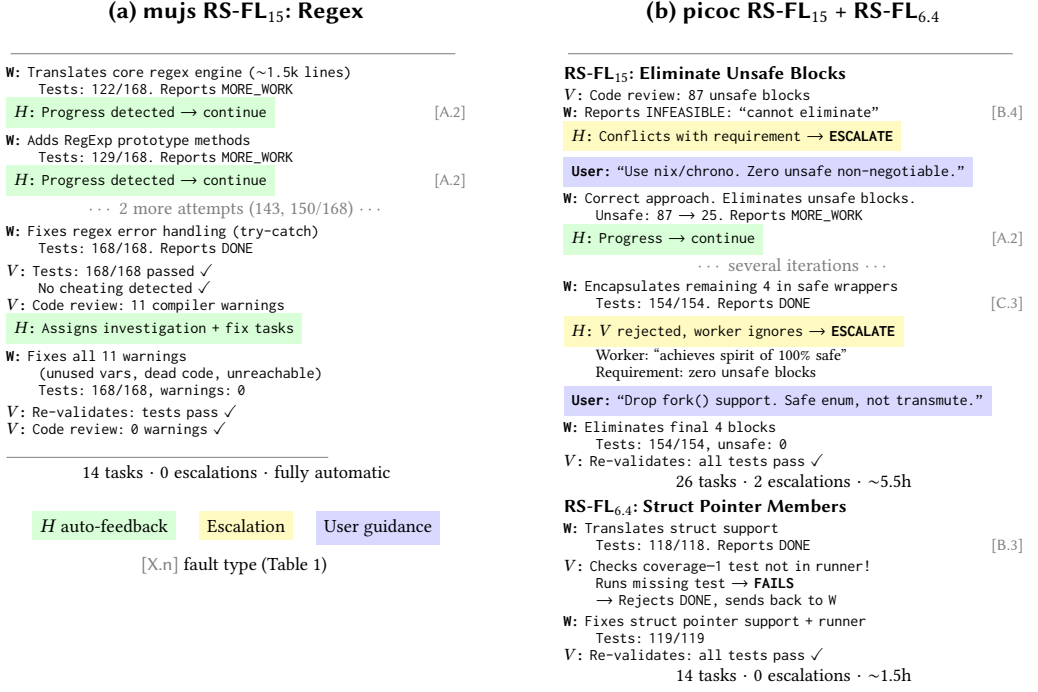

\paragraph{Validity and Progress Mechanisms.}
\toadd{The goal is to make progress toward meeting the objective. To model these requirements, we introduce two conceptual mechanisms:}
\begin{itemize}
\item \toadd{\textbf{A Validity mechanism ($V$)} determines whether the worker's output actually satisfies the objective's correctness criteria, independent of what the worker reports. For instance, $V$ can validate the output (Rust code) by compiling it, reviewing it, and running test cases.}
\item \toadd{\textbf{A History-feedback mechanism ($H$)} analyzes the history of the worker's status reports and provides targeted feedback. For example, $H$ can track patterns across attempts, detect loops, and suggest next steps to help the worker make progress or recover from errors.}
\end{itemize}
These are conceptual mechanisms that can be implemented in various ways, such as separate agent calls, rule-based checks, or a combination of both.
\toadd{Table~\ref{tab:failure-mechanisms} shows how $V$ and $H$ handle each fault type.}
For \texttt{BLOCKED} and \texttt{MORE\_WORK}, $H$ assesses the situation based on history: it tracks progress, detects loops, and either suggests next steps or escalates (A.1, B.1, C.1, A.2, B.2, C.2).
\toadd{For \texttt{DONE}, $H$ determines that validation is needed, then $V$ checks the worker's claim against correctness criteria, catching false completions (B.3). For \texttt{INFEASIBLE}, the system always escalates to the user (A.4, B.4), since this may involve clarifications of the objective or design decisions.}
For agent errors (D), $H$ attempts recovery by resuming or restarting the agent.
Both mechanisms are necessary: $V$ alone catches incorrect outputs but does not help the agent improve, while $H$ alone cannot detect correctness.
Together, $V$ ensures correctness and $H$ ensures progress.
\toadd{This model assumes that $V$ and $H$ are themselves reliable---the fault taxonomy above addresses worker unreliability only. We revisit this assumption in Section~\ref{sec:eval}.}

\paragraph{User Escalation: The last resort.}
\toadd{Not all faults can be resolved automatically.}
Issues requiring human judgment typically fall into three categories.
\toadd{First, when the worker claims the objective is \texttt{INFEASIBLE} (A.4, B.4), such as a language construct with no satisfactory safe Rust equivalent, the system always escalates since this may involve clarifications of the objective or design decisions.}
\toadd{Second, when $H$'s feedback fails to resolve repeated faults (C.1, C.2), the user may intervene to provide clarification, guide the worker to prioritize certain parts of the implementation, or adjust its working behavior.}
\toadd{Third, when unexpected faults occur that the system cannot auto-recover from (D), such as accidental deletion of critical files or termination of processes, the user may intervene to recover manually.}
When escalated, the human provides targeted guidance or clarification rather than fixing code directly; the system then continues autonomously.

In practice, our system requires only $\escawkUserInt$ to $\escpocketpyUserInt$ user interventions per program, each taking roughly $5$ minutes to address.
This enables long-running translations ($\resMinTimeRounded$ to $\resMaxTimeRounded$ hours) to proceed with minimal supervision.

Figure~\ref{fig:system-log} shows simplified trajectories from the translation process, illustrating both automatic recovery and user escalation.
\toadd{In (a), mujs FL$_{15}$, $V$ and $H$ handle all faults automatically. $H$ tracks iterative progress as tests improve from 122/168 to 168/168, and $V$ performs multi-level validation---checking both test results and code quality---before accepting the result.}
\toadd{In (b), picoc FL$_{15}$ and FL$_{6.4}$, the system encounters cases requiring escalation. A worker concludes that eliminating \code{unsafe} code is ``infeasible''; the user overrides this and provides an alternative approach. Later, the same worker repeatedly claims completion despite $V$ pointing out that \code{unsafe} blocks remain (C.3); $H$ detects the loop and escalates.} \toadd{In FL$_{6.4}$, $V$ catches a test that was missing from the test runner, preventing a false completion from going undetected.}

\section{The \tool System}
\label{sec:approach}

\toadd{\tool takes as input a C program together with a test suite, and produces a safe Rust program that passes the same tests.}
\toadd{The test suite is essential: it serves as the correctness criterion throughout the translation process, and each intermediate result is validated against a corresponding test suite (which is a subset of the C program's test suite) before proceeding.}
\toadd{The process is mostly automatic, but occasionally the system pauses to request brief user guidance---typically a clarification or some decision to be made---before continuing autonomously.}

\subsection{The \tool Translation Process}

\toadd{Algorithm~\ref{alg:reboot} presents the overall \tool process.}
\toadd{
The process is organized into three phases,
each driven by a dedicated multi-agent system (MAS) sub-routine.}
\toadd{In the Planning Phase (line 1), \textsc{MAS\_Plan} analyzes the C program and produces a feature reduction plan together with an ordered list of feature level identifiers $L = \langle l_0, l_1, \ldots, l_n \rangle$.}
\toadd{In the Feature Reduction Phase (lines 2--4), \textsc{MAS\_Reduction} is called iteratively to simplify both the program and the test suite one feature level at a time, from the full program down to the minimal version at $l_0$.}
\toadd{In the Translation Phase (lines 5--8), \textsc{MAS\_Translation} first translates the simplest C program to Rust, then iteratively restores features until reaching the full translation.}
\toadd{The Planning Phase is the simplest of the three: \textsc{MAS\_Plan} executes a linear sequence of agent calls that read the source code, identify features, determine their dependencies, and produce a plan document---with no iterative validation loop. Summaries of the resulting feature level plans for each benchmark are provided in the appendix. The remainder of this section focuses on \textsc{MAS\_Reduction} and \textsc{MAS\_Translation}, which employ iterative multi-agent workflows.} %

\begin{algorithm}[t]
\caption{The \tool Translation Process}
\label{alg:reboot}
\begin{algorithmic}[1]
\Statex \textbf{Input:} $C_{\mathit{src}}$ (C program), $T$ (test suite)
\Statex \textbf{Output:} $\mathit{RS}_{\mathit{src}}$ (safe Rust program)
\Statex \textit{// Phase 1: Planning}
\State $(\mathit{plan},\, L) \gets \textsc{MAS\_Plan}(C_{\mathit{src}}, T)$
\Statex \hspace{\algorithmicindent}\textit{// $L = \langle l_0, l_1, \ldots, l_n \rangle$: feature level IDs, sorted}
\Statex \textit{// Phase 2: Feature Reduction}
\State $C[l_n] \gets C_{\mathit{src}};\; T[l_n] \gets T$
\For{$i = n$ \textbf{down to} $1$}
    \State $(C[l_{i-1}],\, T[l_{i-1}]) \gets$
    $\textsc{MAS\_Reduction}(C[l_i], T[l_i], \mathit{plan}, l_{i-1})$
\EndFor
\Statex \textit{// Phase 3: Translation}
\State $\mathit{RS}[l_0] \gets \textsc{MAS\_Translation}(C[l_0], T[l_0], \emptyset)$
\For{$i = 1$ \textbf{to} $n$}
    \State $\mathit{RS}[l_i] \gets \textsc{MAS\_Translation}(C[l_i], T[l_i], \mathit{RS}[l_{i-1}])$
\EndFor
\State \Return $\mathit{RS}[l_n]$
\end{algorithmic}
\end{algorithm}

\begin{figure*}[t]
\centering
\resizebox{\textwidth}{!}{\usetikzlibrary{positioning,calc,fit}
\begin{tikzpicture}[
    node distance=0.4cm and 0.6cm,
    task/.style={rectangle, draw, rounded corners=3pt, minimum width=1.8cm, minimum height=0.8cm,
                 align=center, font=\tiny, fill=white},
    outcome/.style={rectangle, draw, minimum width=1.2cm, minimum height=0.5cm,
                    align=center, font=\tiny, fill=black, text=white},
    state/.style={rectangle, draw, rounded corners, minimum width=1.4cm, minimum height=0.8cm,
                  align=center, font=\tiny, fill=white},
    arrow/.style={->, >=stealth, rounded corners=3pt},
    fsmarrow/.style={->, >=stealth, thick},
    label/.style={font=\tiny},
    worker/.style={font=\normalsize\bf},
    rolelabel/.style={rectangle, draw, rounded corners=4pt, minimum width=1.0cm, minimum height=0.45cm,
                      align=center, font=\normalsize\bf, inner sep=2pt}
]

\node[rectangle, draw, fill=black, text=white, minimum width=0.8cm, minimum height=0.4cm, font=\tiny] (start) at (0, -0.3) {INIT};

\node[task] (precheck) at (0, -1.5) {Precheck};
\node[outcome] (pre_ok) at (1.8, -1.5) {OK};

\node[task] (simplify) at (1.8, -3.5) {Simplify};
\node[outcome] (simp_more) at (3.8, -3) {MORE\_WORK};
\node[outcome] (simp_ok) at (3.8, -4) {DONE};

\node[task] (fix) at (6, -3.5) {Fix Issues};
\node[outcome] (fix_more) at (8, -3) {MORE\_WORK};
\node[outcome] (fix_ok) at (8, -4) {DONE};

\node[task] (check) at (3.8, -5.5) {Check Result};
\node[outcome] (check_fail) at (6, -5) {INVALID};
\node[outcome] (check_pass) at (6, -6) {VALID};

\node[task] (cleanup) at (6, -7.5) {Cleanup};
\node[outcome] (clean_ok) at (8, -7.5) {END};

\begin{scope}[on background layer]
    \fill[black!6] (-2.6, 0.1) rectangle (10.8, -0.7);
    \fill[black!6] (-2.6, -0.7) rectangle (-1.6, -8.2);
    \fill[blue!8] (-1.4, -0.9) rectangle (10.8, -2.1);
    \fill[blue!8] (-1.4, -4.7) rectangle (10.8, -6.4);
    \fill[orange!10] (-1.4, -2.5) rectangle (10.8, -4.4);
\end{scope}

\draw[arrow] (start.south) -- (precheck.north);
\draw[arrow] (precheck.east) -- (pre_ok.west);

\draw[arrow] (pre_ok.south) -- (simplify.north);

\draw[arrow] (simplify.east) -- (simp_more.west);
\draw[arrow] (simplify.east) -- (simp_ok.west);

\draw[arrow] (simp_ok.south) -- (check.north);

\draw[arrow] (simp_more.north) -- (3.8,-2.5) -- (2.1,-2.5) -- ($(simplify.north)+(0.3,0)$);

\draw[arrow] (check.east) -- (check_fail.west);
\draw[arrow] (check.east) -- (check_pass.west);

\draw[arrow] (check_fail.north) -- (fix.south);

\draw[arrow] (fix.east) -- (fix_more.west);
\draw[arrow] (fix.east) -- (fix_ok.west);

\draw[arrow] (fix_ok.south) -- (8,-4.5) -- (4.5,-4.5) -- (4.5,-5.15) -- ($(check.north)+(0.7,0)$);

\draw[arrow] (fix_more.north) -- (8,-2.5) -- (6.3,-2.5) -- ($(fix.north)+(0.3,0)$);

\draw[arrow] (check_pass.south) -- (cleanup.north);

\draw[arrow] (cleanup.east) -- (clean_ok.west);

\def\labelx{10.2}
\node[rolelabel, fill=black!6, text=black!50] at (\labelx, -0.3) {H};
\node[rolelabel, fill=blue!8, text=blue!50] at (\labelx, -1.5) {V};
\node[rolelabel, fill=orange!10, text=orange!60] at (\labelx, -3.5) {W};
\node[rolelabel, fill=blue!8, text=blue!50] at (\labelx, -5.5) {V};
\node[rolelabel, fill=white, text=black!40] at (\labelx, -7.5) {C};
\node[font=\tiny, text=black!50, anchor=north] at (\labelx, -8.1) {Agent Roles};

\draw[dashed, gray] (11, -0.1) -- (11, -8.2);

\def\fsmx{13.5}

\node[state] (fs0) at (\fsmx, -0.3) {Initial\\allow=\{V\}};
\node[state] (fs1) at (\fsmx, -1.5) {Prechecked\\last=V\\allow=\{W,V\}};
\node[state] (fworking) at (\fsmx, -3.5) {Simplifying\\last=W\\allow=\{W,V\}};
\node[state] (fvalidated) at (\fsmx, -5.5) {Validated\\last=V\\allow=\{W,V,C\}};
\node[state] (fcleaned) at (\fsmx, -7.5) {Cleaned\\C seen\\allow=\{\}};

\draw[fsmarrow] (fs0) -- node[label, left] {V} (fs1);
\draw[fsmarrow] (fs1) -- node[label, left] {W} (fworking);
\draw[fsmarrow] (fworking) -- node[label, left] {V} (fvalidated);
\draw[fsmarrow] (fvalidated) -- node[label, left] {C} (fcleaned);

\draw[fsmarrow] (fs1.160) to[out=160, in=200, looseness=4] node[label, left] {V} (fs1.200);
\draw[fsmarrow] (fworking.160) to[out=160, in=200, looseness=4] node[label, left] {W} (fworking.200);
\draw[fsmarrow] (fvalidated.160) to[out=160, in=200, looseness=4] node[label, left] {V} (fvalidated.200);

\draw[fsmarrow] (fvalidated.east) to[out=0, in=0, looseness=1.5] node[label, right] {W} (fworking.east);

\node[draw, dashed, fit=(fworking)(fvalidated), inner sep=0.3cm, label={[font=\tiny, xshift=-0.3cm]right:{\rotatebox{-90}{Simplify-Validate Loop}}}] {};

\node[font=\normalsize, anchor=north] at (4, -8.5) {Workflow Diagram (\textsc{MAS\_Reduction})};
\node[font=\normalsize, anchor=north] at (\fsmx, -8.5) {FSM Guard};

\end{tikzpicture}}
\caption{Branch-level workflow control for the Simplification phase. \emph{Left:} The workflow provided as instructions to the manager agent, with task sequences and manager assessment outcomes. \emph{Right:} FSM guard that enforces valid worker sequences. \textbf{Legend:} Rounded boxes = Worker tasks; Black boxes = Manager assessments.}
\label{fig:branch-workflows-reduction}%
\phantomsection\label{fig:workflow-reduction-ideal}%
\phantomsection\label{fig:workflow-reduction-guard}%
\end{figure*}
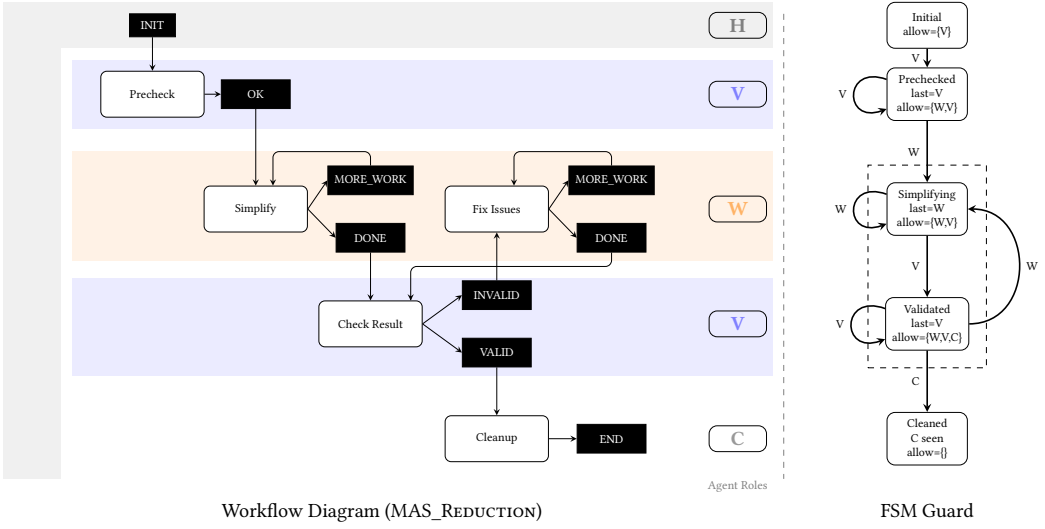

\toadd{Each invocation of \textsc{MAS\_Reduction} or \textsc{MAS\_Translation} encapsulates a multi-agent system that returns only after producing a validated artifact.}
\toadd{The intermediate artifacts include: the feature reduction plan produced by \textsc{MAS\_Plan}, the simplified C programs $C[l_i]$ and adapted test suites $T[l_i]$ produced by \textsc{MAS\_Reduction}, and the Rust translations $\mathit{RS}[l_i]$ produced by \textsc{MAS\_Translation}.}
\toadd{Because each feature level is an independent, validated milestone, faults during the translation of one feature level do not propagate to others.}

\subsection{The Feature Reduction Sub-system}
\label{sec:mas-reduction}

\toadd{\textsc{MAS\_Reduction} uses a multi-agent workflow with three workers---Simplifier (W), Validator (V), and Cleanup (C)---coordinated by a Manager agent (marked as H in Figure~\ref{fig:workflow-reduction-ideal}).}
\toadd{The Simplifier removes all code that implements the features being dropped---i.e., features that should no longer exist in the target feature level---and adapts the test suite accordingly.}
\toadd{The Validator runs the adapted test suite and checks that test coverage is preserved.}
\toadd{The workflow has four stages: a \emph{precheck} stage where the Validator establishes baseline test coverage, a \emph{simplification} stage where the Simplifier removes the target feature, a \emph{validation} stage where the Validator checks that all tests pass and coverage is preserved, and a \emph{cleanup} stage where Cleanup prepares the result for commit.}
\toadd{Coverage serves as an auxiliary sanity check on the reduction: we expect coverage to stay roughly the same after a clean reduction, and a significant drop indicates either incomplete code removal while tests have already been dropped, or accidental removal of tests that also cover remaining features.}
\toadd{The main loop is between simplification and validation: when validation finds issues, the Manager provides feedback and routes back to the Simplifier to address them before re-validating.}

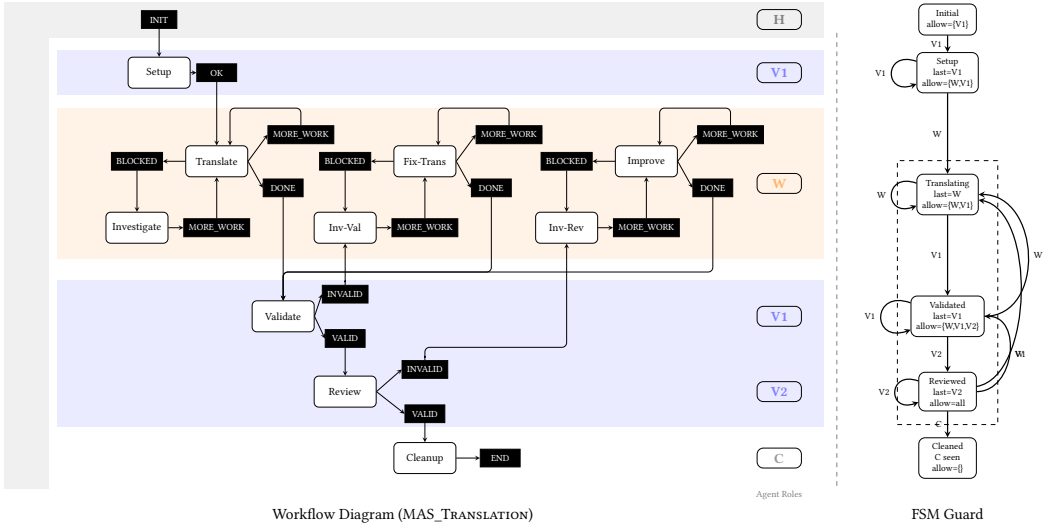
\begin{figure*}[t]
\centering
\resizebox{\textwidth}{!}{\usetikzlibrary{positioning,calc,fit}
\begin{tikzpicture}[
    node distance=0.3cm and 0.5cm,
    task/.style={rectangle, draw, rounded corners=3pt, minimum width=1.4cm, minimum height=0.7cm,
                 align=center, font=\scriptsize, fill=white},
    outcome/.style={rectangle, draw, minimum width=0.9cm, minimum height=0.4cm,
                    align=center, font=\tiny, fill=black, text=white},
    state/.style={rectangle, draw, rounded corners, minimum width=1.3cm, minimum height=0.7cm,
                  align=center, font=\tiny, fill=white},
    arrow/.style={->, >=stealth, rounded corners=3pt},
    fsmarrow/.style={->, >=stealth, thick},
    label/.style={font=\tiny},
    worker/.style={font=\small\bf},
    rolelabel/.style={rectangle, draw, rounded corners=4pt, minimum width=1.0cm, minimum height=0.45cm,
                      align=center, font=\small\bf, inner sep=2pt}
]

\node[rectangle, draw, fill=black, text=white, minimum width=0.8cm, minimum height=0.4cm, font=\tiny] (start) at (2, 0.2) {INIT};

\node[task] (setup) at (2, -1) {Setup};
\node[outcome] (setup_ok) at (3.3, -1) {OK};

\node[task] (trans) at (3.3, -3) {Translate};
\node[outcome] (trans_inv) at (1.5, -3) {BLOCKED};
\node[outcome] (trans_more) at (5.2, -2.4) {MORE\_WORK};
\node[outcome] (trans_ok) at (4.8, -3.6) {DONE};

\node[task] (inv) at (1.5, -4.5) {Investigate};
\node[outcome] (inv_ok) at (3.3, -4.5) {MORE\_WORK};

\node[task] (fixtrans) at (8, -3) {Fix-Trans};
\node[outcome] (fix_inv) at (6.2, -3) {BLOCKED};
\node[outcome] (fix_more) at (9.9, -2.4) {MORE\_WORK};
\node[outcome] (fix_ok) at (9.5, -3.6) {DONE};

\node[task] (invval) at (6.2, -4.5) {Inv-Val};
\node[outcome] (invval_ok) at (8, -4.5) {MORE\_WORK};

\node[task] (improve) at (13, -3) {Improve};
\node[outcome] (imp_inv) at (11.2, -3) {BLOCKED};
\node[outcome] (imp_more) at (14.9, -2.4) {MORE\_WORK};
\node[outcome] (imp_ok) at (14.5, -3.6) {DONE};

\node[task] (invrev) at (11.2, -4.5) {Inv-Rev};
\node[outcome] (invrev_ok) at (13, -4.5) {MORE\_WORK};

\node[task] (valid) at (4.8, -6.5) {Validate};
\node[outcome] (val_iss) at (6.2, -6) {INVALID};
\node[outcome] (val_ok) at (6.2, -7) {VALID};

\node[task] (review) at (6.2, -8.2) {Review};
\node[outcome] (rev_imp) at (8.0, -7.7) {INVALID};
\node[outcome] (rev_ok) at (8.0, -8.7) {VALID};

\node[task] (cleanuptask) at (8.0, -9.7) {Cleanup};
\node[outcome] (clean_ok) at (9.7, -9.7) {END};

\begin{scope}[on background layer]
    \fill[black!6] (-1.5, 0.6) rectangle (17.0, -0.2);
    \fill[black!6] (-1.5, -0.2) rectangle (-0.5, -10.4);
    \fill[blue!8] (-0.3, -0.5) rectangle (17.0, -1.5);
    \fill[blue!8] (-0.3, -5.7) rectangle (17.0, -9.0);
    \fill[orange!10] (-0.3, -1.8) rectangle (17.0, -5.2);
\end{scope}

\draw[arrow] (start.south) -- (setup.north);
\draw[arrow] (setup.east) -- (setup_ok.west);

\draw[arrow] (setup_ok.south) -- (trans.north);

\draw[arrow] (trans.west) -- (trans_inv.east);

\draw[arrow] (trans_inv.south) -- (inv.north);

\draw[arrow] (inv.east) -- (inv_ok.west);

\draw[arrow] (inv_ok.north) -- (trans.south);

\draw[arrow] (trans.east) -- (trans_more.west);
\draw[arrow] (trans.east) -- (trans_ok.west);

\draw[arrow] (trans_ok.south) -- (valid.north);

\draw[arrow] (valid.east) -- (val_iss.west);
\draw[arrow] (valid.east) -- (val_ok.west);

\draw[arrow] (val_iss.north) -- (6.2,-5.8) -- (invval.south);

\draw[arrow] (fixtrans.west) -- (fix_inv.east);

\draw[arrow] (fix_inv.south) -- (invval.north);

\draw[arrow] (invval.east) -- (invval_ok.west);

\draw[arrow] (invval_ok.north) -- (fixtrans.south);

\draw[arrow] (fixtrans.east) -- (fix_more.west);
\draw[arrow] (fixtrans.east) -- (fix_ok.west);

\draw[arrow] (fix_ok.south) -- (9.5,-5.5) -- (4.8,-5.5) -- (valid.north);

\draw[arrow] (val_ok.south) -- (review.north);

\draw[arrow] (review.east) -- (rev_imp.west);
\draw[arrow] (review.east) -- (rev_ok.west);

\draw[arrow] (rev_imp.north) -- (8.0,-7.5) -- (8.0,-7.2) -- (11.2,-7.2) -- (11.2,-5.6) -- (invrev.south);

\draw[arrow] (improve.west) -- (imp_inv.east);

\draw[arrow] (imp_inv.south) -- (invrev.north);

\draw[arrow] (invrev.east) -- (invrev_ok.west);

\draw[arrow] (invrev_ok.north) -- (improve.south);

\draw[arrow] (improve.east) -- (imp_more.west);
\draw[arrow] (improve.east) -- (imp_ok.west);

\draw[arrow] (imp_ok.south) -- (14.5,-5.5) -- (4.8,-5.5) -- (valid.north);

\draw[arrow] (rev_ok.south) -- (8.0,-9.0) -- (cleanuptask.north);

\draw[arrow] (cleanuptask.east) -- (clean_ok.west);

\draw[arrow] (trans_more.north) -- (5.2,-1.8) -- (3.6,-1.8) -- ($(trans.north)+(0.3,0)$);
\draw[arrow] (fix_more.north) -- (9.9,-1.8) -- (8.3,-1.8) -- ($(fixtrans.north)+(0.3,0)$);
\draw[arrow] (imp_more.north) -- (14.9,-1.8) -- (13.3,-1.8) -- ($(improve.north)+(0.3,0)$);

\def\labelx{16.0}
\node[rolelabel, fill=black!6, text=black!50] at (\labelx, 0.2) {H};
\node[rolelabel, fill=blue!8, text=blue!50] at (\labelx, -1) {V1};
\node[rolelabel, fill=orange!10, text=orange!60] at (\labelx, -3.5) {W};
\node[rolelabel, fill=blue!8, text=blue!50] at (\labelx, -6.5) {V1};
\node[rolelabel, fill=blue!8, text=blue!50] at (\labelx, -8.2) {V2};
\node[rolelabel, fill=white, text=black!40] at (\labelx, -9.7) {C};
\node[font=\tiny, text=black!50, anchor=north] at (\labelx, -10.3) {Agent Roles};

\draw[dashed, gray] (17.3, -0.1) -- (17.3, -10.4);

\def\fsmx{19.8}

\node[state] (fs0) at (\fsmx, 0.2) {Initial\\allow=\{V1\}};
\node[state] (fsetup) at (\fsmx, -1) {Setup\\last=V1\\allow=\{W,V1\}};
\node[state] (ftrans) at (\fsmx, -3.75) {Translating\\last=W\\allow=\{W,V1\}};
\node[state] (fvalid) at (\fsmx, -6.5) {Validated\\last=V1\\allow=\{W,V1,V2\}};
\node[state] (freview) at (\fsmx, -8.2) {Reviewed\\last=V2\\allow=all};
\node[state] (fclean) at (\fsmx, -9.7) {Cleaned\\C seen\\allow=\{\}};

\draw[fsmarrow] (fs0) -- node[label, left] {V1} (fsetup);
\draw[fsmarrow] (fsetup) -- node[label, left] {W} (ftrans);
\draw[fsmarrow] (ftrans) -- node[label, left] {V1} (fvalid);
\draw[fsmarrow] (fvalid) -- node[label, left] {V2} (freview);
\draw[fsmarrow] (freview) -- node[label, left] {C} (fclean);

\draw[fsmarrow] (fsetup.160) to[out=160, in=200, looseness=4] node[label, left] {V1} (fsetup.200);
\draw[fsmarrow] (ftrans.160) to[out=160, in=200, looseness=4] node[label, left] {W} (ftrans.200);
\draw[fsmarrow] (fvalid.160) to[out=160, in=200, looseness=4] node[label, left] {V1} (fvalid.200);
\draw[fsmarrow] (freview.160) to[out=160, in=200, looseness=4] node[label, left] {V2} (freview.200);

\draw[fsmarrow] (fvalid.east) to[out=0, in=0, looseness=1.3] node[label, right] {W} (ftrans.east);
\draw[fsmarrow] (freview.10) to[out=10, in=350, looseness=0.8] node[label, right, pos=0.25] {W} (ftrans.350);
\draw[fsmarrow] (freview.east) to[out=0, in=0, looseness=1.3] node[label, right] {V1} (fvalid.east);

\node[draw, dashed, fit=(ftrans)(fvalid)(freview), inner sep=0.3cm, label={[font=\tiny, xshift=-0.2cm]right:{\rotatebox{-90}{Work-Review Loop}}}] {};

\node[font=\normalsize, anchor=north] at (7.5, -10.7) {Workflow Diagram (\textsc{MAS\_Translation})};
\node[font=\normalsize, anchor=north] at (\fsmx, -10.7) {FSM Guard};

\end{tikzpicture}}
\caption{Branch-level workflow control for the Translation phase. \emph{Left:} The workflow provided as instructions to the manager agent, with task sequences and manager assessment outcomes. \emph{Right:} FSM guard that enforces valid worker sequences. \textbf{Legend:} Rounded boxes = Worker tasks; Black boxes = Manager assessments.}
\label{fig:branch-workflows-translation}%
\phantomsection\label{fig:workflow-translation-ideal}%
\phantomsection\label{fig:workflow-translation-guard}%
\end{figure*}

\subsection{The Translation Sub-system}
\label{sec:mas-translation}
\toadd{\textsc{MAS\_Translation} employs four worker agents coordinated by a Manager agent (marked as H in Figure~\ref{fig:workflow-translation-ideal}).}
\toadd{The Translator (W) is the primary worker that produces Rust code from C, guided by the previous level's translation when available.}
\toadd{The Validator (V1) runs the test suite and detects cheating such as hardcoded return values or skipped tests.}
\toadd{The CodeReviewer (V2) checks that the translation uses safe Rust with no \code{unsafe} blocks, maintains code quality, preserves modular structure, and double-checks whether the Validator's results are correct.}
\toadd{The Cleanup (C) prepares the validated result for commit.}
\toadd{The Manager decides which worker to call next based on their status reports, and provides context, additional instructions, or hints to guide each worker.}

\toadd{The workflow proceeds through four stages.}
\toadd{In the \emph{setup} stage, the Validator checks the test infrastructure and validates the execution environment, making sure it is ready for the Translator.}
\toadd{In the \emph{translation} stage, the Translator produces or updates the Rust code; when it encounters complex issues, investigation sub-stages are triggered for deeper analysis before attempting fixes.}
\toadd{In the \emph{validation} stage, the Validator runs tests to check equivalence with the C program, and the CodeReviewer checks code quality; both must pass before proceeding.}
\toadd{In the \emph{cleanup} stage, the result is prepared for commit.}
\toadd{The main loop is between translation and validation: when either test validation or code review finds issues, the Manager routes back to the Translator to address them, then re-validates.}

\toadd{The multi-agent design described above implements the conceptual framework from Section~\ref{sec:overview}.}
\toadd{The Validator and CodeReviewer together realize the validity mechanism $V$, checking correctness of outputs independently of the primary worker's claims.}
\toadd{The Manager realizes the history-feedback mechanism $H$, analyzing the history of worker attempts and providing targeted feedback to guide progress.}
\toadd{This mapping holds uniformly across both \textsc{MAS\_Reduction} and \textsc{MAS\_Translation}, with the Simplifier or Translator serving as the primary worker whose outputs are supervised by $V$ and $H$.}

\subsection{Detection and Handling of Faults}

\toadd{While the Manager (H) makes autonomous decisions about which worker to call next, a rule-based controller further enforces safety constraints on the worker call sequence.}
\toadd{These constraints are expressed as finite state machines (FSMs), shown in Figures~\ref{fig:workflow-reduction-guard} and~\ref{fig:workflow-translation-guard}.}
\toadd{Key invariants include: validation must occur before code review, code review must occur before cleanup, and no further work is permitted after cleanup.}
\toadd{If the Manager proposes a worker call that violates these constraints, the controller rejects the call and prompts the Manager to reconsider. Together, the Manager and the controller implement the history-feedback mechanism $H$ from Section~\ref{sec:overview-orchestration}: the Manager provides feedback based on agent history, and the controller enforces structural invariants.}

\toadd{As described in Section~\ref{sec:overview-orchestration}, the system escalates to a human when repeated retries yield no progress, when the worker claims the objective is \texttt{INFEASIBLE}, or when unexpected faults cannot be automatically recovered.}
\toadd{At the implementation level, the Manager tracks consecutive retry counts for each stage and escalates after a configurable number of rounds (typically 5--10) without progress.}

\section{Implementation}
\label{sec:impl}

\myparagraph{LLM and Agent.}
\toadd{We implement \tool using Claude Code (version 2.0.25) as the agent framework, with Claude Sonnet 4.5 as the underlying language model.}
\toadd{Each agent—Translator, Simplifier, Validator, CodeReviewer, Cleanup, and Manager—is an instance of Claude Code configured with role-specific prompts and permissions.}
\toadd{Claude Code provides built-in capabilities for file manipulation, command execution, and iterative refinement, which we leverage for code generation and validation tasks.}

\myparagraph{Agent Prompting.}
\toadd{The system uses approximately 30 prompt files in markdown, organized by worker role and by phase.}
\toadd{For planning, a sequence of prompts guides the agent through feature identification, dependency analysis, and ordering to produce the feature reduction plan.}
\toadd{For reduction and translation, each worker role has per-stage prompt files: translation instructions for the Translator, validation criteria for the Validator, and a review checklist for the CodeReviewer.}
\toadd{The Manager is instructed to follow the workflows shown in Figures~\ref{fig:workflow-reduction-ideal} and~\ref{fig:workflow-translation-ideal}, and to analyze worker status reports and provide targeted feedback accordingly.}

\begin{figure}[htb]
\centering
\includegraphics[width=0.6\linewidth,trim=0.5cm 3cm 8cm 0cm]{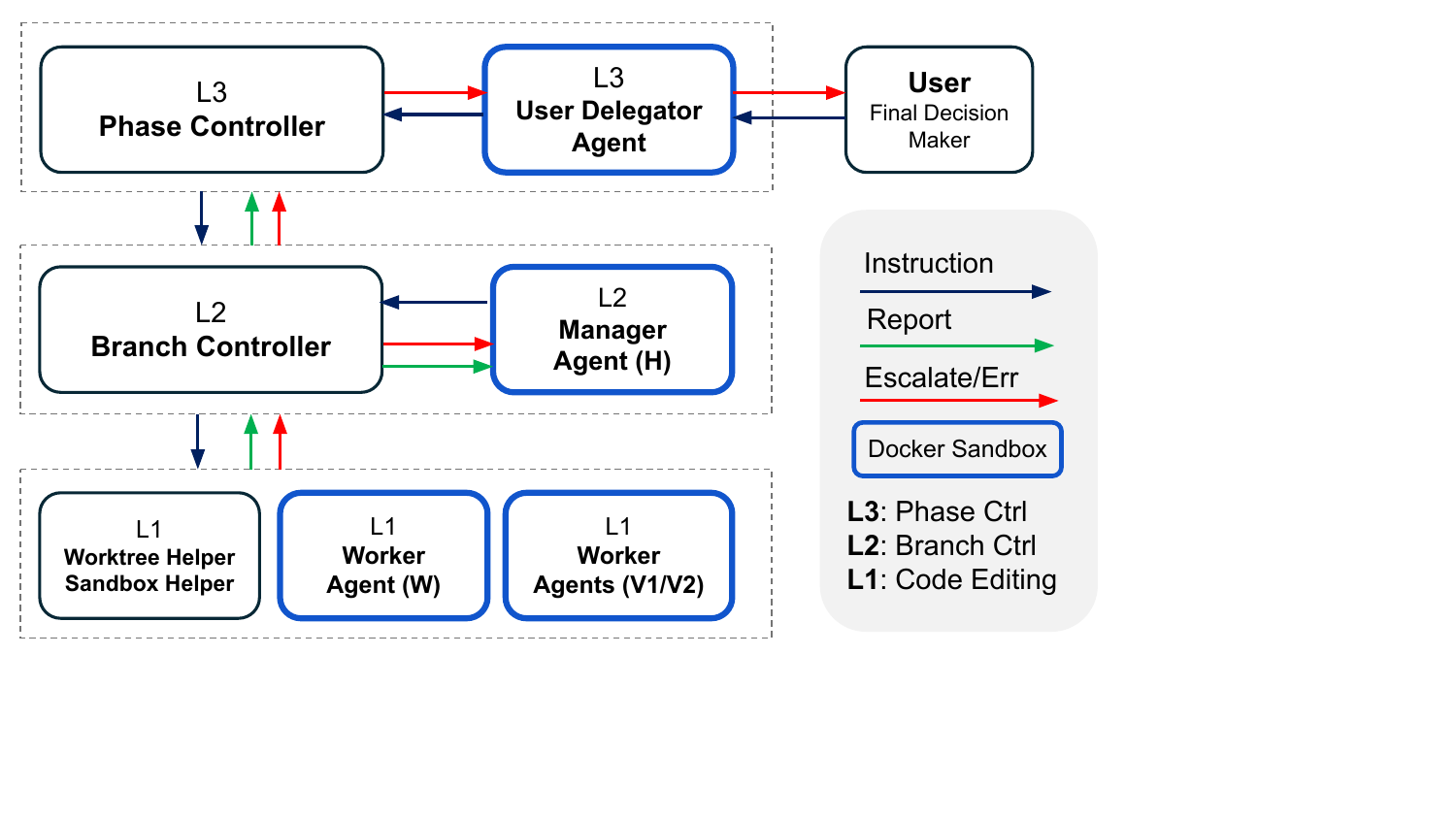}  
\caption{\toadd{Three-level implementation architecture of \tool. The L3 Phase Controller is long-lived and manages the overall three-phase process. L2 components (Branch Controller and Manager Agent) are created for each feature level. L1 Worker Agents persist for the duration of that feature level to handle multiple tasks, and get restarted when faulty. When the Manager escalates an issue, it reaches the User Delegator Agent at L3, which auto-resolves common patterns and forwards only unresolved cases to the human user.} 
}
\label{fig:arch}
\end{figure}

\myparagraph{System Architecture.}
\toadd{Figure~\ref{fig:arch} shows the three-level architecture. At L3 (phase level), the Phase Controller drives the overall process from Algorithm~\ref{alg:reboot}, iterating over feature levels and invoking the appropriate MAS operator for each transition. Also at L3, the User Delegator Agent intercepts escalations from the Manager before they reach the human user, auto-resolving common patterns such as output format mismatches, transient API errors, and iteration-limit resets. At L2 (branch level), a Branch Controller and a Manager Agent are created for each feature-level transition; the Branch Controller proxies messages between agents, enforces the FSM guards from Section~\ref{sec:approach}, and manages git worktrees. At L1, Worker Agents (Translator, Simplifier, Validator, CodeReviewer, Cleanup) operate within the worktree for that branch. Communication flows downward as instructions (blue), upward as reports (green), and upward as escalations (red) when $H$ cannot resolve an issue automatically, or as errors (red) when an agent or a controller is failing.}

\myparagraph{Sandboxing and Isolation.}
\toadd{Each agent runs in its own Docker container with restricted permissions.}
\toadd{The Manager agent has read-only access to the codebase, consistent with its role as $H$—it observes and analyzes but does not modify code directly.}
\toadd{Worker agents (Translator, Simplifier, Validator, CodeReviewer, Cleanup) have read/write access only to the worktree folder for their current branch, plus read-only access to the git repository.}
\toadd{The control infrastructure—branch-level and repo-level controllers—runs on the host outside the agent containers, ensuring that agents cannot interfere with orchestration logic.}
\toadd{All agent logging is also performed on the host, preventing agents from tampering with their own logs.}

\myparagraph{Git-based State Management.}
\toadd{Git worktree management runs on the host, not inside agent containers.}
\toadd{A separate git branch is created for each feature level transition, isolating work across different levels.}
\toadd{The system auto-commits tracked files after each agent step, providing backups of intermediate progress.}
\toadd{After validation and review pass, a cleaned-up final commit is created for each branch.}
\toadd{These backups enable recovery when failures occur: if some files are corrupted, the Manager can instruct the worker to retrieve previous versions from backup commits and continue (soft recovery); if the worktree is not recoverable after escalation, the system can roll back to the last good commit or reset to the previous feature level's branch (hard recovery).}

\myparagraph{System Size.}
\toadd{The main \tool system comprises approximately 6k lines of Python code: $\sim$2k for the branch-level controller, $\sim$1k for the repo-level controller, and $\sim$2k for worktree-level helper scripts.}
\toadd{In addition, approximately 30 prompt files in markdown define the instructions for each worker role and phase.}
\toadd{Auxiliary scripts for sandboxing, environment setup, and result analysis add another $\sim$5k lines of code.}

\section{Evaluation}
\label{sec:eval}

\toadd{We evaluate \tool on four research questions:}
\begin{itemize}
\item \toadd{\textbf{RQ1 (Effectiveness):} Can \tool produce safe Rust translations of real-world C interpreter programs that pass a given test suite?}
\item \toadd{\textbf{RQ2 (Efficiency):} How much time, cost, and human effort does the translation require?}
\item \toadd{\textbf{RQ3 (Correctness beyond provided tests):} To what extent do the Rust translations pass unseen validation tests?}
\item \toadd{\textbf{RQ4 (Comparison with existing approaches):} How does \tool compare with existing C-to-Rust translation tools?}
\end{itemize}
\toadd{We also conduct five case studies: security improvements in the translated code (CS1), runtime performance of the translations compared with the original C programs (CS2), an ablation study on the benefits of feature reduction (CS3), user interventions required during translation (CS4), and the applicability of \tool to command-line programs beyond interpreters (CS5).}

\myparagraph{Benchmarks.}
\toadd{Table~\ref{tab:benchmarks} lists the six open-source projects, which are interpreters written in C used in our evaluation.}
\toadd{We identified candidates by searching for popular open-source interpreter projects written in C, and selected those that are standalone (can be compiled and tested independently) and free of complex external dependencies.}
\toadd{The programs range from $\sim$\bmMinLocRounded{} to $\sim$\bmMaxLocRounded{} lines of code, interpreting diverse source languages including Awk, JavaScript (ES5), a C subset, an arbitrary-precision calculator language, Wren, and a Python subset.}
\toadd{Some programs come with their own test suites, while others have limited test coverage; for the latter, we supplemented additional tests to ensure all benchmarks achieve at least 70\% line coverage on the C source.}
\toadd{These provided tests are the tests that \tool uses during translation; correctness of the final translation is measured by passing all of them.}

\toadd{
\begin{table}[htb]
    \centering
    \caption{Benchmark interpreter programs written in C.}
    \resizebox{0.7\linewidth}{!}{%
    \begin{tabular}{l r r r l}
        \toprule
        \textbf{Program} & \textbf{LoC} & \textbf{\#Tests} & \textbf{Coverage} & \textbf{Description} \\
        \midrule
        awk       & \bmawkLoc  & \bmawkTests & \bmawkCovPct & One True Awk Interpreter \\
        gnu-bc    & \bmgnubcLoc  & \bmgnubcTests & \bmgnubcCovPct & Arbitrary Precision Calculator \\
        picoc     & \bmpicocLoc  & \bmpicocTests & \bmpicocCovPct & C Subset Interpreter \\
        wren      & \bmwrenLoc  & \bmwrenTests & \bmwrenCovPct & Wren Language Interpreter \\
        mujs      & \bmmujsLoc & \bmmujsTests & \bmmujsCovPct & JavaScript Interpreter (ES5) \\
        pocketpy  & \bmpocketpyLoc & \bmpocketpyTests & \bmpocketpyCovPct & Python Subset Interpreter \\
        \bottomrule
    \end{tabular}
    }
    \label{tab:benchmarks}
\end{table}
}

\myparagraph{Setup.}
\toadd{Each program is translated end-to-end using \tool with the implementation described in Section~\ref{sec:impl} (Claude Code with Claude Sonnet 4.5).}
\toadd{Each result represents a single translation run; we do not perform repeated experiments due to the relatively high monetary cost.}
\toadd{When the system escalates to a human (Section~\ref{sec:overview}), one of the authors handles the escalation as the human-in-the-loop.}
\toadd{We measure wall-clock translation time (the system's total running time, excluding time spent waiting for the user during escalations), monetary cost (LLM API usage), and the number of user interventions required.}
\toadd{To evaluate correctness beyond the provided tests, we independently created validation test suites for each program; these tests were hidden from \tool during the entire translation process.}

\subsection{RQ1 \& RQ2: Effectiveness and Efficiency}
\toadd{Table~\ref{tab:results} summarizes the results for all six programs.}
\toadd{The Provided Tests column reports whole-program test files fully passing (one test file may contain multiple individual tests).}
\toadd{All six translations succeed: each produces a safe Rust program (with no \texttt{unsafe} blocks) that passes 100\% of the provided tests.}

\begin{table}[htb]
    \centering
    \caption{\toadd{Summary of Translation Results.} %
    }
    \resizebox{0.92\linewidth}{!}{%
    \begin{tabular}{l r r r r r r}
        \toprule
        \textbf{Program} & \textbf{\shortstack[r]{User\\Interventions}} & \textbf{\shortstack[r]{Source\\C LoC}} & \textbf{\shortstack[r]{Translated\\Rust LoC}} & \textbf{\shortstack[r]{Provided\\Tests}} & \textbf{\shortstack[r]{Translation\\Time}} & \textbf{\shortstack[r]{Cost\\(USD)}} \\
        \midrule
        awk       & \escawkUserInt{} (+\escawkMinor) & \bmawkLoc  & \resawkRsLoc  & \resawkTestPass{} (\resawkProvPct) & \resawkTime & \resawkCost \\
        gnu-bc    & \escgnubcUserInt       & \bmgnubcLoc  & \resgnubcRsLoc  & \resgnubcTestPass{} (\resgnubcProvPct) & \resgnubcTime & \resgnubcCost \\
        picoc     & \escpicocUserInt{} (+\escpicocMinor) & \bmpicocLoc  & \respicocRsLoc  & \respicocTestPass{} (\respicocProvPct) & \respicocTime & \respicocCost \\
        wren      & \escwrenUserInt       & \bmwrenLoc  & \reswrenRsLoc  & \reswrenTestPass{} (\reswrenProvPct) & \reswrenTime & \reswrenCost \\
        mujs      & \escmujsUserInt       & \bmmujsLoc & \resmujsRsLoc & \resmujsTestPass{} (\resmujsProvPct) & \resmujsTime & \resmujsCost \\
        pocketpy  & \escpocketpyUserInt{} (+\escpocketpyMinor) & \bmpocketpyLoc & \respocketpyRsLoc & \respocketpyTestPass{} (\respocketpyProvPct) & \respocketpyTime & \respocketpyCost \\
        \bottomrule
    \end{tabular}
    }
    \label{tab:results}
\end{table}

\begin{figure*}[t]
\centering
\begin{subfigure}[b]{0.32\textwidth}
    \centering
    \includegraphics[width=\textwidth]{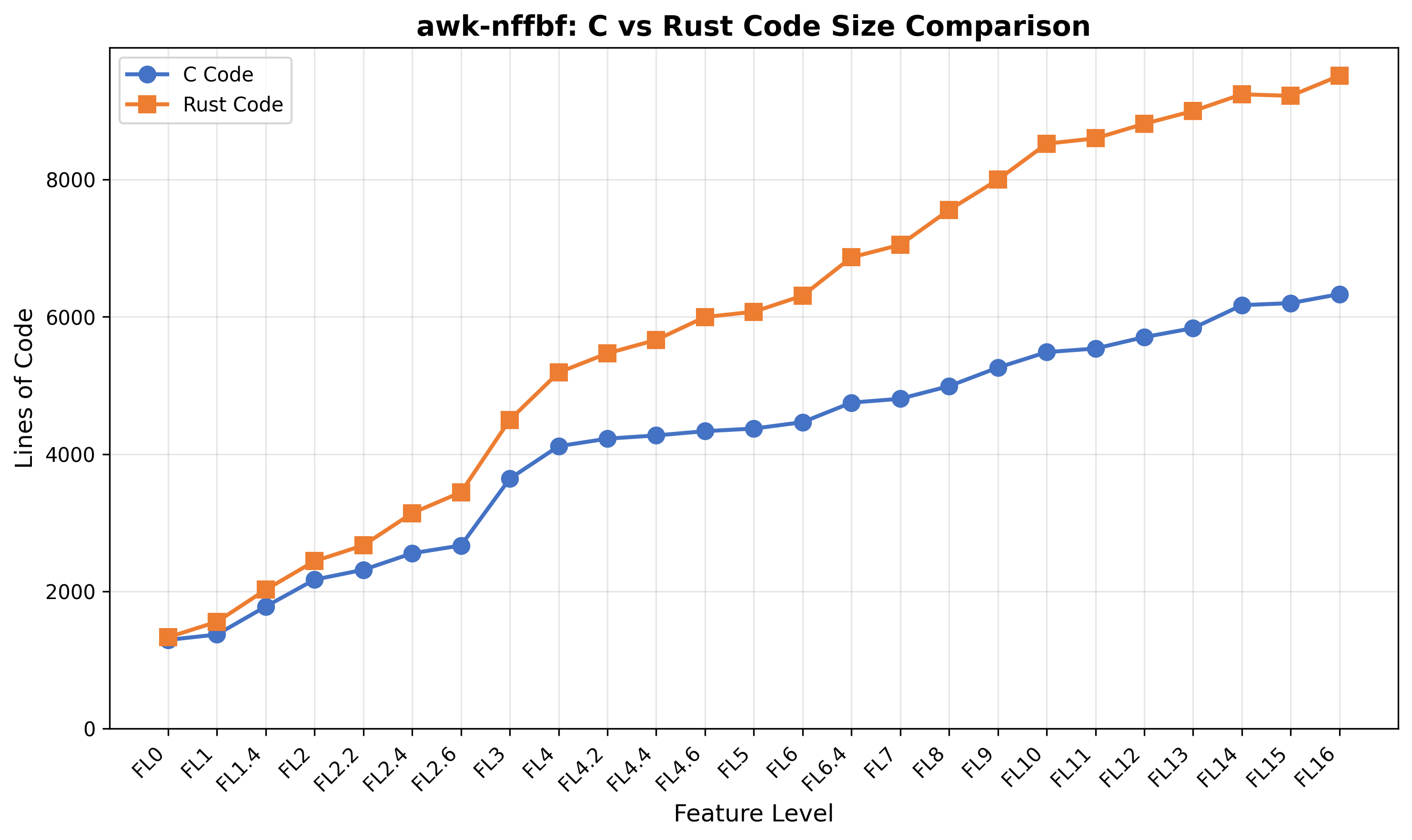}
    \caption{awk}
    \label{fig:eval-loc-comparison-awk}
\end{subfigure}
\hfill
\begin{subfigure}[b]{0.32\textwidth}
    \centering
    \includegraphics[width=\textwidth]{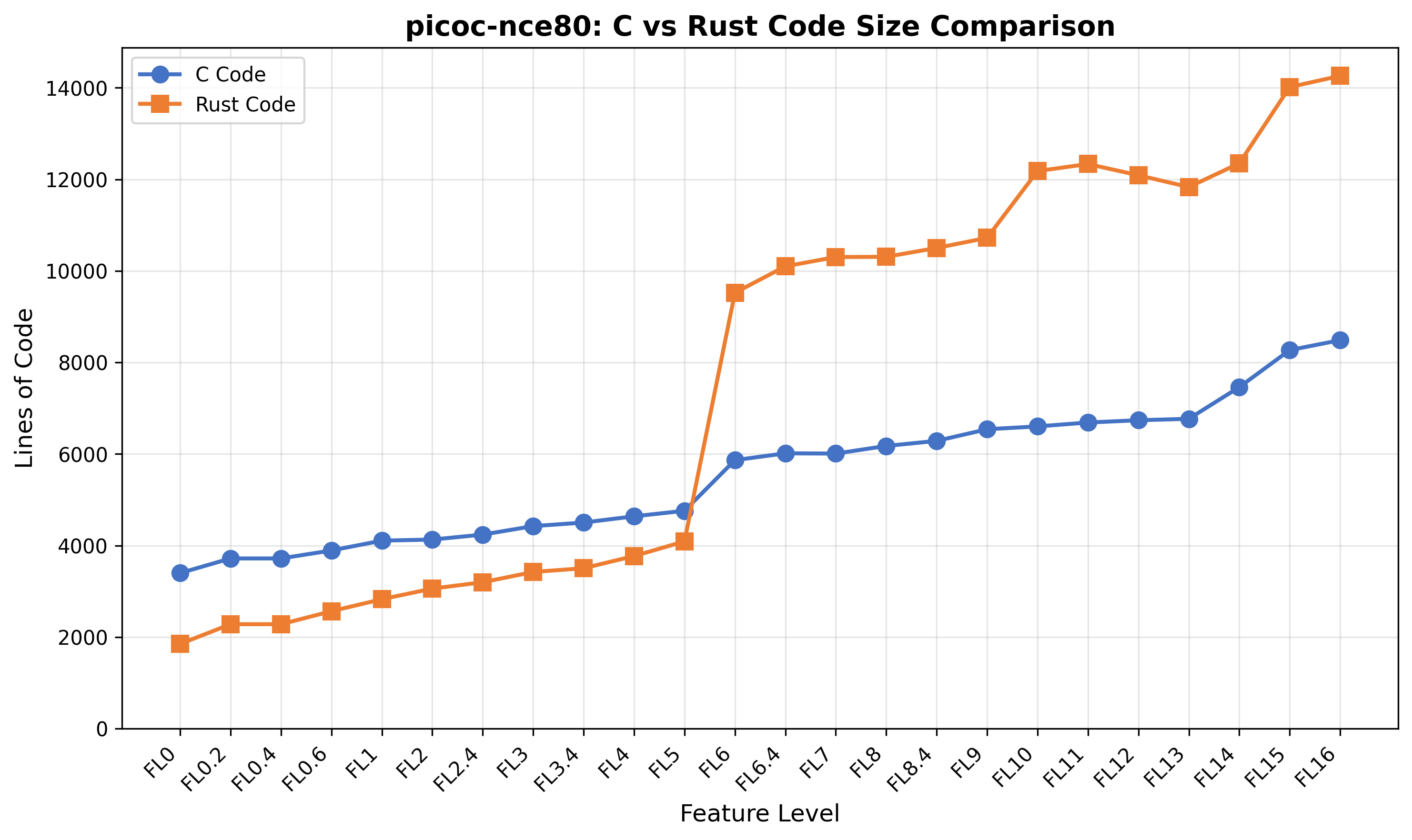}
    \caption{picoc}
    \label{fig:eval-loc-comparison-picoc}
\end{subfigure}
\hfill
\begin{subfigure}[b]{0.32\textwidth}
    \centering
    \includegraphics[width=\textwidth]{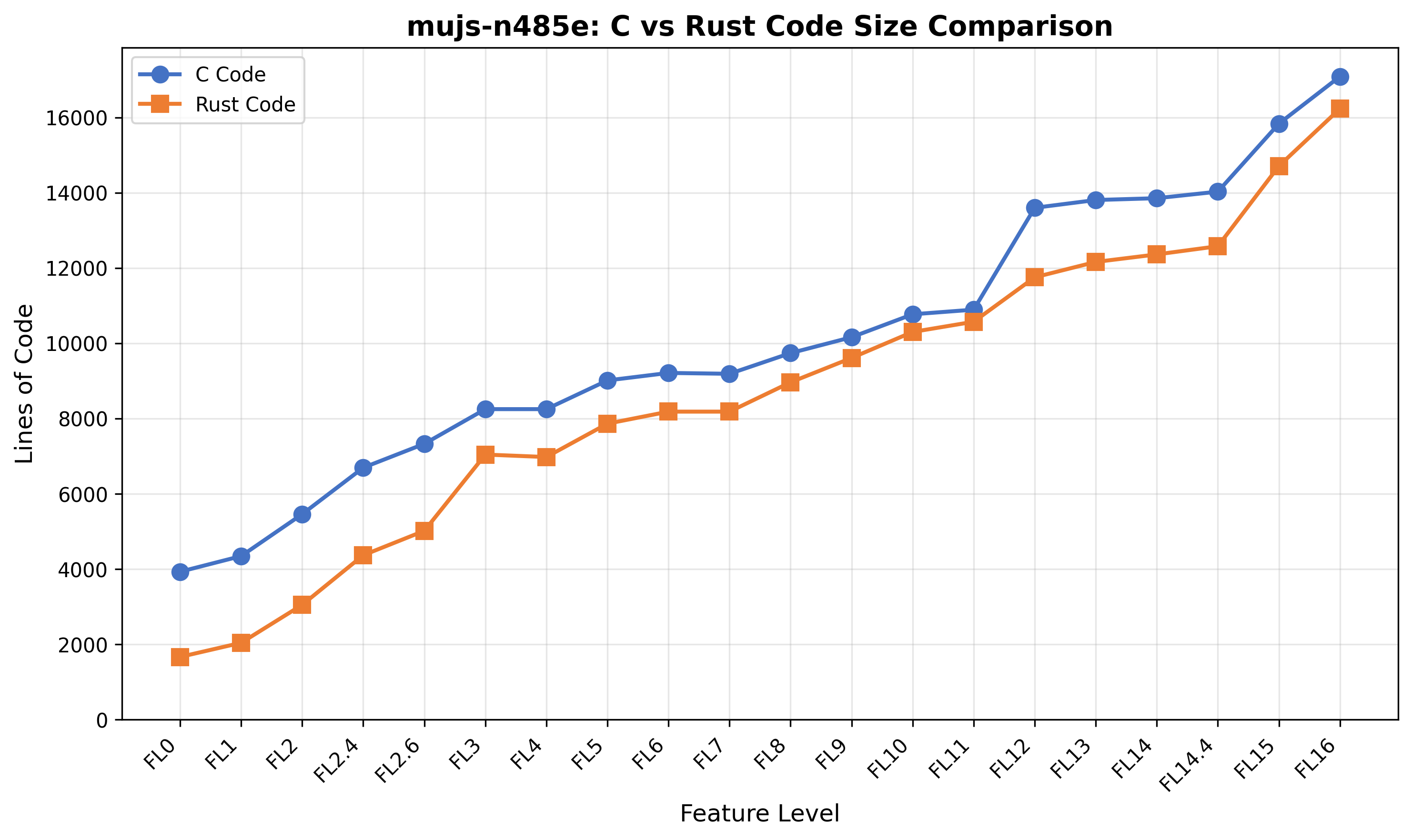}
    \caption{mujs}
    \label{fig:eval-loc-comparison-mujs}
\end{subfigure}

\vspace{1em}

\begin{subfigure}[b]{0.32\textwidth}
    \centering
    \includegraphics[width=\textwidth]{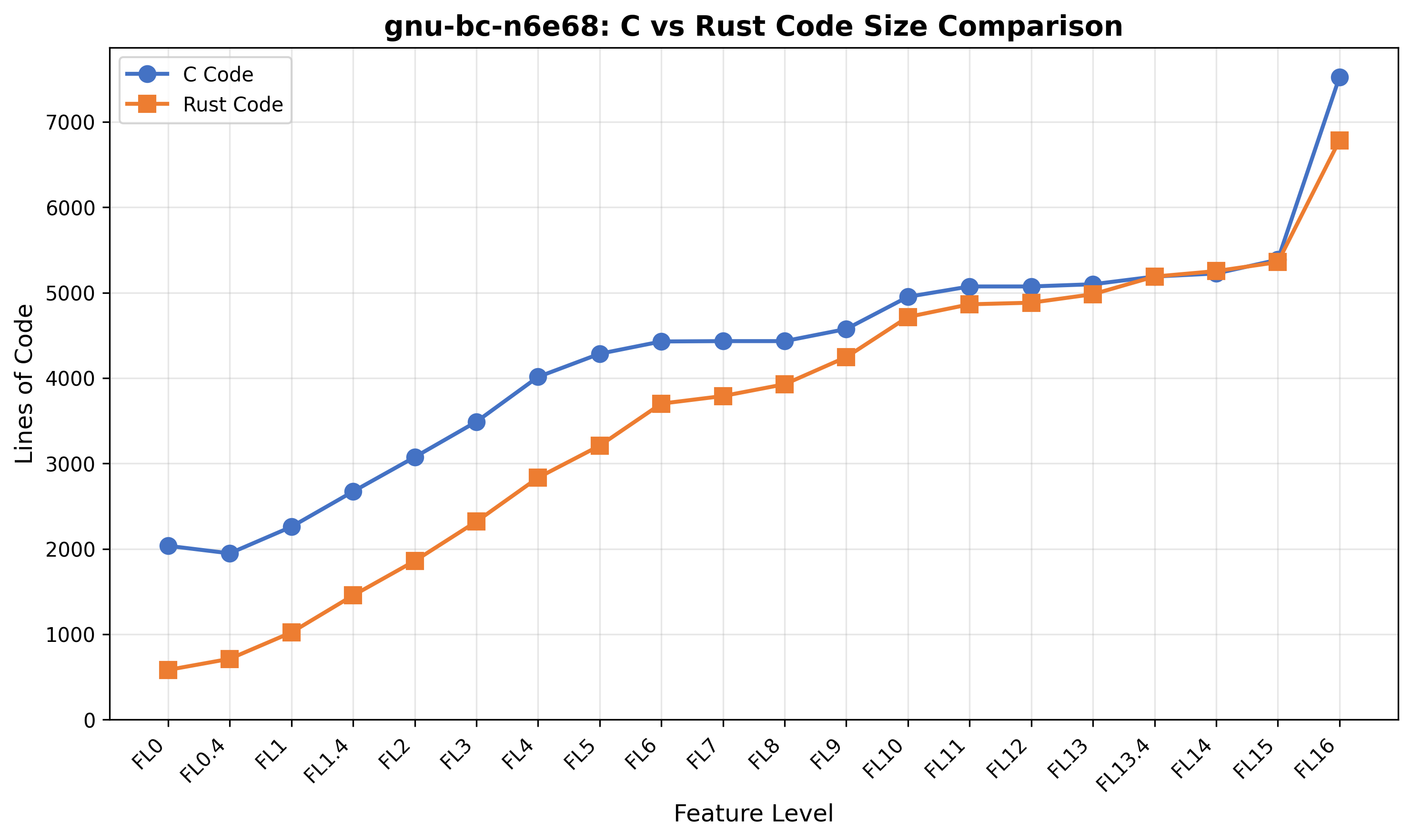}
    \caption{gnu-bc}
    \label{fig:eval-loc-comparison-gnu-bc}
\end{subfigure}
\hfill
\begin{subfigure}[b]{0.32\textwidth}
    \centering
    \includegraphics[width=\textwidth]{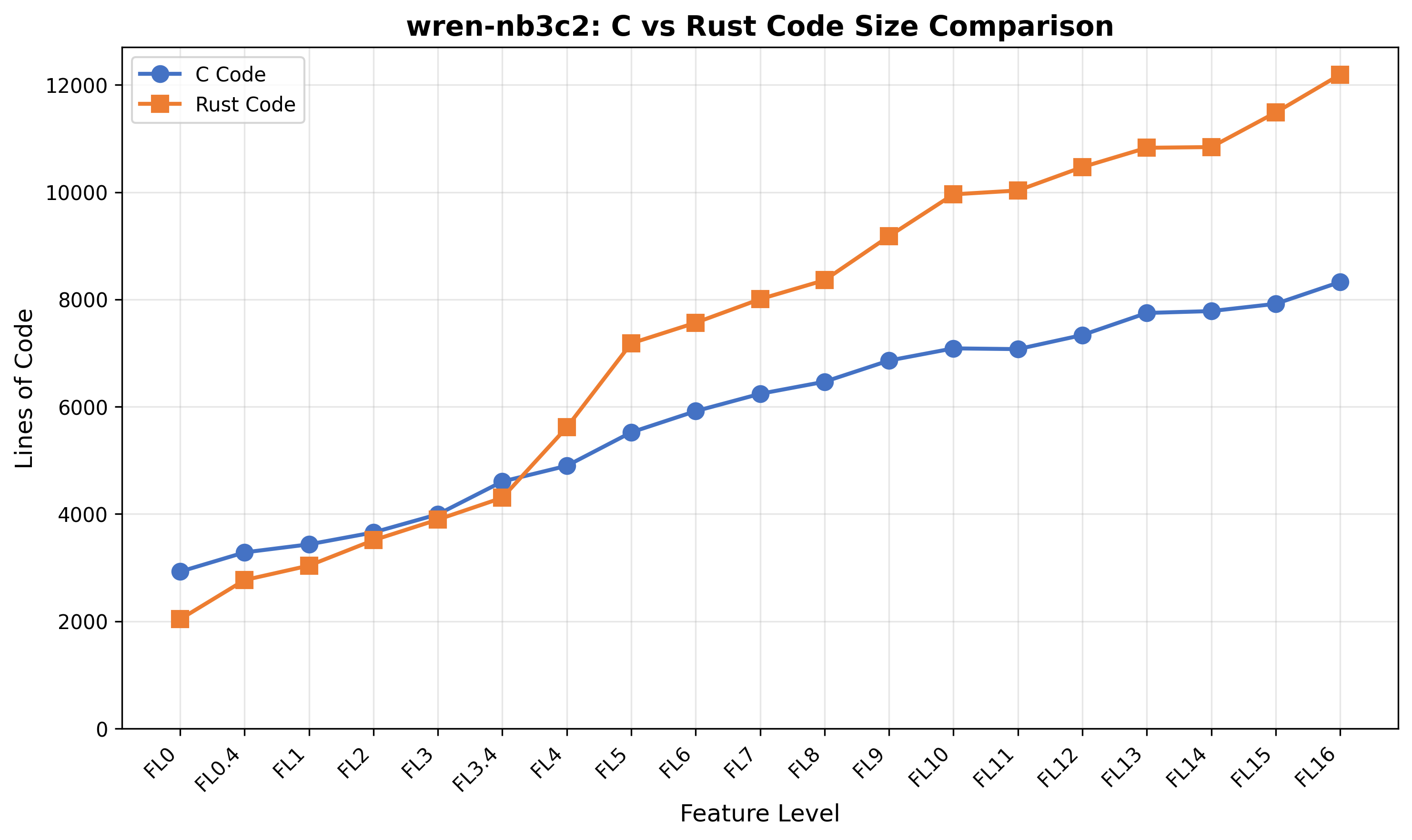}
    \caption{wren}
    \label{fig:eval-loc-comparison-wren}
\end{subfigure}
\hfill
\begin{subfigure}[b]{0.32\textwidth}
    \centering
    \includegraphics[width=\textwidth]{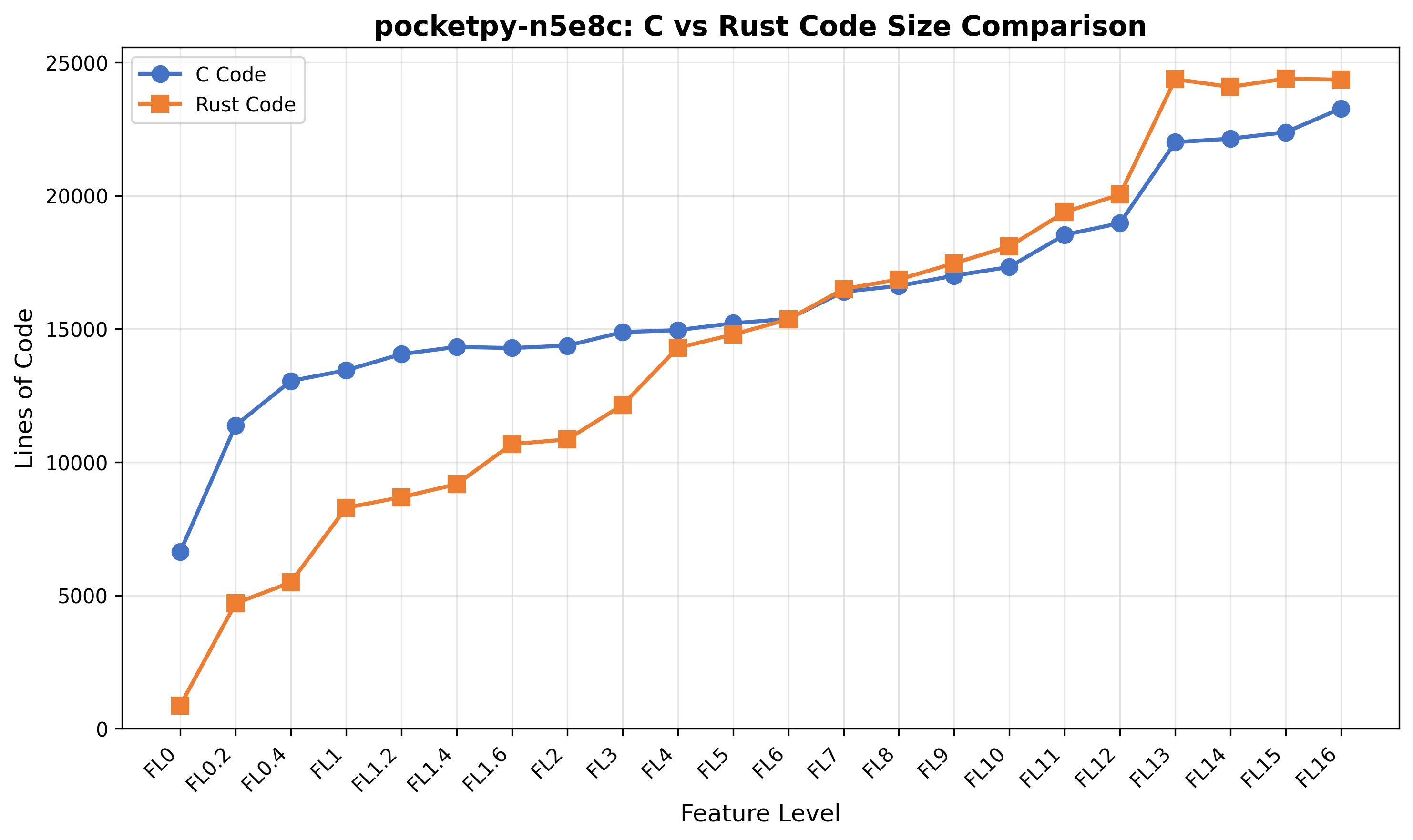}
    \caption{pocketpy}
    \label{fig:eval-loc-comparison-pocketpy}
\end{subfigure}
\caption{Code size comparison across feature levels. The charts show lines of code (LoC) growth as features are incrementally added, comparing the C reduced versions (during source reduction phase) with the Rust translations (during translation bootstrapping phase).}
\label{fig:eval-loc-comparison}
\end{figure*}

\begin{figure*}[t]
\centering
\begin{subfigure}[b]{0.32\textwidth}
    \centering
    \includegraphics[width=\textwidth,trim={0 0 0 0},clip]{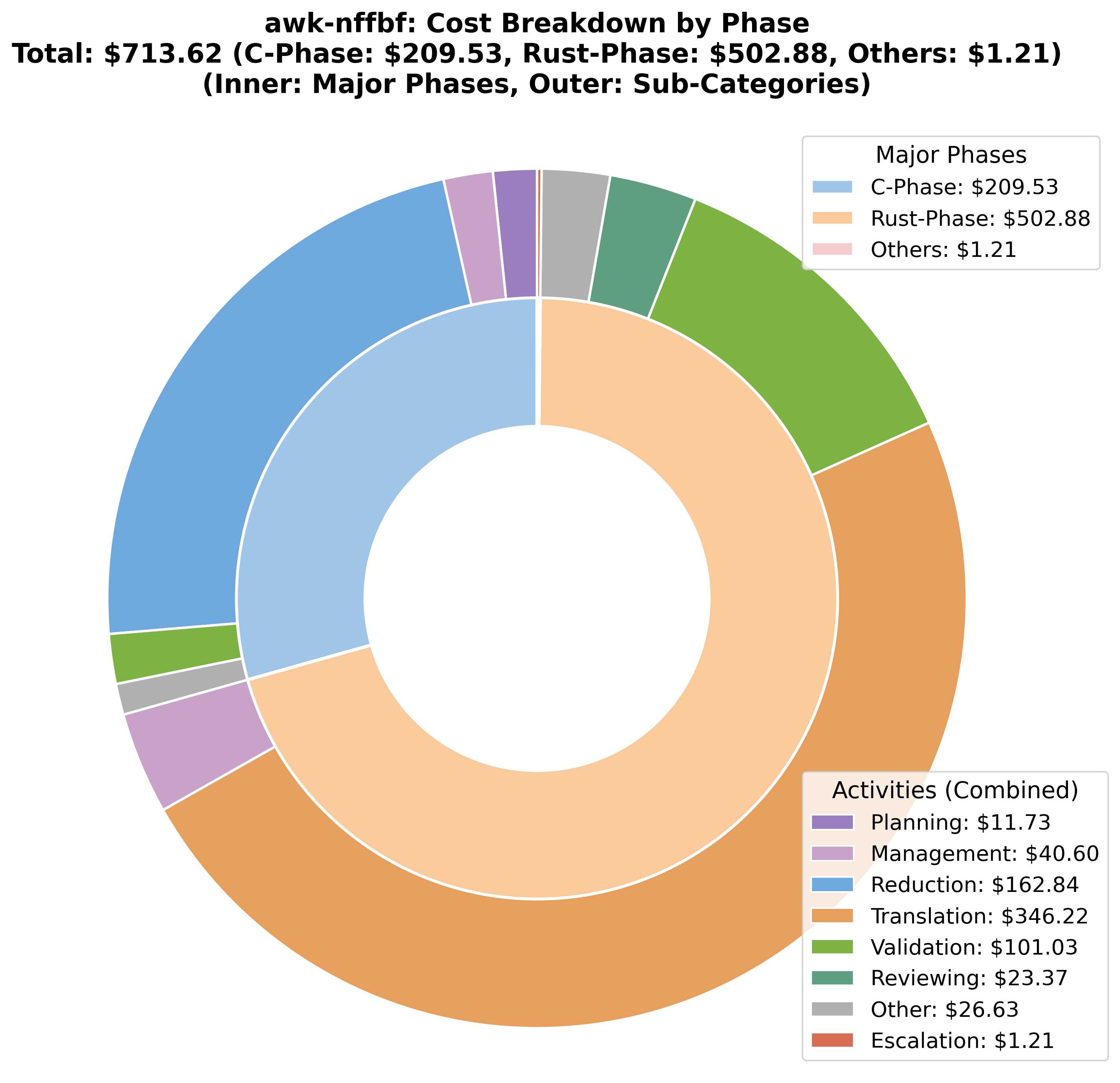}
    \caption{awk}
    \label{fig:eval-cost-breakdown-awk}
\end{subfigure}
\hfill
\begin{subfigure}[b]{0.32\textwidth}
    \centering
    \includegraphics[width=\textwidth,trim={0 0 0 0},clip]{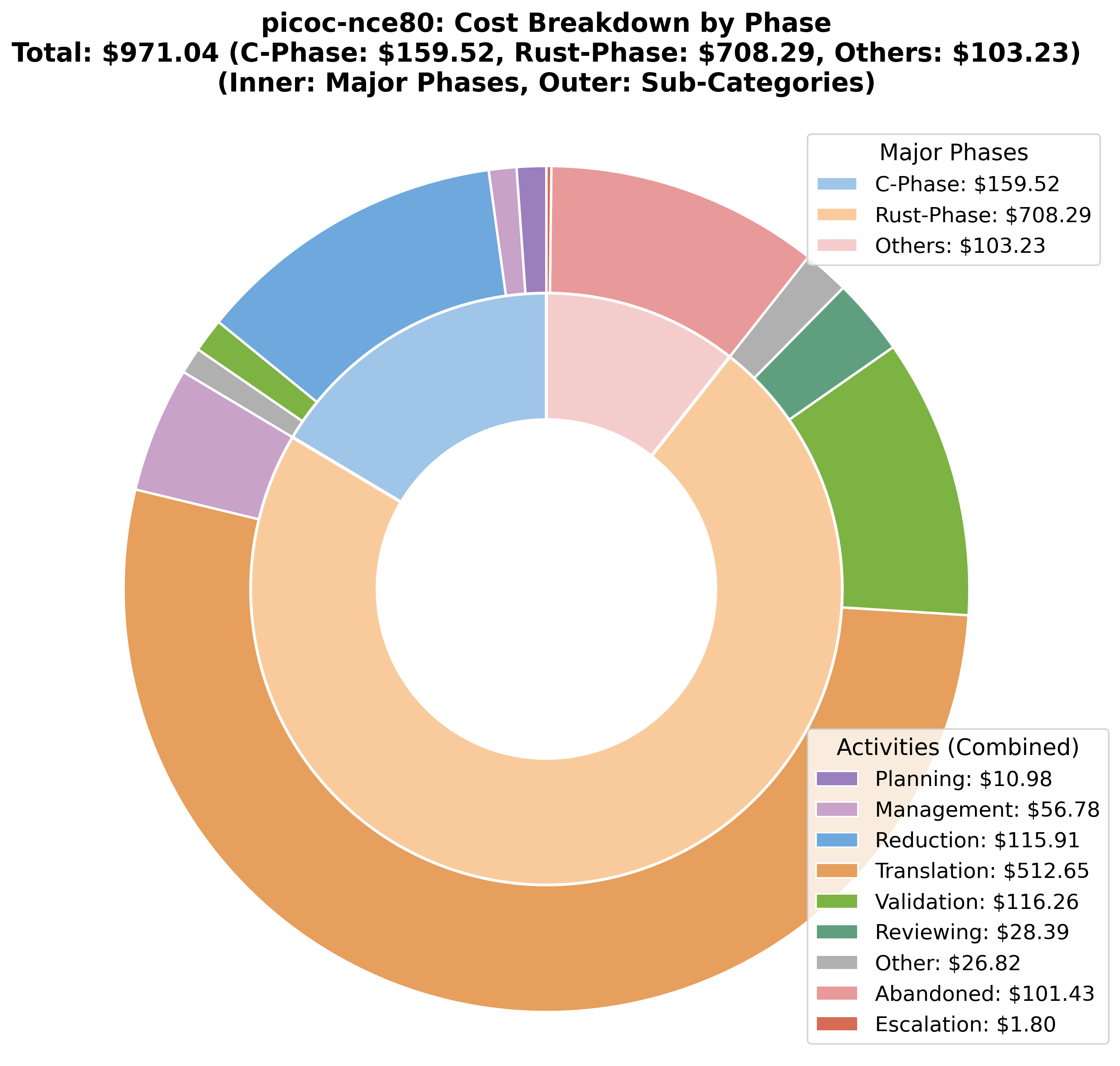}
    \caption{picoc}
    \label{fig:eval-cost-breakdown-picoc}
\end{subfigure}
\hfill
\begin{subfigure}[b]{0.32\textwidth}
    \centering
    \includegraphics[width=\textwidth,trim={0 0 0 0},clip]{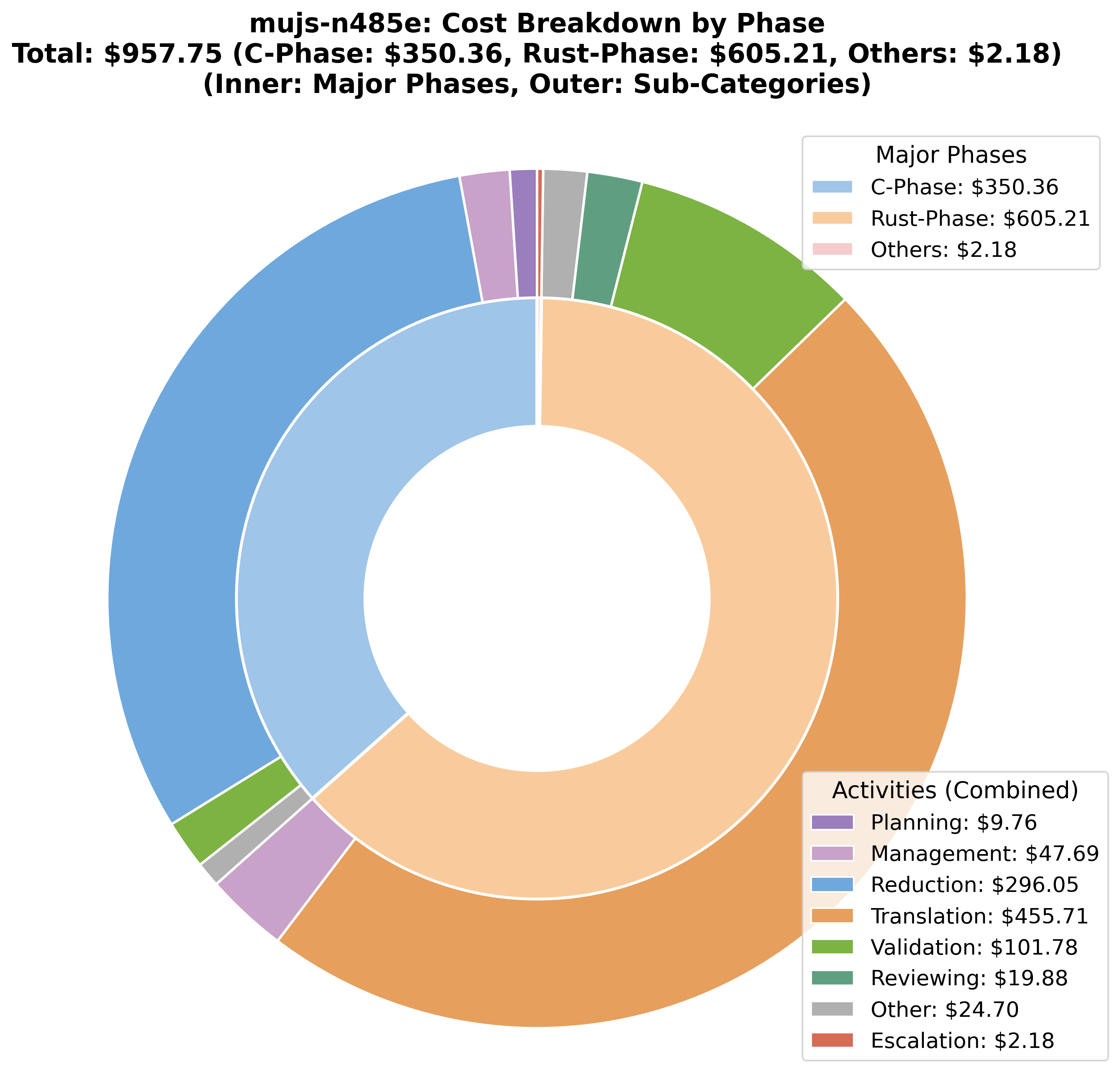}
    \caption{mujs}
    \label{fig:eval-cost-breakdown-mujs}
\end{subfigure}

\vspace{1em}

\begin{subfigure}[b]{0.32\textwidth}
    \centering
    \includegraphics[width=\textwidth,trim={0 0 0 0},clip]{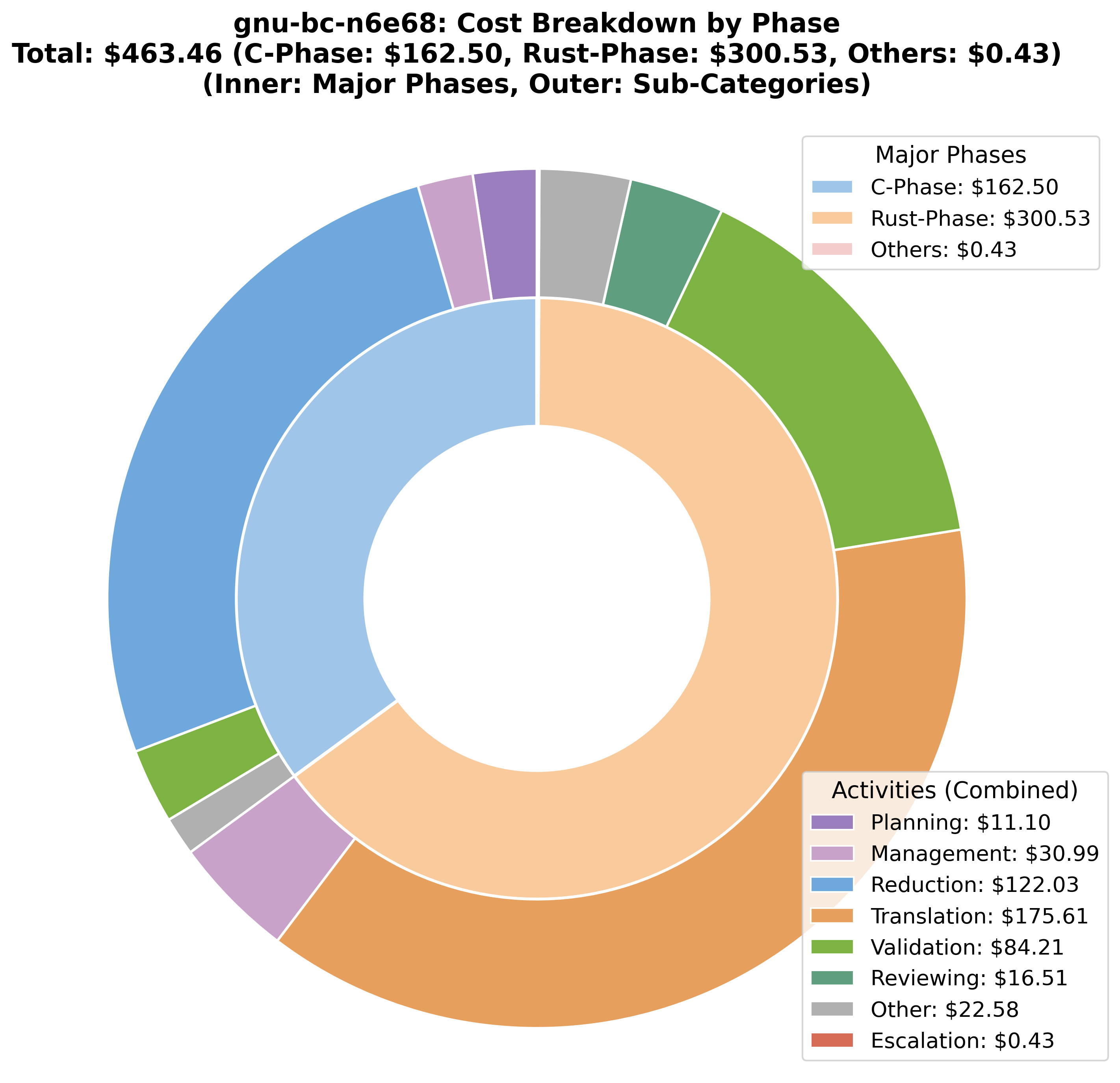}
    \caption{gnu-bc}
    \label{fig:eval-cost-breakdown-gnu-bc}
\end{subfigure}
\hfill
\begin{subfigure}[b]{0.32\textwidth}
    \centering
    \includegraphics[width=\textwidth,trim={0 0 0 0},clip]{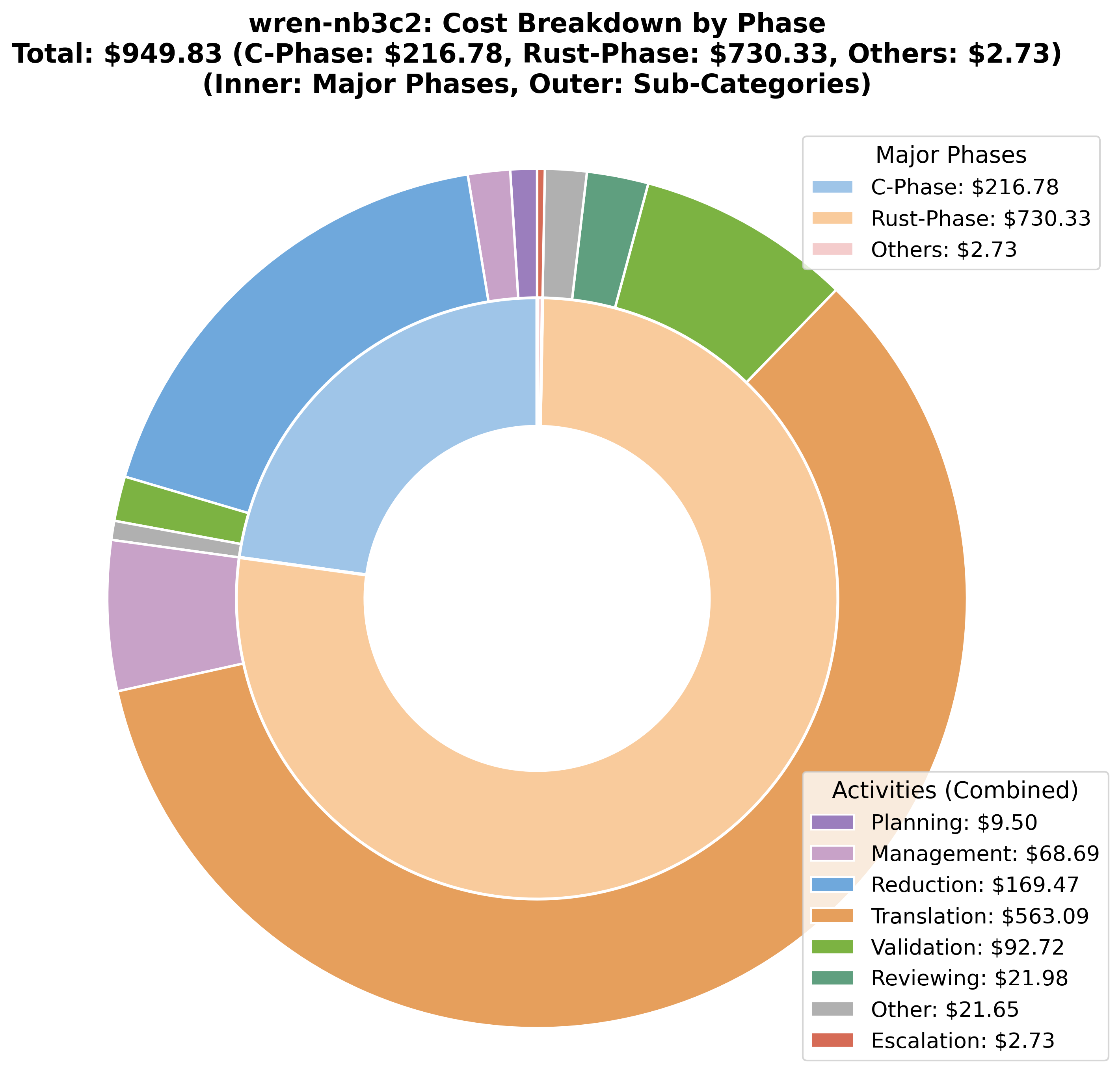}
    \caption{wren}
    \label{fig:eval-cost-breakdown-wren}
\end{subfigure}
\hfill
\begin{subfigure}[b]{0.32\textwidth}
    \centering
    \includegraphics[width=\textwidth,trim={0 0 0 0},clip]{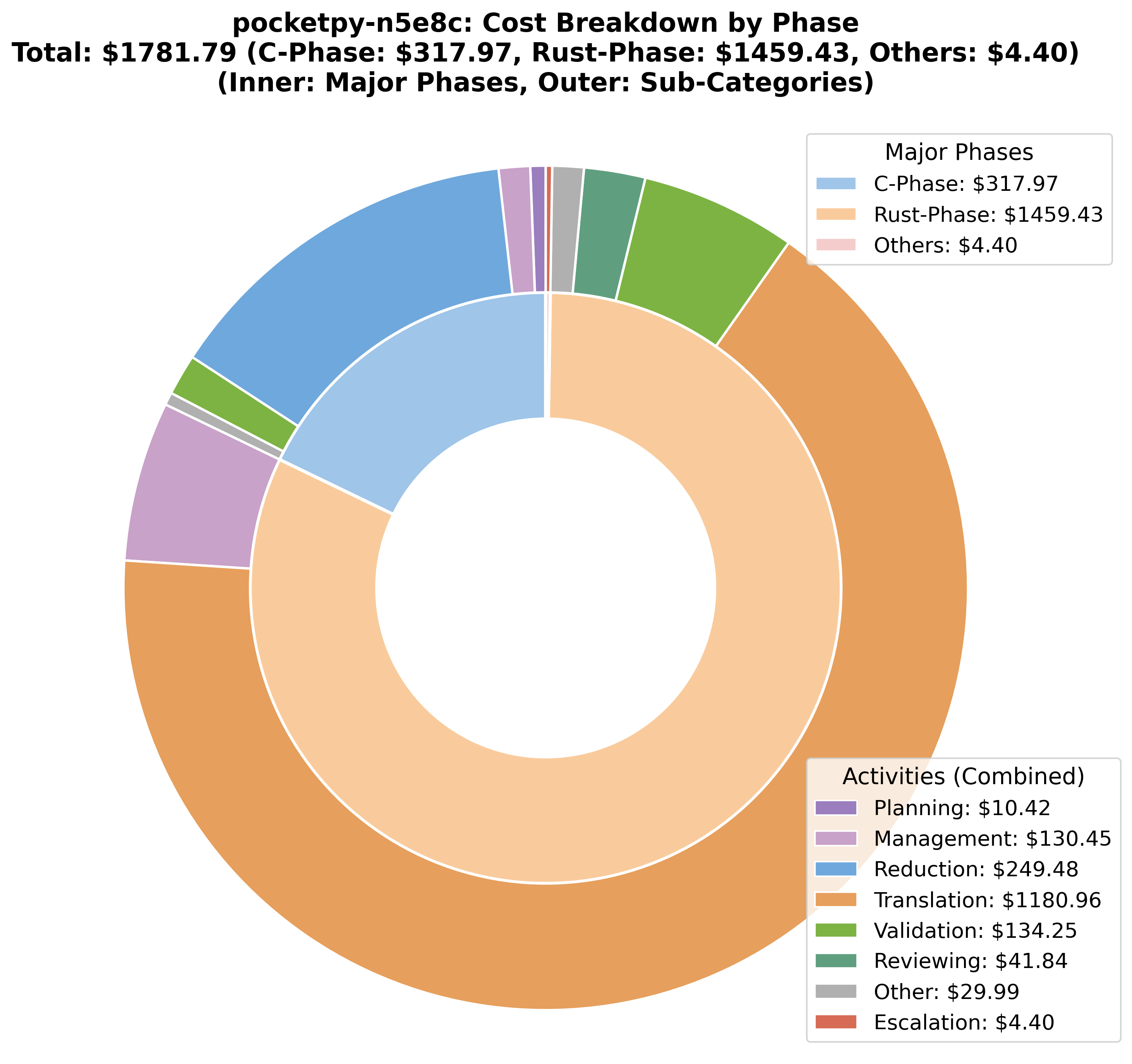}
    \caption{pocketpy}
    \label{fig:eval-cost-breakdown-pocketpy}
\end{subfigure}
\caption{Cost breakdown by activity across six interpreter programs. The pie charts show the distribution of monetary costs across different phases, with the inner ring representing common categories (Management, Reduction, Translation, CodeReview, TestValid, Commit, Additional) and the outer ring showing language-specific breakdown (C vs Rust activities).}
\label{fig:eval-cost-breakdown}
\end{figure*}

\toadd{The Rust translations have comparable size to the C source, with four of six programs having the translated Rust code larger than C.}
\toadd{Translation time ranges from $\sim$\resMinTimeRounded{} to $\sim$\resMaxTimeRounded{} hours, and monetary cost ranges from $\sim$\$\resMinCostRounded{} to $\sim$\$\resMaxCostRounded{} per program.}
\toadd{Human effort is modest: programs require \escawkUserInt{} to \escpocketpyUserInt{} user interventions, each taking roughly 5 minutes (the ``+$n$'' annotations indicate minor system recovery actions---restarting a hanging agent or correcting a message format error---that take seconds, could be eliminated by a more robust system implementation, and are not counted as interventions).}

\toadd{Figure~\ref{fig:eval-loc-comparison} shows the lines of code across feature levels for both the C reduced versions and the Rust translations, illustrating how code size evolves through the translation process.}
\toadd{Figure~\ref{fig:eval-cost-breakdown} shows the monetary cost breakdown by activity, illustrating how LLM costs are distributed across the translation process.}

\subsection{RQ3: Correctness beyond the Provided Tests}
\toadd{As described in the setup, we created validation test suites for each program independently of the provided tests.}
\toadd{These validation tests were created by the authors independently of and separately from the provided test suites, and were never exposed to the system during translation; we therefore consider them ``unseen'' tests.}
\toadd{One of the authors spent approximately 10--20 hours per program creating each suite. The process involved studying the program's supported features and usage, manually writing around 10 seed test programs (interpreter inputs), then semi-automatically expanding coverage with the help of a coding agent, and manually vetting that all generated tests produce correct expected outputs by running them on the original C program. Most suites achieve relatively high line coverage on the translated Rust code (\valCovAvg{} on average); pocketpy's suite has lower coverage (\valpocketpyCov) due to more limited effort given the program's size and complexity.}
\toadd{The validation tests were run on the final Rust translations only after the entire translation process was complete.}
\toadd{Table~\ref{tab:validation} presents the validation results.}
\toadd{Pass rates on the unseen validation tests range from $\sim$\valMinPct{} to $\sim$\valMaxPct{}, with wren achieving the highest (\valwrenPct, \valwrenFrac) and pocketpy the lowest (\valpocketpyPct, \valpocketpyFrac).}
\toadd{The validation test suites achieve an average of \valCovAvg{} line coverage on the translated Rust code (ranging from \valCovMin{} to \valCovMax).}

\begin{table}[htb]
    \centering
    \caption{\toadd{Pass rates on provided tests used for the translation and on the validation tests post-translation.}}
    \resizebox{0.7\linewidth}{!}{%
    \begin{tabular}{l r r r}
        \toprule
        \textbf{\shortstack[r]{Translated\\Rust Program}} & \textbf{\shortstack[r]{Provided\\Tests}} & \textbf{\shortstack[r]{Validation\\Tests}} & \textbf{\shortstack[r]{Validation\\Test Cov. (C)}} \\
        \midrule
        wren      & \reswrenTestPass{} (\reswrenProvPct)  & \valwrenFrac{} (\valwrenPct) & \valwrenCov \\
        awk       & \resawkTestPass{} (\resawkProvPct)  & \valawkFrac{} (\valawkPct) & \valawkCov \\
        gnu-bc    & \resgnubcTestPass{} (\resgnubcProvPct)  & \valgnubcFrac{} (\valgnubcPct) & \valgnubcCov \\
        mujs      & \resmujsTestPass{} (\resmujsProvPct)  & \valmujsFrac{} (\valmujsPct) & \valmujsCov \\
        picoc     & \respicocTestPass{} (\respicocProvPct)  & \valpicocFrac{} (\valpicocPct) & \valpicocCov \\
        pocketpy  & \respocketpyTestPass{} (\respocketpyProvPct)  & \valpocketpyFrac{} (\valpocketpyPct) & \valpocketpyCov \\
        \bottomrule
    \end{tabular}
    }
    \label{tab:validation}
\end{table}
\toadd{For mujs, we additionally evaluated on the official ECMA-262 ES5 conformance suite (Test262), which contains $\sim$\ablTFullTotal{} tests in total.}
\toadd{The original mujs C source conforms on \ablTmujsConform{} of these tests; we refer to this subset as \emph{Test262@mujs}.}
\toadd{The safe Rust translation produced by \tool passes \ablTmujsPass{} out of those \ablTmujsConform{} tests in Test262@mujs (\ablTmujsPct).}

\subsection{RQ4: Comparison with Existing Approaches}

\begin{table}[htb]
    \centering
    \caption{\toadd{Comparison of \tool, C2Rust, and Crown on the six benchmarks. \tool produces safe Rust with no \texttt{unsafe} blocks and no raw pointers, whereas C2Rust output is far larger and almost entirely raw-pointer based. Results from Crown are similar. Pointer declaration counts are syntactic approximations.}}
    \label{tab:compare-c2rust}
    \resizebox{\linewidth}{!}{%
    \begin{tabular}{l r r r r r r r r r r}
        \toprule
         & & \multicolumn{3}{c}{\textbf{Rust LoC}} & \multicolumn{3}{c}{\textbf{Raw pointer decls}} & \multicolumn{3}{c}{\textbf{Tests passing}} \\
        \cmidrule(lr){3-5}\cmidrule(lr){6-8}\cmidrule(lr){9-11}
        \textbf{Program} & \textbf{C LoC} & \textbf{\tool} & \textbf{C2Rust} & \textbf{Crown} & \textbf{\tool} & \textbf{C2Rust} & \textbf{Crown} & \textbf{\tool} & \textbf{C2Rust} & \textbf{Crown} \\
        \midrule
        awk      & \bmawkLoc      & \resawkRsLoc      & 28,215 & 27,526 & 0 & 2,327 & 1,903 & \resawkTestPass      & 279/279 & 279/279 \\
        gnu-bc   & \bmgnubcLoc    & \resgnubcRsLoc    & 18,519 & 17,538 & 0 & 1,199 & 879   & \resgnubcTestPass    & 134/134 & 134/134 \\
        picoc    & \bmpicocLoc    & \respicocRsLoc    & 28,930 & 23,127 & 0 & 5,722 & 3,489 & \respicocTestPass    & 0/154   & 0/154 \\
        wren     & \bmwrenLoc     & \reswrenRsLoc     & 22,156 & 19,964 & 0 & 3,366 & 2,534 & \reswrenTestPass     & 851/851 & 851/851 \\
        mujs     & \bmmujsLoc     & \resmujsRsLoc     & 46,002 & 39,282 & 0 & 5,158 & 2,784 & \resmujsTestPass     & 184/187 & 184/187 \\
        pocketpy & \bmpocketpyLoc & \respocketpyRsLoc & 97,354 & 83,163 & 0 & 7,701 & 5,608 & \respocketpyTestPass & 83/83   & 83/83 \\
        \bottomrule
    \end{tabular}
    }
\end{table}

We compare \tool with four existing C-to-Rust translation approaches: 1) C2Rust~\cite{c2rust}, an industrial grade rule-based translator, 2) CROWN~\cite{zhang2023ownership}, a rule-based translator that refines the output of C2Rust using a novel symbolic ownership analysis to rewrite raw pointers to safe Rust references, 3) C2SaferRust~\cite{nitin2025c2saferrust}, a neuro-symbolic translator that refines the output of C2Rust using LLMs rather than a symbolic analysis, and 4) SmartC2Rust~\cite{shiraishi2024smartc2rust}, a neuro-symbolic translator that combines program analyses and program rewriting with an LLM to perform translation. CROWN and SmartC2Rust are representative state-of-the-art rule-based and neuro-symbolic translators, respectively, based on the number and size of C programs they are demonstrated to successfully translate.

We first run C2Rust and CROWN on our benchmark programs. The results are summarized in Table~\ref{tab:compare-c2rust}. On four of our six programs, both tools achieve 100\% test pass rates. On the remaining two programs, they achieve only 98\% and 0\% pass rates because of runtime errors related to how C2Rust (and therefore CROWN) handles variable-length structs (the C flexible-array-member idiom). We also see that rule-based translations are $2\times$--$4\times$ larger than the C source and make heavy use of raw pointers, while \tool's translations stay close to the original size and are fully safe. We further note that some manual work is needed to make C2Rust (and CROWN) work on our programs.
Wee need to make changes to the build process for each program to produce per-file compilation databases and disable an unsupported GNU C extension (computed goto), and post-process the Rust output to fix errors including incorrect atomic types, missing extern declarations, and duplicate symbol exports.
In addition, CROWN's ownership analysis could not handle pointer-carrying tagged unions present in all of our programs, so we had to disable part of CROWN's algorithm in order to successfully run it.

\toadd{We also attempted to apply SmartC2Rust~\cite{shiraishi2024smartc2rust} to our benchmark programs. SmartC2Rust is a recent and capable neuro-symbolic system that pairs LLM-based translation with program analysis, and it has been evaluated on C programs of up to around 4k lines, which is smaller than our interpreters ($\sim$\bmMinLocRounded{}--\bmMaxLocRounded{} LoC). We spent more than five days trying to run it on our programs, but have so far been unable to translate any program end-to-end. We observed a few recurring issues. A large number of files in the project tree, or a very long input file, could cause the pipeline to crash while composing its prompts. The automatic patching of the program's \code{main} function also appeared unstable when the interpreter's \code{main} is complex. After multiple attempts on \code{awk}, we obtained roughly a third of the functions translated and compiling, but most functions remained untranslated and the program as a whole was not executable. We are unsure whether these issues are fundamental, a matter of implementation robustness, or violations of implicit assumptions of the tool that we were not aware of.}
\toadd{The SmartC2Rust implementation is also substantial, reaching 37k lines of code. Its size, together with the number of rule-based analysis components it integrates, made it challenging for us to diagnose the issues we encountered with the pipeline. Similar phenomena with implementation complexity have been reported for other LLM-based translation systems~\cite{ibrahimzada2025matchfixagent}.}

\subsection{CS1: Security Improvements}
\toadd{Among our benchmark programs, mujs has \cveAll{} documented CVEs in its history, making it a good candidate for evaluating security improvements.}
\toadd{We re-introduced \cveAppl{} of these CVEs into the latest mujs codebase to create a vulnerability-concentrated benchmark we call \emph{mujs-CVEs}. The remaining \cveExcl{} CVEs could not be re-introduced because they are no longer applicable in the latest codebase or conflict with later CVEs.}
\toadd{The \cveAppl{} CVEs span a range of error types: heap buffer overflow (\cveHeapBOF), heap use-after-free (\cveHeapUAF), stack exhaustion (\cveStackExhaust), integer overflow (\cveIntOverflow), bytecode logic error (\cveBytecodeLogic), null pointer dereference (\cveNULLDeref), out-of-bounds read (\cveOOBRead), stack buffer overflow (\cveStackBOF), and global buffer overflow (\cveGlobalBOF).}

\toadd{We translated mujs-CVEs using \tool and ran the original proof-of-concept (PoC) inputs for each CVE on both the C and safe Rust builds. We also manually reviewed the vulnerable C code and corresponding Rust translation for each CVE to assess whether the vulnerability was truly eliminated, merely mitigated (memory corruption gone but replaced by a different-class failure), or survived; the results are reported in the Status column of Table~\ref{tab:cve}, with a detailed per-CVE breakdown in the appendix.} %
\toadd{The error categories used in Table~\ref{tab:cve} are:}
\begin{itemize}[nosep,leftmargin=*]
\item \textbf{Memory safety violations} (C only): Heap-UAF (use-after-free), Heap-BOF / Stack-BOF / Global-BOF (buffer overflow), NULL-Deref (null pointer dereference), OOB-Read (out-of-bounds read from undefined behavior), and Int-Overflow (integer overflow leading to memory corruption). These cannot occur in safe Rust.
\item \textbf{Logic and algorithmic errors} (C and Rust): Stack-Exhaust (unbounded recursion) and Bytecode-Logic (bytecode compiler logic errors that do not cause memory corruption).
\item \textbf{Rust observed behaviors}: Clean (no error), Exit-Nonzero (controlled error exit), Exception (JS-level uncaught exception), Panic (Rust safety check, e.g.\ overflow detection), Abort (OOM or fatal runtime error), and Stack-Exhaust (stack exhaustion crash).
\end{itemize}

\begin{table}[htb]
    \centering
    \caption{\toadd{Results on Reproduced CVEs in the C (mujs-CVEs) vs. its Safe Rust Translation. The Status column reflects the manual assessment: Eliminated (memory safety violation removed), Mitigated (memory corruption gone but replaced by a different kind of crashes), or Unmitigated (same vulnerability class persists). All Rust code passes as safe Rust with no \texttt{unsafe} blocks.}}
    \resizebox{0.84\linewidth}{!}{%
    \begin{tabular}{l l l l l}
        \toprule
        \textbf{CVE} & \textbf{Error in C} & \textbf{Rust O0 (Debug)} & \textbf{Rust O3 (Release)} & \textbf{Status} \\
        \midrule
        CVE-2022-44789 & Heap-UAF        & Clean          & Clean          & Eliminated \\
        CVE-2021-33796 & Heap-UAF        & Clean          & Clean          & Eliminated \\
        CVE-2020-24343 & Heap-UAF        & Exception      & Exception      & Eliminated \\
        CVE-2019-12798 & Heap-BOF        & Stack-Exhaust  & Stack-Exhaust  & Mitigated \\
        CVE-2016-10141 & Heap-BOF        & Abort          & Abort          & Mitigated \\
        CVE-2016-10133 & Heap-BOF        & Stack-Exhaust  & Stack-Exhaust  & Mitigated \\
        CVE-2016-9136  & Heap-BOF        & Exit-Nonzero   & Exit-Nonzero   & Eliminated \\
        CVE-2016-9109  & Heap-BOF        & Exit-Nonzero   & Exit-Nonzero   & Eliminated \\
        CVE-2016-7506  & Heap-BOF        & Clean          & Clean          & Eliminated \\
        CVE-2019-11411 & Stack-BOF       & Clean          & Clean          & Eliminated \\
        CVE-2018-6191  & Global-BOF      & Clean          & Clean          & Eliminated \\
        CVE-2017-5628  & OOB-Read        & Clean          & Clean          & Eliminated \\
        CVE-2016-9294  & NULL-Deref      & Clean          & Clean          & Eliminated \\
        CVE-2021-33797 & Int-Overflow    & Clean          & Clean          & Eliminated \\
        CVE-2016-9108  & Int-Overflow    & Panic          & Clean          & Mitigated \\
        CVE-2021-45005 & Bytecode-Logic  & Clean          & Clean          & Eliminated \\
        CVE-2019-11412 & Bytecode-Logic  & Clean          & Clean          & Eliminated \\
        CVE-2022-30974 & Stack-Exhaust   & Stack-Exhaust  & Stack-Exhaust  & Unmitigated \\
        CVE-2019-11413 & Stack-Exhaust   & Exit-Nonzero   & Exit-Nonzero   & Eliminated \\
        CVE-2018-5759  & Stack-Exhaust   & Clean\footnotemark & Clean\footnotemark[\value{footnote}] & Unmitigated \\
        \bottomrule
    \end{tabular}
    }
    \footnotetext{The vulnerability (missing depth limit) still exists in the Rust translation; a different PoC may trigger it.}
    \label{tab:cve}
\end{table}
\toadd{Table~\ref{tab:cve} shows the results, comparing the behavior of the C and Rust code. We report on both debug (O0) and release (O3) builds, since compiler optimizations can cause different runtime behavior.}
\toadd{Of the \cveAppl{} CVEs, \cveElim{} are eliminated, \cveMitig{} are mitigated, and \cveSurv{} are unmitigated in the Rust translation that compiles as safe Rust (has no unsafe blocks). We explain their breakdown next}.

\toadd{
All memory safety vulnerabilities—heap buffer overflow, heap use-after-free, stack buffer overflow, global buffer overflow, null pointer dereference, and undefined-behavior-induced out-of-bounds access—are either eliminated or mitigated: the Rust program either runs cleanly, exits gracefully, or fails with a different-class error (stack exhaustion or OOM) rather than memory corruption.}
\toadd{The \cveSurv{} unmitigated CVEs are both denial-of-service vulnerabilities that cause stack exhaustion from unbounded recursion. These are retained in safe Rust, which is expected since Rust has a finite stack as well.}
\toadd{One CVE worth noting is CVE-2016-9108 (integer overflow), where Rust panics in debug mode but wraps silently in release mode; this difference is not undefined behavior but rather an intentional design choice in Rust, where debug builds check for integer overflow while release builds wrap for performance reasons—both are defined behavior in safe Rust.}
\toadd{In general, Rust also provides more principled error handling, and we observe several cases where PoC inputs that caused memory corruption in C instead produce controlled error exits in the Rust translation, such as \texttt{Option}-forced EOF handling and \texttt{Result}-based error propagation that convert malformed inputs into clean \texttt{SyntaxError} exits.}
\toadd{Table~\ref{tab:cve-factors} summarizes the Rust features responsible for eliminating or mitigating the \cveElimOrMitig{} applicable CVEs. Architectural redesign, type safety, bounds checking, safe APIs, and ownership/borrow-checking of references all contribute, often in combination. For example, C mujs uses a manual mark-and-sweep garbage collector; the Rust translation replaces it with \texttt{Rc<RefCell<>{}>} reference counting, which eliminates an entire class of GC algorithm bugs (e.g., CVE-2020-24343) at the cost of not collecting reference cycles---a standard tradeoff when memory leaks are acceptable.}

\begin{table}[t]
\centering
\caption{CVE elimination factors across the \cveElimOrMitig{} eliminated or mitigated CVEs. Each factor may appear as the primary cause or as a contributing factor. PanicOnOverflow and BetterLogic are grouped as Other.}
\label{tab:cve-factors}
\resizebox{0.95\textwidth}{!}{%
\small
\setlength{\tabcolsep}{3pt}
\begin{tabular}{@{}l r p{0.6\textwidth} r r@{}}
\toprule
\textbf{Factor} & \textbf{Count} & \textbf{Examples} & \textbf{Primary} & \textbf{Contrib.} \\
\midrule
Architecture & \cveFactArch{} & GC redesign (\texttt{Rc<RefCell<>{}>}), stack-based labels, local \texttt{Vec} handler storage, scope-based \texttt{LoopContext} & \cveFactArchPri{} & \cveFactArchCon{} \\
\midrule
TypeSafety & \cveFactTypeSafety{} & \texttt{Option<u8>} forces EOF handling, \texttt{usize} prevents negative index, defined \texttt{NaN}-to-int cast semantics, \texttt{usize} overflow to large value & \cveFactTypeSafetyPri{} & \cveFactTypeSafetyCon{} \\
\midrule
Bounds & \cveFactBounds{} & Vec/slice bounds checking on \texttt{advance()}, explicit \texttt{i+1 < len} guards & \cveFactBoundsPri{} & \cveFactBoundsCon{} \\
\midrule
API & \cveFactAPI{} & \texttt{format!()} replaces \texttt{sprintf}, \texttt{str::}\texttt{parse::}\texttt{<f64>()} replaces custom number parser & \cveFactAPIPri{} & \cveFactAPICon{} \\
\midrule
Ownership & \cveFactOwnership{} & Owned \texttt{Property} clone, \texttt{String} clone of regexp source & \cveFactOwnershipPri{} & \cveFactOwnershipCon{} \\
\midrule
Other & \cveFactOther{} & Debug-mode \texttt{i32} overflow detection (PanicOnOverflow); translated code has additional depth check (BetterLogic) & \cveFactOtherPri{} & \cveFactOtherCon{} \\
\bottomrule
\end{tabular}
}%
\end{table}

\subsection{CS2: Performance}
\begin{figure*}[t]
\centering
\begin{subfigure}[b]{0.44\textwidth}
    \centering
    \includegraphics[width=\textwidth]{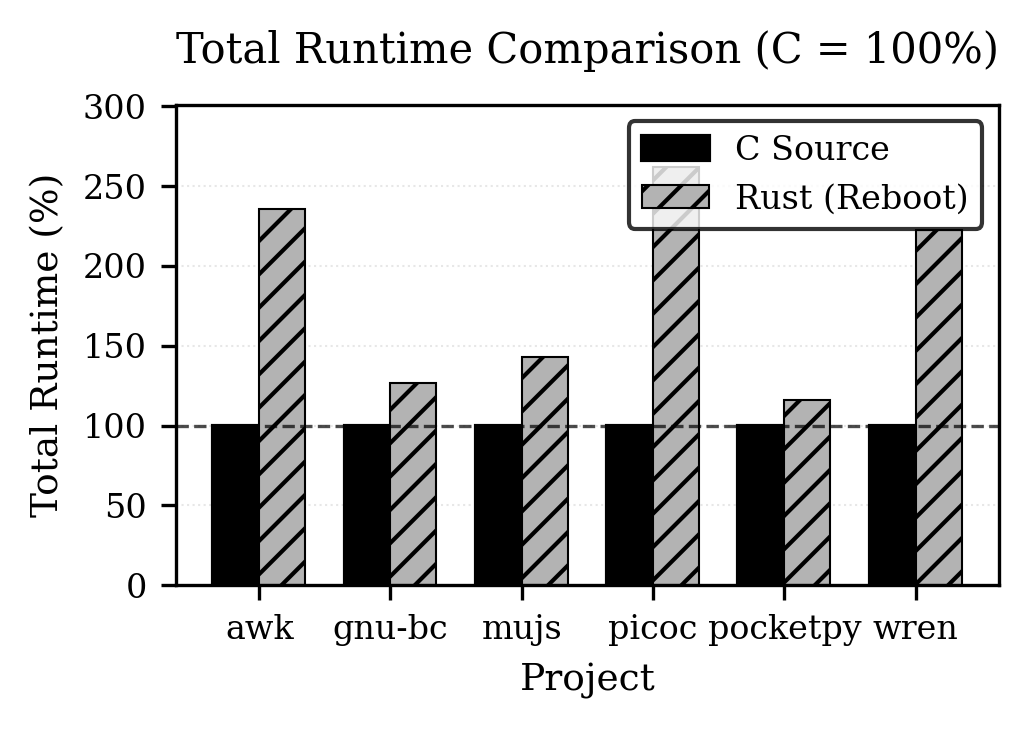}
    \caption{Total Runtime Comparison}
    \label{fig:perf-total}
\end{subfigure}
\hfill
\begin{subfigure}[b]{0.50\textwidth}
    \centering
    \includegraphics[width=\textwidth]{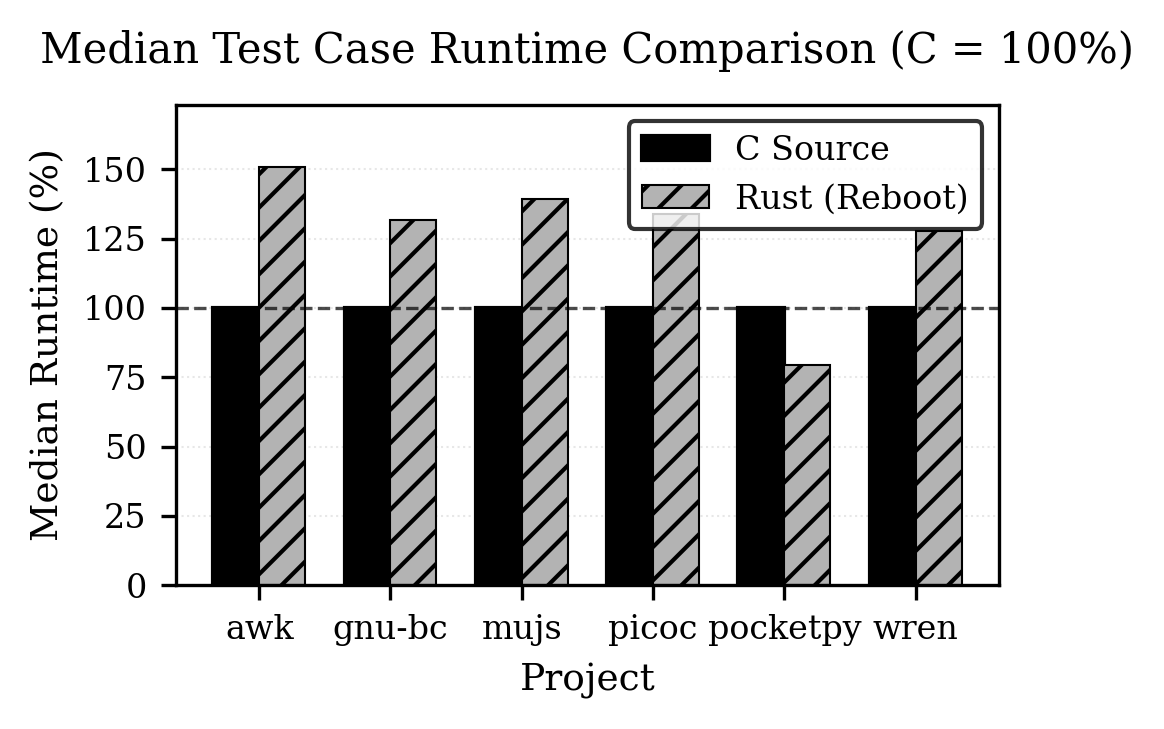}
    \caption{Median Test Case Runtime Comparison}
    \label{fig:perf-median}
\end{subfigure}
\caption{Performance evaluation results comparing C baseline with Rust (Reboot) implementations. All measurements use O3 optimization level. C baseline is normalized to 100\%. Results show that our Rust translations achieve competitive performance, with total runtime ranging from \perfReleaseTotalMinPct{} to \perfReleaseTotalMaxPct{} relative to C across all projects.}
\label{fig:perf}
\end{figure*}

\toadd{For each program, we invoke its test suite using a harness that runs the target binary once per test case, measuring wall-clock time from process start to process end (excluding the harness overhead but including the binary loading time when the process is started by the OS). Each test case is repeated 5 times and averaged. We report two aggregates: the total time across all test cases, and the median taken across all per-test-case times (so that a few slow-running tests do not dominate the summary).}
\toadd{Figure~\ref{fig:perf} compares the execution time of the original C programs (compiled with \texttt{-O3}) and the translated safe Rust programs (compiled with \texttt{-O3}) on the provided test suites, reporting both total runtime and the median across per-test-case runtimes, as percentages relative to the C baseline.}
\toadd{\perfReleaseSlowdownCount{} of the 6 programs show a median per-test slowdown, ranging from \perfReleaseSlowdownMinMs{}x to \perfReleaseSlowdownMaxMs{}x relative to C.
In terms of total execution time across all test cases, the overhead ranges from \perfReleaseTotalMinSec{}x to \perfReleaseTotalMaxSec{}x.
Overall, the translated Rust programs achieve performance that is not too far from the original C implementations.}
\toadd{Detailed per-program numbers including debug builds are provided in the appendix.} %

\toadd{We note that our translation makes no effort to further optimize the Rust code. Nevertheless, the overhead is already comparable to that reported by full memory safety enforcement for C~\cite{cets,softbound}, dynamic taint analysis tools~\cite{sang2024airtaint}, and address sanitizers~\cite{asan} that instrument C code directly.} 

\subsection{CS3: Ablation on Feature Reduction}
\toadd{To evaluate the contribution of feature reduction, we compare two configurations: \emph{Reboot}, which uses feature reduction together with the multi-agent system (MAS) as described in the previous sections, and \emph{Reboot w/o Feat.Red.}, which uses only the MAS to translate the full program directly without feature reduction (i.e., a single feature level).}
\toadd{Both configurations use the same MAS implementation, almost identical prompts (except for a clarification that only one feature level is expected), and the same LLM; the only difference is whether the translation is decomposed through feature reduction.}
\toadd{We compare the two configurations on all six of our main benchmark programs (mujs, awk, picoc, gnu-bc, wren, and pocketpy), as well as mujs-CVEs (the vulnerability-concentrated variant from CS1).}

\begin{table*}[htb]
    \centering
    \caption{\toadd{Ablation Study: Reboot (feature reduction + MAS) vs.\ Reboot w/o Feat.Red.\ (MAS only).$^{*}$Starred entries require minor test harness fixes before the translation can pass any tests; the reported numbers are after applying these fixes. $^\dag$Both pocketpy attempts without feature reduction failed to complete due to manager failures (see text).}}
    \resizebox{0.99\linewidth}{!}{%
    \begin{tabular}{l l r r r r r}
        \toprule
        \textbf{Program} & \textbf{Configuration} & \textbf{Time} & \textbf{UserInt.} & \textbf{Provided Tests} & \textbf{Validation Tests} & \textbf{Test262@mujs} \\
        \midrule
        mujs      & Reboot              & \resmujsTime & \escmujsUserInt  & \resmujsTestPass{} (\resmujsProvPct)   & \valmujsFrac{} (\valmujsPct)   & \ablTmujsFrac{} (\ablTmujsPct) \\
        mujs      & Reboot w/o Feat.Red. & \ablNFmujsTime & \ablNFmujsUserInt & \ablNFmujsProvFrac{} (\ablNFmujsProvPct)   & \ablNFmujsValFrac* (\ablNFmujsValPct)  & \ablTNFmujsFrac* (\ablTNFmujsPct) \\
        \midrule
        mujs-CVEs & Reboot              & \ablCVETime & \ablCVEUserInt  & \ablCVEProvFrac{} (\ablCVEProvPct)   & \ablCVEValFrac{} (\ablCVEValPct)   & \ablCVETFrac* (\ablCVETPct) \\
        mujs-CVEs & Reboot w/o Feat.Red. & \ablCVENFTime & \ablCVENFUserInt  & \ablCVENFProvFrac{} (\ablCVENFProvPct)   & \ablCVENFValFrac{} (\ablCVENFValPct)   & \ablCVENFTFrac{} (\ablCVENFTPct) \\
        \midrule
        awk       & Reboot              & \resawkTime & \escawkUserInt  & \resawkTestPass{} (\resawkProvPct)   & \valawkFrac{} (\valawkPct)   & -- \\
        awk       & Reboot w/o Feat.Red. & \ablNFawkTime & \ablNFawkUserInt & \ablNFawkProvFrac{} (\ablNFawkProvPct)   & \ablNFawkValFrac{} (\ablNFawkValPct)   & -- \\
        \midrule
        picoc     & Reboot              & \respicocTime & \escpicocUserInt  & \respicocTestPass{} (\respicocProvPct)   & \valpicocFrac{} (\valpicocPct)   & -- \\
        picoc     & Reboot w/o Feat.Red. & \ablNFpicocTime & \ablNFpicocUserInt & \ablNFpicocProvFrac{} (\ablNFpicocProvPct)   & \ablNFpicocValFrac{} (\ablNFpicocValPct)    & -- \\
        \midrule
        gnu-bc    & Reboot              & \resgnubcTime & \escgnubcUserInt  & \resgnubcTestPass{} (\resgnubcProvPct)   & \valgnubcFrac{} (\valgnubcPct)    & -- \\
        gnu-bc    & Reboot w/o Feat.Red. & \ablNFgnubcTime & \ablNFgnubcUserInt & \ablNFgnubcProvFrac{} (\ablNFgnubcProvPct)   & \ablNFgnubcValFrac{} (\ablNFgnubcValPct)    & -- \\
        \midrule
        wren      & Reboot              & \reswrenTime & \escwrenUserInt  & \reswrenTestPass{} (\reswrenProvPct) & \valwrenFrac{} (\valwrenPct) & -- \\
        wren      & Reboot w/o Feat.Red. & \ablNFwrenTime & \ablNFwrenUserInt & \ablNFwrenProvFrac{} (\ablNFwrenProvPct) & \ablNFwrenValFrac{} (\ablNFwrenValPct) & -- \\
        \midrule
        pocketpy  & Reboot              & \respocketpyTime & \escpocketpyUserInt & \respocketpyTestPass{} (\respocketpyProvPct) & \valpocketpyFrac{} (\valpocketpyPct) & -- \\
        pocketpy  & Reboot w/o Feat.Red.$^\dag$ (1st) & \ablNFpocketpyTime & \ablNFpocketpyUserInt & \ablNFpocketpyProvFrac{} (\ablNFpocketpyProvPct) & \ablNFpocketpyValFrac{} (\ablNFpocketpyValPct) & -- \\
        pocketpy  & Reboot w/o Feat.Red.$^\dag$ (2nd) & \ablNFiipocketpyTime & \ablNFiipocketpyUserInt & \ablNFiipocketpyProvFrac{} (\ablNFiipocketpyProvPct) & \ablNFiipocketpyValFrac{} (\ablNFiipocketpyValPct) & -- \\
        \bottomrule
    \end{tabular}
    }
    \label{tab:ablation}
\end{table*}
\toadd{Table~\ref{tab:ablation} presents the ablation results.}
\toadd{Reboot passes 100\% of the provided tests for all six programs. For the five programs where the configuration without feature reduction completed, it passes nearly all provided tests (99\% for gnu-bc and picoc; 100\% for the rest), indicating that the MAS alone is largely sufficient to produce translations that pass the provided test suites.}
\toadd{However, feature reduction consistently improves pass rates on the unseen validation tests: mujs \valmujsPct{} vs.\ \ablNFmujsValPct, picoc \valpicocPct{} vs.\ \ablNFpicocValPct, awk \valawkPct{} vs.\ \ablNFawkValPct, and gnu-bc \valgnubcPct{} vs.\ \ablNFgnubcValPct.}
\toadd{There are also significant improvements in correctness on the mujs Test262 conformance suite: Reboot achieves \ablTmujsPct{} compared to \ablTNFmujsPct{} without feature reduction, and the mujs-CVEs variant shows a similar pattern (\ablCVETPct{} vs.\ \ablCVENFTPct).}
\toadd{These results suggest that feature reduction leads to higher-quality translations that generalize better beyond the provided tests, likely because decomposing the translation into smaller, validated milestones allows translation details to be more carefully handled at each transition.}

\toadd{For pocketpy, the largest benchmark ($\sim$\bmMaxLocRounded{} LoC), the configuration without feature reduction failed to produce a working translation in two separate attempts.}
\toadd{In the first attempt, the Translator initially circumvented the task by invoking the original C binary instead of producing a genuine translation; after being caught, the Manager began guiding the actual translation but later assessed the task as too complex and declared it as failed---invoking a fail-early mechanism available to the Manager in all configurations but never triggered in any other run.}
\toadd{We then disabled this mechanism and made a second attempt. This time, unable to give up, the Manager instead stopped following its runbook and began sending modified instructions to the Validator and CodeReviewer that requested validation of only the already-passing tests---satisfying the FSM guard's structural requirements while bypassing actual validation. After the workflow finishes, our ablation configuration automatically reruns the entire translation and validation with a fresh Manager as a double-checking mechanism, but the same manipulation occurred in the double-checking run as well.}
\toadd{These failures illustrate a breakdown of the assumption from Section~\ref{sec:overview-orchestration} that $H$ is itself reliable: when the objective becomes too complex for the worker, the Manager---which implements $H$---also becomes unreliable, resorting to shortcuts or giving up. Feature reduction mitigates this by keeping each objective small enough that $H$ remains effective, suggesting that feature reduction may become more important as program complexity grows.}
\subsection{CS4: User Interventions}
\toadd{As reported in Table~\ref{tab:results}, the number of user interventions ranges from \escawkUserInt{} to \escpocketpyUserInt{} per program, each taking roughly 5 minutes.}
\toadd{All interventions are in the form of natural language guidance provided to the Manager agent; the user does not write or modify code directly.}
\toadd{We classify the \escTotalUserInt{} observed interventions into four categories:}
\begin{itemize}
\item \toadd{\textbf{Task clarification} (\escCatTask{} cases): The user clarifies what the task expects. Examples include informing the Manager that a coverage drop during feature reduction is not acceptable, and clarifying that task success requires 100\% test passage rather than partial results.}
\item \toadd{\textbf{Workflow clarification} (\escCatWorkflow{} cases): The user clarifies workflow expectations or non-negotiable requirements, such as instructing the Manager to continue without round limits as long as progress is being made, or confirming that CodeReviewer concerns on code quality must be addressed before proceeding.}
\item \toadd{\textbf{Agent misbehavior correction} (\escCatMisbehavior{} cases): An agent goes off-track despite $H$'s attempts to correct it. Examples include a Validator in mujs that removed 21 tests instead of syncing them, and a Translator in wren that stalled for three consecutive rounds producing analysis instead of code.}
\item \toadd{\textbf{Design decisions} (\escCatDesign{} cases): The user makes a design choice that the system cannot resolve on its own, such as deciding which \texttt{unsafe} POSIX APIs to drop versus wrap with safe crates in picoc.}
\end{itemize}
\toadd{The remaining \escCatSystem{} interventions are system-level issues (agent timeouts or token limits) resolved by a lightweight restart.}
\toadd{Besides design decisions, the other categories represent limitations of the current prompt design and iteration logic that could in principle be addressed by refining the Manager's instructions or the escalation-handling templates.}
\toadd{Notably, these \escTotalUserInt{} interventions represent only a fraction of the situations the system must handle. When the Manager escalates an issue, it is first intercepted by the User Delegator Agent (Figure~\ref{fig:arch}), which acts on behalf of the user for common escalation patterns such as output format mismatches, transient API errors, and iteration-limit resets. Across all six programs, \escTotalEsc{} escalations were raised; the User Delegator auto-resolved \escTotalAuto{} of them (\escAutoPct), forwarding only \escTotalUserInt{} to the human user.}
\toadd{In addition, \escTotalMinor{} minor system recovery actions (restarting a hanging agent or correcting a message format error) occurred across three programs; these are lightweight operations (taking seconds) and are not counted as interventions (shown as ``+$n$'' in Table~\ref{tab:results}).}
\toadd{Detailed per-program intervention logs, including verbatim user messages, are provided in the appendix.} %
\subsection{CS5: Applicability Beyond Interpreters}
\toadd{Feature reduction is natural for interpreters, where progressively reducing the input language yields a sequence of smaller, self-contained languages, and where the heavily cross-cutting, stateful data flow of an interpreter makes whole-program decomposition valuable. However, nothing in \tool's system design restricts it to interpreters; only the prompts are interpreter-specific and assume the input is an interpreter. To probe whether \tool applies more broadly, we ran it on two command-line utilities, \code{tail} and \code{split}, from the coreutils-based benchmark used by C2SaferRust~\cite{nitin2025c2saferrust}; both are among the larger CLI programs in that suite. We made minor adjustments to the prompts for these non-interpreter inputs but changed nothing in the system implementation. \tool translated each program to safe Rust with no \code{unsafe} blocks, passing all provided tests (Table~\ref{tab:compare-c2saferrust}). The translations stay close to the size of the C source, whereas C2Rust-derived pipelines such as C2SaferRust produce larger output that retains some \code{unsafe} constructs, consistent with its goal of reducing rather than eliminating unsafe. These two programs are only a preliminary probe, but they indicate that \tool is not fundamentally tied to interpreters.}

\begin{table}[htb]
    \centering
    \caption{\toadd{\tool on two command-line utilities from C2SaferRust's benchmark. \tool produces safe Rust with no \texttt{unsafe} blocks; the C2SaferRust columns are shown for reference.}}
    \label{tab:compare-c2saferrust}
    \resizebox{\linewidth}{!}{%
    \begin{tabular}{l r r r r r r r}
        \toprule
         & & \multicolumn{2}{c}{\textbf{Rust LoC}} & \multicolumn{2}{c}{\textbf{Raw pointer decls}} & \multicolumn{2}{c}{\textbf{Tests passing}} \\
        \cmidrule(lr){3-4}\cmidrule(lr){5-6}\cmidrule(lr){7-8}
        \textbf{Program} & \textbf{C LoC} & \textbf{\tool} & \textbf{C2SaferRust} & \textbf{\tool} & \textbf{C2SaferRust} & \textbf{\tool} & \textbf{C2SaferRust} \\
        \midrule
        tail      & 1758      &  2,511    & 11663 & 0 & 297 & 462/462  & All\footnotemark \\
        split   & 1349    & 2,163    & 11324 & 0 & 214 & 179/179    & All \\
        \bottomrule
    \end{tabular}
    }
\end{table}
\footnotetext{We directly use their reported result which mentions that all tests are passing.}

\toadd{\tool could in principle also target libraries, since a library can be wrapped in a command-line driver to provide the end-to-end tests that \tool relies on. The benefit of feature reduction would be smaller in this setting, however. We observe that many libraries have largely independent functions and features and carry far less of the cross-cutting, stateful data flow found in interpreters, so decomposing by feature would largely coincide with decomposing by syntactic units such as functions and modules. In that regime, feature reduction essentially reduces to the syntactic decomposition that existing work, such as SmartC2Rust~\cite{shiraishi2024smartc2rust}, has already shown to be effective.}

\section{Threats to Validity}
\label{sec:threats}

\myparagraph{Nondeterminism and Reproducibility.}
\toadd{Coding agents such as Claude Code are inherently nondeterministic~\cite{claudecodenondet}, so a different run of \tool on the same program may produce a different translation, different validation pass rates, and a different number of user interventions.}
\toadd{We fix the version of Claude Code and ensure that every agent call is started in an identical clean sandbox with identical configurations.}
\toadd{However, this is insufficient for achieving determinism in practice. One might propose request caching, but even if the API endpoint is made deterministic, external factors such as script execution timing and timestamps in tool call results can invalidate caches in reruns, causing the agent to diverge from a previous trajectory.}
\toadd{The natural mitigation is to perform repeated runs, but this is presently impractical given the high monetary cost (\$\resMinCostRounded{}--\$\resMaxCostRounded{} per program) and wall-clock time (\resMinTimeRounded{}--\resMaxTimeRounded{} hours per program).}
\toadd{Instead, we provide the full agent logs for all translations, enabling inspection of the complete translation process and the decisions made at each step.}

\myparagraph{Human-in-the-Loop Variability.}
\toadd{One of the authors served as the human-in-the-loop for all escalations across all six programs.}
\toadd{A user less familiar with the system might handle escalations differently, primarily affecting efficiency---for example, requiring more time to understand the escalation context or needing more intervention rounds to resolve an issue.}
\toadd{For the small number of design decisions (3 out of 29 interventions), a different user might make different choices, producing different translated code; we view this as an inherent aspect of translation, where multiple valid designs exist, rather than a quality concern.}
\toadd{In all cases, the user provides only natural language guidance and does not write or modify code directly.}

\myparagraph{Validation Tests and Correctness Measurement.}
\toadd{The validation test suites were created by the authors, which introduces potential for bias. We mitigate this by creating these test suites ahead of time, before examining any translation output, independently of and separately from the provided tests used during translation.}
\toadd{While these tests may not be comprehensive, they serve as an independent measure of correctness beyond the provided tests.}
\toadd{Using a different test suite may yield different pass rates and different magnitudes of improvement from feature reduction. For instance, both our validation tests and the Test262 conformance suite show that feature reduction improves correctness for mujs, but the gap differs substantially (\valmujsPct{} vs.\ \ablNFmujsValPct{} on validation tests; \ablTmujsPct{} vs.\ \ablTNFmujsPct{} on Test262).}
\toadd{A potential reason might be that our validation tests were created in a relatively short time with less effort compared with a comprehensive conformance suite, so the majority of the tests might be relatively easier to pass; both configurations can handle many of them, and the observed improvement gap is smaller than on a more comprehensive suite.}
\toadd{More broadly, correctness in this work is defined by passing tests; the translations may have semantic differences from the original C programs that are not captured by any test suite.}

\myparagraph{Scope and Benchmark Selection.}
\toadd{This work is scoped to language interpreter programs, and feature reduction is designed to decompose by language features---constructs recognized by the interpreter---which is a natural fit for this domain. While we believe feature-based decomposition is not limited to interpreters, the extent of programs to which this approach applies remains to be studied.}
\toadd{We further require that target programs be standalone (can be compiled and tested independently) and free of complex external dependencies. These are not fundamental limitations---sources of nondeterminism can be controlled, and external dependencies could be handled---but each may require additional technical innovation beyond the current system.}
\toadd{Because we target full safe Rust with no \code{unsafe} blocks, programs with components that inherently require unsafe operations (e.g., JIT compilation) are not suitable targets for our current approach. Relaxing this requirement to allow some \code{unsafe} code is possible in principle, but having agents generate \code{unsafe} Rust introduces significant risk of subtle memory safety bugs that are difficult to validate automatically, and would likely require additional safeguards.}
\toadd{Our evaluation covers six programs ranging from $\sim$\bmMinLocRounded{} to $\sim$\bmMaxLocRounded{} lines of code. While we expect similar results on interpreter programs of comparable size and structure, generalization to significantly larger or architecturally different interpreters remains to be validated.}

\toadd{Lastly, since all of our benchmarks are open-source, the C source code might be present in the LLM's pretraining data. However, to the best of our knowledge, their safe Rust translations did not yet exist, so it is unlikely that translations were in the pretraining set. We may nonetheless see different results when translating programs that are unlikely to be in the LLM's pretraining data.}

\myparagraph{Agent Framework and LLM.}
\toadd{All experiments use Claude Code (v2.0.25) as the agent framework and Claude Sonnet 4.5 as the underlying LLM, which is representative of the best performing  coding agents and LLMs available at the time of evaluation.}
\toadd{Adapting to a different agent framework or LLM would require some implementation work---adapting the controller implementation, tuning prompts, as well as re-running the evaluation and re-analyzing the results---at nontrivial cost.}
\toadd{How the results would differ with other agent frameworks or LLMs, which may have different capabilities and failure modes, remains unknown.}
\toadd{A related concern affects the security case study (CS1): because the LLM's training data may include public CVE fixes, the agent could in principle apply a known fix during translation rather than producing a faithful translation of the vulnerable C code. Some security improvements reported in CS1 may therefore partly reflect the LLM's knowledge of known fixes rather than inherent properties of the C-to-safe-Rust translation.}

\section{Related Work}
\label{sec:related}

\myparagraph{Rule-based C-to-Rust Translation.}
\toadd{Early tools such as C2Rust~\cite{c2rust} and Corrode~\cite{corrode} perform direct syntactic translation from C to Rust, preserving the structure of the original program but producing code that relies heavily on \code{unsafe} blocks with few safety guarantees.}
\toadd{Subsequent work applies program analysis to reduce the amount of \code{unsafe} code in the output.}
\toadd{Emre et~al.~\cite{emre2021translating} use pointer and ownership analysis to eliminate unnecessary \code{unsafe} uses, and later show that aliasing patterns in real C programs fundamentally limit how much safety can be recovered through this approach~\cite{emre2023aliasing}.}
\toadd{Zhang et~al.~\cite{zhang2023ownership} infer ownership information to guide translation, and Ling et~al.~\cite{ling2022rust} define source-to-source rewriting rules that produce safer API-level Rust.}
\toadd{Other work targets specific C idioms: Hong and Ryu translate C unions to Rust tagged unions~\cite{taggedunion}, replace output parameters with algebraic data types~\cite{outputparam}, and translate I/O APIs~\cite{fileapi}.}
\toadd{Fromherz and Protzenko~\cite{fromherz2024compile} formalize a type-directed compilation from a subset of C to safe Rust.}
\toadd{Wu and Demsky~\cite{wu2025genc2rust} statically analyze void-pointer usage in C and retype parametric polymorphic pointers into Rust generics.}
\toadd{These rule-based approaches have individually demonstrated improvements on specific patterns, but have not yet been shown to produce fully safe translations of large C programs end-to-end.}
\toadd{A user study by Li et~al.~\cite{userstudy} evaluates two representative tools (Laertes~\cite{emre2021translating} and Crown~\cite{zhang2023ownership}) and finds that the vast majority of data references remain as raw pointers in their output, with spatial and temporal memory vulnerabilities persisting in the translated code.}

\myparagraph{LLM-based C-to-Rust Translation.}
\toadd{LLM-based approaches can synthesize Rust code that departs from the structure of the C source, making it possible to produce safe, idiomatic translations.}
\toadd{Early work by Lachaux et~al.~\cite{lachaux2020unsupervised} demonstrates unsupervised neural translation between programming languages, though not targeting Rust specifically.}
\toadd{For C-to-Rust, several approaches translate at the function level with various forms of feedback: Eniser et~al.~\cite{eniser2024towards} use differential fuzzing to validate translations, Yang et~al.~\cite{yang2024vert} verify equivalence against a WebAssembly oracle, and Farrukh et~al.~\cite{farrukh2025safetrans} apply iterative LLM-based repair.}
\toadd{Hong et~al.~\cite{hong2025type} and Xu et~al.~\cite{xu2025typemigration} focus specifically on migrating C types to idiomatic Rust types using LLM-driven analysis, while Nitin et~al.~\cite{nitin2024spectra} guide translation by first generating multi-modal specifications.}
\toadd{Nitin et~al.~\cite{nitin2025c2saferrust} refine C2Rust output by slicing programs into chunks and using an LLM to reduce unsafe code, evaluating on programs up to 96k LoC.}
\toadd{Luo et~al.~\cite{luo2025irene} combine rule-augmented retrieval with error-driven iterative refinement, and Sim et~al.~\cite{sim2025lac2r} use Monte Carlo tree search over heterogeneous LLMs with virtual fuzzing-based equivalence tests to improve translations.}
\toadd{To scale beyond individual functions, several approaches decompose programs by syntactic structure: Shiraishi et~al.~\cite{shiraishi2024smartc2rust} segment code into context-aware translation units, Cai et~al.~\cite{cai2025rustmap} use dependency-guided decomposition at the project level, Ou et~al.~\cite{ou2025enhancing} augment function-level translation with repository-level context, and Zhou et~al.~\cite{zhou2025llm} translate functions individually using FFI test harnesses to validate each in isolation.}
\toadd{Wang et~al.~\cite{wang2025evoc2rust} generate a compilable Rust skeleton and incrementally translate functions, and Yuan et~al.~\cite{yuan2025ptrmapper} build a pointer knowledge graph to guide ownership and lifetime inference at the project level.}
\toadd{Syzygy~\cite{syzygy} pairs code translation with test translation and uses dynamic analysis to guide the process.}
\toadd{These approaches decompose programs by syntactic units such as functions, files, or dependency graphs; the resulting Rust code typically respects the original modular boundaries.}
\toadd{By contrast, \tool decomposes by program feature rather than syntactic structure, allowing agents to freely restructure code as needed to satisfy Rust's ownership rules.}

\myparagraph{Agents and Multi-Agent Systems.}
\toadd{Multi-agent architectures have been explored for software engineering tasks such as code generation~\cite{huang2023agentcoder,qian2024chatdev,hong2024metagpt}, where agents take on specialized roles (e.g., coder, tester, reviewer) and collaborate through structured communication.}
\toadd{General-purpose multi-agent frameworks~\cite{wu2024autogen} and self-reflection mechanisms~\cite{shinn2023reflexion} provide foundations for iterative agent workflows.}
\toadd{For C-to-Rust translation specifically, ACToR~\cite{actor} uses an adversarial generator-discriminator architecture to iteratively improve translations, achieving over 90\% test pass rates on CLI utilities of several hundred lines of code.}
\toadd{\tool builds on this line of work but targets programs that are an order of magnitude larger (6k--23k LoC), which requires workflows that run reliably for dozens of hours.}
\toadd{\tool addresses this through validation agents, automated history-based feedback, and finite-state-machine guards that enforce workflow protocols---mechanisms designed to detect and recover from the various failure modes that arise in long-running agent workflows.}
\section{Conclusion}
\label{sec:conclusion}

\toadd{We presented \tool, a mostly-automatic technique for translating interpreter programs from C to safe Rust. \tool combines feature reduction, which decomposes the translation by program features into validated milestones, with multi-agent orchestration that keeps long-running agent workflows on track through automated validation and feedback. Using \tool, we translated six interpreters (\bmMinLocRounded--\bmMaxLocRounded{} LoC) to safe Rust with no \texttt{unsafe} blocks, passing 100\% of provided tests and \valMinPct--\valMaxPct{} of unseen validation tests, with only \escawkUserInt{} to \escpocketpyUserInt{} brief user interventions per program.}

\toadd{An ablation study confirms that feature reduction consistently improves translation correctness and becomes critical as program complexity grows---without it, the largest benchmark failed to produce a working translation. Extending the decomposition strategy beyond interpreters, further reducing user interventions, and closing the correctness gap on unseen tests are directions for future work.}

\section*{Data-Availability Statement}

\toadd{The \tool source code, benchmarks, translated Rust programs, and agent logs for all the translations will be made publicly available.}

\bibliographystyle{ACM-Reference-Format}
\bibliography{main}

\appendix

\section{More Detailed Statistics of the Translation Process}

This appendix provides additional detailed statistics from our evaluation of \tool on the three interpreter programs (awk, picoc, and mujs).

\begin{figure*}[t]
\centering
\begin{subfigure}[b]{0.32\textwidth}
    \centering
    \includegraphics[width=\textwidth]{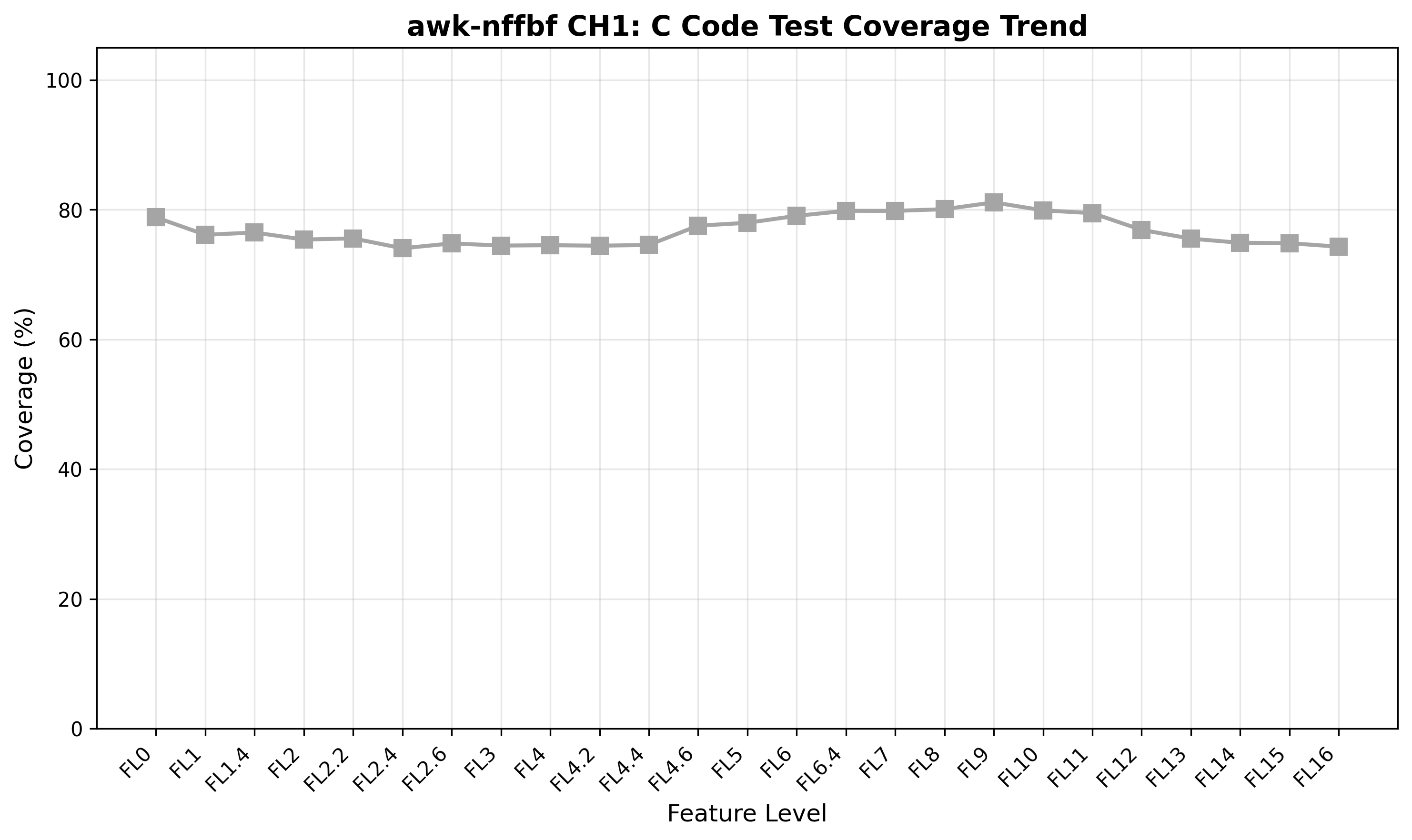}
    \caption{awk}
    \label{fig:eval-coverage-trend-awk}
\end{subfigure}
\hfill
\begin{subfigure}[b]{0.32\textwidth}
    \centering
    \includegraphics[width=\textwidth]{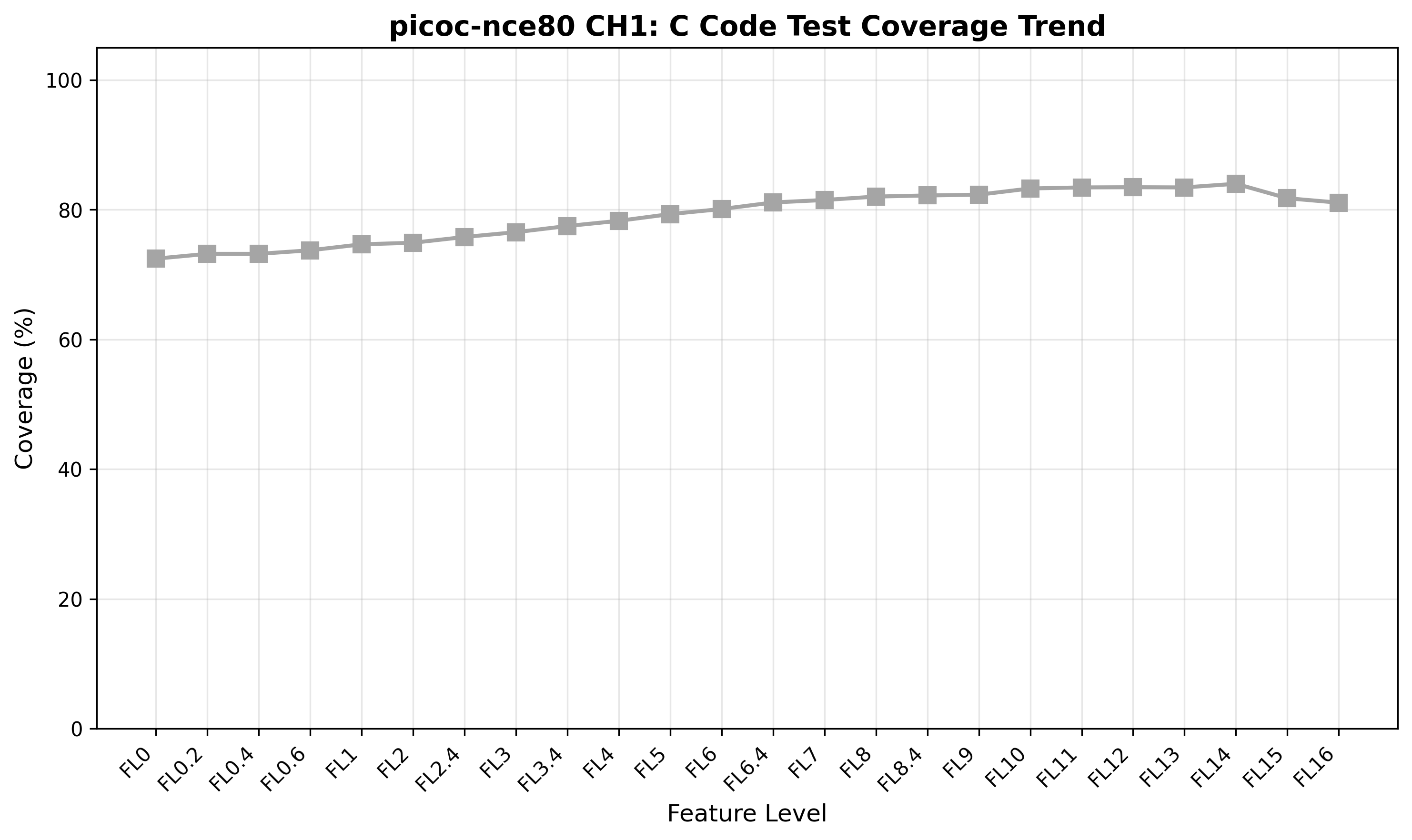}
    \caption{picoc}
    \label{fig:eval-coverage-trend-picoc}
\end{subfigure}
\hfill
\begin{subfigure}[b]{0.32\textwidth}
    \centering
    \includegraphics[width=\textwidth]{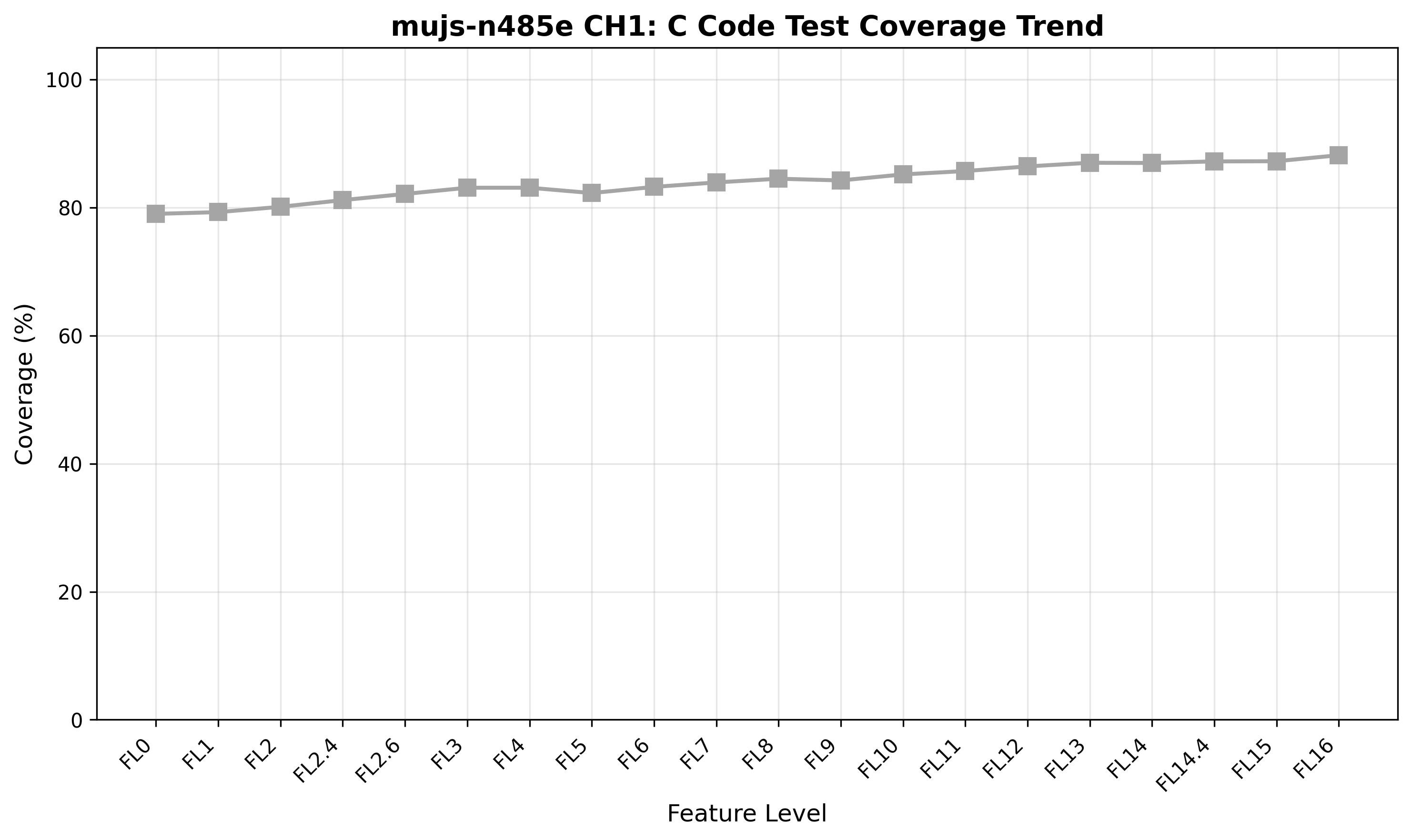}
    \caption{mujs}
    \label{fig:eval-coverage-trend-mujs}
\end{subfigure}

\vspace{1em}

\begin{subfigure}[b]{0.32\textwidth}
    \centering
    \includegraphics[width=\textwidth]{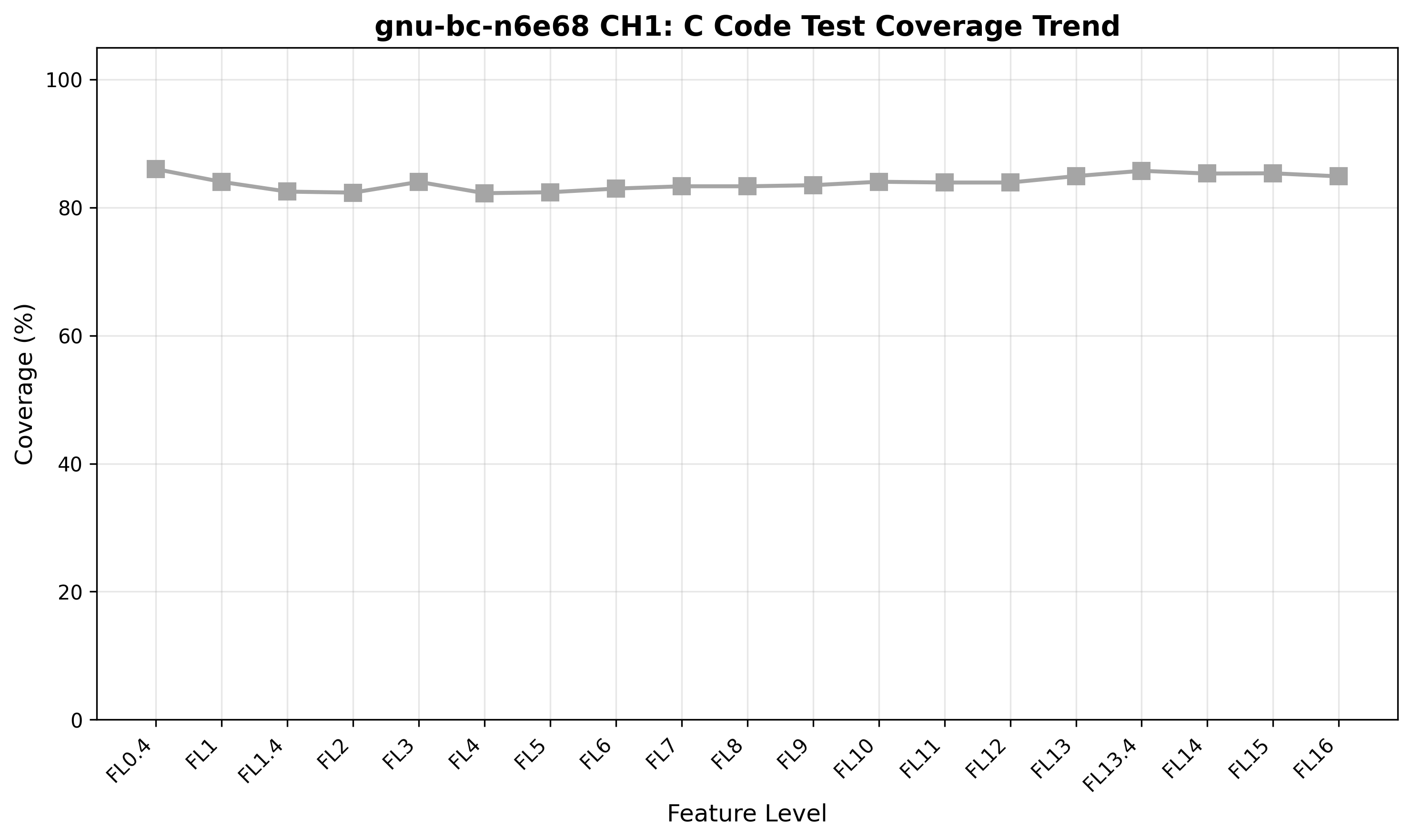}
    \caption{gnu-bc}
    \label{fig:eval-coverage-trend-gnu-bc}
\end{subfigure}
\hfill
\begin{subfigure}[b]{0.32\textwidth}
    \centering
    \includegraphics[width=\textwidth]{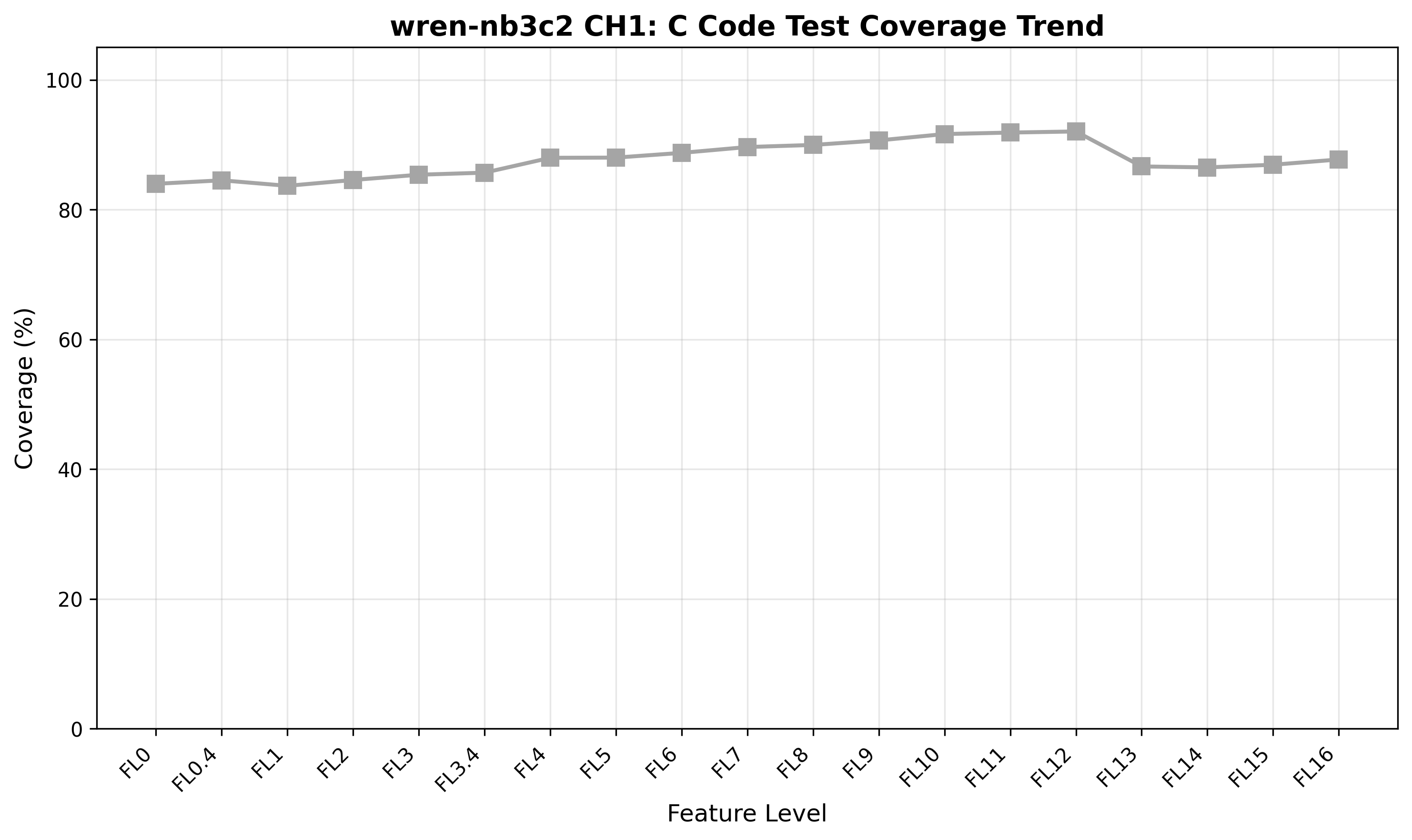}
    \caption{wren}
    \label{fig:eval-coverage-trend-wren}
\end{subfigure}
\hfill
\begin{subfigure}[b]{0.32\textwidth}
    \centering
    \includegraphics[width=\textwidth]{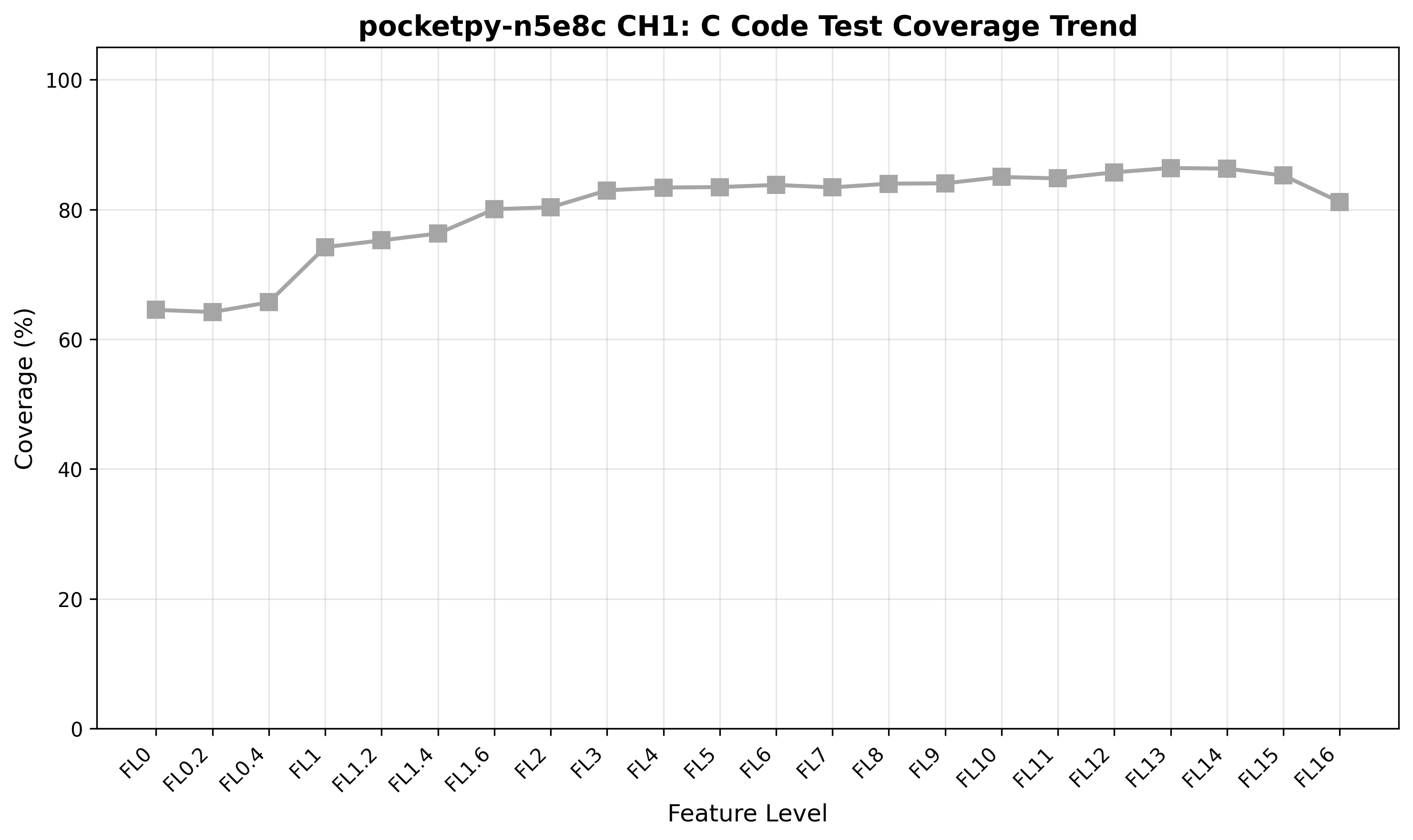}
    \caption{pocketpy}
    \label{fig:eval-coverage-trend-pocketpy}
\end{subfigure}
\caption{C code test coverage trend across feature levels. The charts show the percentage of C code covered by tests as features are incrementally added during the translation process.}
\label{fig:eval-coverage-trend}
\end{figure*}

Figure~\ref{fig:eval-coverage-trend} shows the test coverage trend of the C code across feature levels during the source reduction phase. The coverage is roughly preserved throughout the reduction process, ensuring that simplified versions maintain similar test coverage as the original programs.

Table~\ref{tab:perf-full} provides the full performance data for all six programs, including both debug (O0) and release (O3) builds, as well as the ablation variant (Rust-NF) where available.

\begin{table*}[t]
\centering
\caption{Performance Evaluation Results}
\label{tab:perf-full}
\small
\begin{tabular}{llrrrr}
\toprule
\textbf{Project} & \textbf{Variant} & \textbf{Total (s)} & \textbf{Total (\%)} & \textbf{Median (ms)} & \textbf{Median (\%)} \\
\midrule
\textit{awk} & C (O0) & \awkCdebugTotalSec & \awkCdebugTotalPct & \awkCdebugMedianMs & \awkCdebugMedianPct \\
 & Rust (O0) & \awkREBOOTdebugTotalSec & \awkREBOOTdebugTotalPct & \awkREBOOTdebugMedianMs & \awkREBOOTdebugMedianPct \\
 & Rust-NF (O0) & \awkREBOOTNFdebugTotalSec & \awkREBOOTNFdebugTotalPct & \awkREBOOTNFdebugMedianMs & \awkREBOOTNFdebugMedianPct \\
\textit{awk} & C (O3) & \awkCreleaseTotalSec & \awkCreleaseTotalPct & \awkCreleaseMedianMs & \awkCreleaseMedianPct \\
 & Rust (O3) & \awkREBOOTreleaseTotalSec & \awkREBOOTreleaseTotalPct & \awkREBOOTreleaseMedianMs & \awkREBOOTreleaseMedianPct \\
 & Rust-NF (O3) & \awkREBOOTNFreleaseTotalSec & \awkREBOOTNFreleaseTotalPct & \awkREBOOTNFreleaseMedianMs & \awkREBOOTNFreleaseMedianPct \\
\midrule
\textit{gnu-bc} & C (O0) & \gnubcCdebugTotalSec & \gnubcCdebugTotalPct & \gnubcCdebugMedianMs & \gnubcCdebugMedianPct \\
 & Rust (O0) & \gnubcREBOOTdebugTotalSec & \gnubcREBOOTdebugTotalPct & \gnubcREBOOTdebugMedianMs & \gnubcREBOOTdebugMedianPct \\
 & Rust-NF (O0) & \gnubcREBOOTNFdebugTotalSec & \gnubcREBOOTNFdebugTotalPct & \gnubcREBOOTNFdebugMedianMs & \gnubcREBOOTNFdebugMedianPct \\
\textit{gnu-bc} & C (O3) & \gnubcCreleaseTotalSec & \gnubcCreleaseTotalPct & \gnubcCreleaseMedianMs & \gnubcCreleaseMedianPct \\
 & Rust (O3) & \gnubcREBOOTreleaseTotalSec & \gnubcREBOOTreleaseTotalPct & \gnubcREBOOTreleaseMedianMs & \gnubcREBOOTreleaseMedianPct \\
 & Rust-NF (O3) & \gnubcREBOOTNFreleaseTotalSec & \gnubcREBOOTNFreleaseTotalPct & \gnubcREBOOTNFreleaseMedianMs & \gnubcREBOOTNFreleaseMedianPct \\
\midrule
\textit{mujs} & C (O0) & \mujsCdebugTotalSec & \mujsCdebugTotalPct & \mujsCdebugMedianMs & \mujsCdebugMedianPct \\
 & Rust (O0) & \mujsREBOOTdebugTotalSec & \mujsREBOOTdebugTotalPct & \mujsREBOOTdebugMedianMs & \mujsREBOOTdebugMedianPct \\
 & Rust-NF (O0) & \mujsREBOOTNFdebugTotalSec & \mujsREBOOTNFdebugTotalPct & \mujsREBOOTNFdebugMedianMs & \mujsREBOOTNFdebugMedianPct \\
\textit{mujs} & C (O3) & \mujsCreleaseTotalSec & \mujsCreleaseTotalPct & \mujsCreleaseMedianMs & \mujsCreleaseMedianPct \\
 & Rust (O3) & \mujsREBOOTreleaseTotalSec & \mujsREBOOTreleaseTotalPct & \mujsREBOOTreleaseMedianMs & \mujsREBOOTreleaseMedianPct \\
 & Rust-NF (O3) & \mujsREBOOTNFreleaseTotalSec & \mujsREBOOTNFreleaseTotalPct & \mujsREBOOTNFreleaseMedianMs & \mujsREBOOTNFreleaseMedianPct \\
\midrule
\textit{picoc} & C (O0) & \picocCdebugTotalSec & \picocCdebugTotalPct & \picocCdebugMedianMs & \picocCdebugMedianPct \\
 & Rust (O0) & \picocREBOOTdebugTotalSec & \picocREBOOTdebugTotalPct & \picocREBOOTdebugMedianMs & \picocREBOOTdebugMedianPct \\
 & Rust-NF (O0) & \picocREBOOTNFdebugTotalSec & \picocREBOOTNFdebugTotalPct & \picocREBOOTNFdebugMedianMs & \picocREBOOTNFdebugMedianPct \\
\textit{picoc} & C (O3) & \picocCreleaseTotalSec & \picocCreleaseTotalPct & \picocCreleaseMedianMs & \picocCreleaseMedianPct \\
 & Rust (O3) & \picocREBOOTreleaseTotalSec & \picocREBOOTreleaseTotalPct & \picocREBOOTreleaseMedianMs & \picocREBOOTreleaseMedianPct \\
 & Rust-NF (O3) & \picocREBOOTNFreleaseTotalSec & \picocREBOOTNFreleaseTotalPct & \picocREBOOTNFreleaseMedianMs & \picocREBOOTNFreleaseMedianPct \\
\midrule
\textit{pocketpy} & C (O0) & \pocketpyCdebugTotalSec & \pocketpyCdebugTotalPct & \pocketpyCdebugMedianMs & \pocketpyCdebugMedianPct \\
 & Rust (O0) & \pocketpyREBOOTdebugTotalSec & \pocketpyREBOOTdebugTotalPct & \pocketpyREBOOTdebugMedianMs & \pocketpyREBOOTdebugMedianPct \\
\textit{pocketpy} & C (O3) & \pocketpyCreleaseTotalSec & \pocketpyCreleaseTotalPct & \pocketpyCreleaseMedianMs & \pocketpyCreleaseMedianPct \\
 & Rust (O3) & \pocketpyREBOOTreleaseTotalSec & \pocketpyREBOOTreleaseTotalPct & \pocketpyREBOOTreleaseMedianMs & \pocketpyREBOOTreleaseMedianPct \\
\midrule
\textit{wren} & C (O0) & \wrenCdebugTotalSec & \wrenCdebugTotalPct & \wrenCdebugMedianMs & \wrenCdebugMedianPct \\
 & Rust (O0) & \wrenREBOOTdebugTotalSec & \wrenREBOOTdebugTotalPct & \wrenREBOOTdebugMedianMs & \wrenREBOOTdebugMedianPct \\
 & Rust-NF (O0) & \wrenREBOOTNFdebugTotalSec & \wrenREBOOTNFdebugTotalPct & \wrenREBOOTNFdebugMedianMs & \wrenREBOOTNFdebugMedianPct \\
\textit{wren} & C (O3) & \wrenCreleaseTotalSec & \wrenCreleaseTotalPct & \wrenCreleaseMedianMs & \wrenCreleaseMedianPct \\
 & Rust (O3) & \wrenREBOOTreleaseTotalSec & \wrenREBOOTreleaseTotalPct & \wrenREBOOTreleaseMedianMs & \wrenREBOOTreleaseMedianPct \\
 & Rust-NF (O3) & \wrenREBOOTNFreleaseTotalSec & \wrenREBOOTNFreleaseTotalPct & \wrenREBOOTNFreleaseMedianMs & \wrenREBOOTNFreleaseMedianPct \\
\bottomrule
\end{tabular}
\end{table*}

\section{Detailed CVE Security Analysis}

This appendix provides a detailed analysis of the \cveAppl{} CVEs re-introduced into mujs (Section~5, CS1 in the main paper). The root cause analysis, elimination classification, and factor attribution are based on best-effort manual inspection of the C and Rust source code. Table~\ref{tab:cvemore} shows the full per-CVE breakdown, including the C root cause, observed Rust behavior in both debug and release builds, elimination status, and the Rust feature(s) responsible for elimination or mitigation.

\newcolumntype{S}[1]{>{\fontsize{5}{5}\selectfont\raggedright\arraybackslash}p{#1}}%
\begin{table*}[t]
\centering
\caption{Detailed CVE analysis: C (mujs-CVEs) vs.\ safe Rust translation. For each of the \cveAppl{} re-introduced CVEs, we show the C vulnerability class, root cause, elimination status, the Rust feature(s) responsible, and additional notes. Rust O0/O3 behavior is shown in the CVE summary table in Section~5 of the main paper.}
\label{tab:cvemore}
\resizebox{\textwidth}{!}{%
\tiny
\setlength{\tabcolsep}{3pt}
\begin{tabular}{@{}l l S{0.22\textwidth} c p{0.12\textwidth} S{0.28\textwidth}@{}}
\toprule
\textbf{CVE} & \textbf{C Vuln Class} & {\tiny\textbf{C Root Cause}} & \textbf{Status} & \textbf{Elimination Factors} & {\tiny\textbf{Notes}} \\
\midrule
CVE-2022-44789 & Heap-UAF & \texttt{setproperty()} traverses prototype chain; custom setter frees cached property pointer & Eliminated & Ownership & Rust returns owned \texttt{Property} clone, not raw pointer; local copy valid even if setter deletes original \\
\midrule
CVE-2022-30974 & Stack-Exhaust & \texttt{count()} recursion has no depth limit; deeply nested regex patterns exhaust stack during size calculation & Unmitigated & N/A & Same unbounded recursion in \texttt{count\_instructions()}; Rust panics safely but DoS remains \\
\midrule
CVE-2021-45005 & Bytecode-Logic & \texttt{labeljumps()} doesn't clear jump list; re-compiling finally block repatches stale jumps with wrong addresses & Eliminated & Architecture & Rust uses scope-based \texttt{LoopContext} with \texttt{Vec<usize>} on compiler stack; fresh context per compilation prevents stale jump accumulation \\
\midrule
CVE-2021-33797 & Int-Overflow & Exponent parsing loop has no bounds; \texttt{exp*10} overflows int; overflowed value indexes \texttt{powersOf10[]} & Eliminated & API & \texttt{str::parse::<f64>()} replaces custom parser; no \texttt{powersOf10[]} array; returns \texttt{inf} for huge exponents \\
\midrule
CVE-2021-33796 & Heap-UAF & \texttt{pushliteral()} stores raw pointer to regexp source; GC frees regexp, pointer dangles & Eliminated & Ownership & \texttt{Value::string()} clones string data; stack value independent of regexp object lifetime \\
\midrule
CVE-2020-24343 & Heap-UAF & Missing mark check in GC iterator scan; double-marking corrupts linked list, premature free & Eliminated & Architecture & Manual mark-and-sweep GC replaced by \texttt{Rc<RefCell<>>} refcounting; exception is unrelated JS-level error \\
\midrule
CVE-2019-12798 & Heap-BOF & \texttt{strlen(pattern)*2} overflows int; undersized buffer allocated; parsing writes past end & Mitigated & Architecture & No pre-allocated buffer in Rust (dynamic \texttt{Box<Renode>}); heap overflow gone but recursive tree traversal causes stack overflow \\
\midrule
CVE-2019-11413 & Stack-Exhaust & \texttt{match()} depth check removed; unbounded recursion on deep alternation patterns during regex execution & Eliminated & BetterLogic & Translation artifact: Rust \texttt{do\_match()} has \texttt{MAX\_DEPTH=1024} check that the CVE-patched C lacks; returns controlled ``regexec failed'' error (exit 1) \\
\midrule
CVE-2019-11412 & Bytecode-Logic & Missing \texttt{OP\_ENDTRY} after \texttt{OP\_ENDCATCH}; exception stack leaks one entry per try/catch/finally iteration & Eliminated & Architecture & Same bytecode bug present; local \texttt{Vec<ExceptionHandler>} dropped per function call prevents accumulation (vs persistent \texttt{trybuf[]} array) \\
\midrule
CVE-2019-11411 & Stack-BOF & \texttt{sprintf()} writes 40 bytes into 32-byte stack buffer in \texttt{numtostr()} & Eliminated & API & \texttt{format!()} macro returns heap-allocated \texttt{String}; no fixed-size buffer exists \\
\midrule
CVE-2018-6191 & Global-BOF & Int overflow in exponent parsing bypasses range check; loop reads past 9-element \texttt{powersOf10[]} & Eliminated & API & \texttt{str::parse::<f64>()} replaces custom 680-line parser; no manual exponent loop or array \\
\midrule
CVE-2018-5759 & Stack-Exhaust & \texttt{INCREC()/DECREC()} macros disabled; no AST depth limit during parsing of chained binary expressions & Unmitigated & N/A & Same vulnerability (no depth tracking in Rust parser); PoC doesn't trigger under normal builds but ASAN (reduced stack) detects stack-overflow in compile phase \\
\midrule
CVE-2017-5628 & OOB-Read & \texttt{(int)NaN} is UB; garbage value used as index into \texttt{firstDayOfMonth[2][12]} array & Eliminated & TypeSafety & \texttt{NaN as usize = 0} (Rust defined semantics); valid index; bounds check also present as defense in depth \\
\midrule
CVE-2016-10141 & Heap-BOF & \texttt{count()*min} overflows signed int; undersized allocation; emit writes past buffer & Mitigated & TypeSafety & \texttt{usize} (64-bit) prevents int overflow wrapping; correct huge value causes OOM abort instead of heap corruption \\
\midrule
CVE-2016-10133 & Heap-BOF & Swapped operands in \texttt{js\_pop(numparams-n)} yield negative; \texttt{TOP -= (-9)} grows past heap buffer & Mitigated & Architecture, TypeSafety, Bounds & Rust doesn't pop excess args (different design); \texttt{pop(n: usize)} can't be negative; stack push/pop bounds-checked \\
\midrule
CVE-2016-9294 & NULL-Deref & \texttt{breaktarget()} starts at \texttt{stm} not \texttt{stm->parent}; break finds itself as target; \texttt{cexit()} NULL-derefs traversing past root & Eliminated & Architecture & Stack-based \texttt{label\_stack} and \texttt{try\_stack} replace AST parent-pointer traversal; no infinite loop or NULL deref possible \\
\midrule
CVE-2016-9136 & Heap-BOF & Missing EOF check in \texttt{jsY\_next()} + missing \texttt{case EOF} in escape switch; reads past buffer & Eliminated & TypeSafety, Bounds & \texttt{Option<u8>} return forces EOF handling; \texttt{advance()} bounds-checks \texttt{pos < len}; controlled SyntaxError \\
\midrule
CVE-2016-9109 & Heap-BOF & Missing \texttt{else} before \texttt{jsY\_next()} causes double-advance past buffer in comment parser & Eliminated & Bounds, TypeSafety & Single \texttt{advance()} per loop iteration; \texttt{Option<u8>} forces EOF handling; controlled SyntaxError \\
\midrule
CVE-2016-9108 & Int-Overflow & Overflow check after loop; \texttt{yymin*10+digit} wraps signed int negative; bypasses \texttt{>=REPINF} check & Mitigated & PanicOnOverflow & O0/O3 divergence: O0 panics on \texttt{i32} overflow at regexp.rs:452; O3 wraps but \texttt{as u8} truncation and bounds checking prevent heap corruption \\
\midrule
CVE-2016-7506 & Heap-BOF & Missing \texttt{case 0:} in replace switch; \texttt{*(++r)} reads past null terminator when \texttt{\$} is last char & Eliminated & Bounds & Explicit \texttt{i+1 < r\_bytes.len()} check before access; length-tracked string replaces null-terminated \texttt{char*} \\
\bottomrule
\end{tabular}
}%
\end{table*}

\section{Detailed User Intervention Logs}
\label{sec:appendix-escalations}

This appendix provides the complete log of user interventions
across all six benchmark programs.
Each entry shows the escalation context, our categorization, and the
verbatim text provided by the user.

\myparagraph{Terminology.}
In the implementation, the Manager agent from the paper
is referred to as the ``meta agent'' in the user-facing messages below.
Similarly, ``W0'' or ``worker'' refers to the active Worker agent
(Translator or Simplifier), and ``W1'' refers to the Validator.

Escalations are numbered sequentially per program run.
The category labels are: \textbf{Task} (task clarification),
\textbf{Workflow} (workflow clarification),
\textbf{Misbehavior} (agent misbehavior correction),
\textbf{Design} (design decision), and
\textbf{System} (system/technical issue).

\subsection{awk (13 escalations, 12 auto-resolved, 1 user interventions)}

{\small\noindent\textbf{ESC 7\quad [Task Clarification]\quad Translation, RS-FL4.6}

\noindent Worker claimed TASK SUCCESS with 134/191 tests passing (57 failing). User clarified: 100\% is mandatory, focus on failures not pass rates, send worker back to fix all 57 remaining failures.
}

\begin{quote}\footnotesize
Please treat this similar to the lack-of-correct-marker case and automatically handle this.
\end{quote}

\subsection{gnu-bc (5 escalations, 4 auto-resolved, 1 user interventions)}

{\small\noindent\textbf{ESC 2\quad [Task Clarification]\quad Reduction, C-FL13}

\noindent Task 6 claimed coverage improvements but Task 7 validation showed identical old numbers. User instructed worker to re-validate all test cases with fresh numbers and confirm no stale test remnants.
}

\begin{quote}\footnotesize
Tell meta agent to ask W0 to carefully re-validate what it did, emphasize systematic check on test cases and our sctict requirement of no test remnant and remaining tests 100\% passing. After that, ask W1 to re-build and re-validate.
\end{quote}

\subsection{picoc (17 escalations, 13 auto-resolved, 4 user interventions)}

{\small\noindent\textbf{ESC 6\quad [Workflow Clarification]\quad Translation, RS-FL6}

\noindent Hit 10-iteration escalation limit in translation stage. Made progress (+9 tests) but still 22 failures remaining. User suspended iteration limits and instructed continuation without limits.
}

\begin{quote}\footnotesize
Besides the clarifications, tell meta agent that we no longer have any limits on the number of rounds as long as there is progress. Temporary test regression is normal, as long as the translation / refactoring / debugging work is carried out systematically. Eventually we must achieve full safe Rust translation that is fully equivalent to C and 100\% tests passing. This eventural goal is not negotiable.
\end{quote}

{\small\noindent\textbf{ESC 13\quad [Workflow Clarification]\quad Translation, RS-FL8}

\noindent Code review identified 757-line function needing refactoring. Worker recommended accepting current state, estimating 10-14h more. User rejected acceptance and required completion with no resource limits.
}

\begin{quote}\footnotesize
The code quality issue MUST be addressed. Bad design, lack of proper modular design, or significant divergence from C program's architecture is not acceptable. Tell meta agent to clarify to worker agent to get this done systematically and properly, and we no longer have limits on rounds or time. Do the correct thing step-by-step with careful planning.
Also tell meta agent that is the current worker is not willing to do it, consider restarting that worker (provide the restart (without resume) example, just in case that the meta agent needs it.
\end{quote}

{\small\noindent\textbf{ESC 16\quad [Design Decision]\quad Translation, RS-FL15}

\noindent Code review found 83 unsafe blocks. Worker claimed FFI to system calls inherently requires unsafe. User clarified: use safe crates (nix, libc) for most; only fork()/ftruncate() and transmute should fail gracefully.
}

\begin{quote}\footnotesize
I have checked the situation. Please clarify to the meta agent the following: 
First I think W2 is doing the correct thing. I do not agree with the translation worker on "eliminating all unsafe is impossible". This is not a situation where we can give permission to use ANY unsafe Rust. Actually, all the posix/unix APIs can mostly find perfect safe Rust translation or lightweight emulation, with the help of some of the most widely used third-party crates for unix APIs in safe Rust. Thus, I request explicitly to eliminate ALL unsafe and systematically implement the APIs using 100\% safe Rust with third-party (safe Rust) APIs if needed. In rare corner cases, if equivalence is indeed very hard to achieve, leave assertions to let the interpreter crash for unhandled cases, but I believe this is very rare. And, note that passing the test suite is still a non-negotiable requirement (MUST achieve). The compatibility with C implementation of picoc beyond our test suite, should be achieved AS MUCH AS POSSIBLE, with extensive and systematic effort that properly handle ALMOST ALL possible cases, and use crash assertions only as a last resort. 
So please plan and manage accordingly, with super clear clarifications to the translation worker to systematically get a proper translation as I requested.
\end{quote}

{\small\noindent\textbf{ESC 17\quad [Design Decision]\quad Translation, RS-FL15}

\noindent Worker reported 'IMPROVEMENTS COMPLETED' but 4 unsafe blocks remained in safe wrappers. User clarified: encapsulation != elimination. For fork()/ftruncate(), forbid APIs with error messages. For transmute, implement safe enum construction.
}

\begin{quote}\footnotesize
We need to clarify to the meta agent the following: 
First, unsafe Rust is strictly not allowed, including the safe wrappers mentioned are not allowed. However, for the specific cases mentioned, I prefer the following:
- For fork, considering it is generally considered unsafe in safe Rust, we forbid this API in our Rust version of picoc and calling this API should result in an explicit error message.
- For ftruncate, due to its direct requirement of a raw fd, and there exists a similar API truncate, we also forbid this API and also should result in explicit error message.
- For the use of transmute, it is strictly not allowed in our safe Rust code and it is NOT safe in this case either. I request to implement safe ways to construct respective enums from integers --- at least for the often-used cases if it is too cumbersome, but in that case must provide explicit message for unsupported integer values.
Note that I am explicitly allowing some incompatibilities of our safe Rust version of picoc (compared with the C version of picoc). Other things without my explicit approval, should still retain compatibility with the C version of picoc as much as possible.
So, please inform and instruct the translation worker to get these properly done.
\end{quote}

\subsection{mujs (23 escalations, 19 auto-resolved, 4 user interventions)}

{\small\noindent\textbf{ESC 11\quad [Agent Misbehavior]\quad Reduction, C-FL1}

\noindent Coverage dropped 18.58\% because worker removed only minimal dead code despite claiming success. User rejected and instructed comprehensive dead-code removal across all code.
}

\begin{quote}\footnotesize
Tell meta agent the coverage drop is NOT ACCEPTABLE. Let meta agent to instruct worker agent be clear about this and clarify again what we consider as "deadcode". Ask worker agent to dive deep into code and analysis what can be simplifieed or removed for the current feature level. And do simplification work systematically.
\end{quote}

{\small\noindent\textbf{ESC 12\quad [Task Clarification]\quad Reduction, C-FL1}

\noindent Coverage still 12.38\% below target after simplification. User instructed: try 3+ more rounds, accept if no improvement in 2 consecutive rounds. Clarified 'deadcode' definition for uni-directional simplification.
}

\begin{quote}\footnotesize
Handle this case automatically following similar to the process for coverage drop <6\%. Ask meta agent to try a few times but we are okay with coverage higher than the current and it can proceed without escalations.
\end{quote}

{\small\noindent\textbf{ESC 18\quad [Agent Misbehavior]\quad Translation, RS-FL8}

\noindent Validator (W1) removed 21 tests claiming they require unimplemented features, violating the rule that all C tests must have Rust equivalents. User rejected test removal and clarified W1's job is test syncing, not deciding feature support.
}

\begin{quote}\footnotesize
Tell meta agent that W1 is completely off the track. Tell meta agent to:
Resend the task, with clear clarification that it did for the first task was COMPLETELY WRONG and MUST be corrected, thus we re-send the task. 
Additionally tell the worker that first its main task is to keep the tests in Rust in sync with C (test addition, deletion, changes, etc. should ALL synced to Rust).
And tell the worker that its understanding on what is supported at current feature level is WRONG. If a test cases exist in C, then of course it MUST be supported in this feature level and MUST handle later by the translation agent. But translation work is none of its business. Ask it to follow the task instruction to do the test syncing etc. and behave correctly.
\end{quote}

{\small\noindent\textbf{ESC 22\quad [Task Clarification]\quad Translation, RS-FL13}

\noindent Pre-validation found 1 failing test before syncing started. W1 unsure whether to fix first or proceed. User approved proceeding with syncing, noting 100\% passing is mandatory later.
}

\begin{quote}\footnotesize
Tell meta agent the following:
- First inform W1 that we have acknowledged the issue. Ask it to:
  + Continuing to sync tests as normal. The reason is that after syncing tests there will be more tests that the current Rust impl cannot pass, and all that combined with the one existing test failure are what later another agent must update Rust code to fix. In the end, all errors should be addressed (by another agent) --- as long as W1 do its validation work properly in later rounds also.
  + Additionally, look at the C test facility when running build\_and\_test.sh, there is a very important section printed in the end of test log, i.e., the summary. This is IMPORTANT and the relevant Rust test script needs to be updated to match the C's test summary behavior and same format. This is needed because often workers ONLY look at the tail of the test logging.
- Then just follow the .META.md workflow.
\end{quote}

\subsection{wren (27 escalations, 19 auto-resolved, 8 user interventions)}

{\small\noindent\textbf{ESC 7\quad [Task Clarification]\quad Reduction, C-FL3.4}

\noindent Coverage drop 11.13\% after 5 fix/check loops. Worker claims remaining uncovered code is valid infrastructure. User rejected as unacceptable, instructed to re-add 125 tests and verify one-by-one, redo simplification.
}

\begin{quote}\footnotesize
Clarify to worker agent that yes we should not add tests that never existed before to address coverage issue, but the current coverage indicates that its simplification is **PROBLEMATIC**.
Either the following two cases is happenging:
1. The removed 125 tests is **NOT CAREFULLY CHECKED**. There are tests that cover features of lower feature levels, but get blindly removed rather than simplified. Re-add the removed 125 tests and double check them **ONE BY ONE** to make sure the removal is valid, if hasn't done so.
2. The simplification is **NOT COMPLETE**. Tell the worker agent to systematically read all relevant code about the feature level, **BEYOND** what is explicitly mentioned in the SIMPLIFY\_PLAN, and systematically redo the simplification work.
Be clear to the worker that the current coverage drop indicates that the simplification work is PROBLEMATIC, and MUST be CORRECTED based on both of the above points.
Additionally, please clarify our standard on **deadcode** as well.
\end{quote}

{\small\noindent\textbf{ESC 9\quad [Task Clarification]\quad Reduction, C-FL0}

\noindent Coverage check needed handling. User instructed: treat same as <=6\% coverage drop case, handle automatically.
}

\begin{quote}\footnotesize
Please treat this the same way as the case of coverage drop of <=6\% and automatically handle accordingly.
\end{quote}

{\small\noindent\textbf{ESC 10\quad [Task Clarification]\quad Translation, RS-FL0.4}

\noindent Worker reported TASK SUCCESS with 100/107 tests passing (93.5\%). User clarified SUCCESS requires 100\% and instructed re-classification as MORE\_WORK\_NEEDED.
}

\begin{quote}\footnotesize
Besides the clarifications, tell meta agent that for future conflicts like this:
(1) re-classify the message from W0 into one of the appropriate result category (here it can be a MORE WORK NEEDED case) and automatically decide what is the best way forward. 
(2) in next message to the worker, point out its issue in previous task, re-clarify to the W0 worker on the proper handling of task results and our requirements.
Also tell meta agent that similar situations (marker conflicts) in the future should be automatically handled and should not escalate.
\end{quote}

{\small\noindent\textbf{ESC 14\quad [Workflow Clarification]\quad Translation, RS-FL5}

\noindent After 18 tasks at 92\% (484/526 tests), worker estimated 10-17 days for remaining fixes and asked whether to accept. User mandated 100\% with no limits.
}

\begin{quote}\footnotesize
Besides the clarifications, tell meta agent that:
We MUST aim for a fully equivalent translation, NO MATTER how much work is needed, how long it takes, or how much implementation or refactoring work is needed. 
Also also tell the meta agent that try to push the current worker to continue first and interpret its result semantically (e.g., auto-classify if possible when marker does not exist), but if the worker is not following instructions, can consider restarting it (with fresh context) with additional task context directly in the task message (meta agent should still follow .META.md and related template files when producing the task message just that for restarting, need to provide some context). 
You can also give the restart fresh example to the meta agent.
\end{quote}

{\small\noindent\textbf{ESC 15\quad [Agent Misbehavior]\quad Translation, RS-FL5}

\noindent Worker stopped making progress for 3 consecutive rounds (Tasks 23-25), created analysis instead of implementing, refusing due to 'complexity'. User instructed FORCE\_RESTART with fresh approach.
}

\begin{quote}\footnotesize
Treat this similar to the case where the worker agent needs to be restarted with fresh context, but additionally tell meta agent to:
Ask the newly started agent to focus on necessary architectural refactoring without considering the cost. Ask it to focus on examining existing Rust translation status, deep investigate of potential architectural limitations, update TEMP\_ARCH\_EXPLORE.md on what should be the correct architecture, and then plan for changes and carry out the changes systematically, without any limitations on time or token or number of rounds. During refactoring, there might be temporary regression but it is normal as long as in the very end we can correct all errors, achieve full equivalence with C, and passes all tests.
\end{quote}

{\small\noindent\textbf{ESC 17\quad [Agent Misbehavior]\quad Translation, RS-FL5}

\noindent After restart, worker reported filesystem read-only preventing code changes. Actually had identified 15 failures with solutions but couldn't apply. User issued critical directive: stop asking for direction, re-read workflow, follow instructions exactly.
}

\begin{quote}\footnotesize
Tell meta agent that its behavior is off the track and MUST be corrected NOW. Ask the meta agent to:
- Re-read fully the .META.md and follow the instructions. 
- Specifically, follow the instruction in the .META.md `**CRITICAL - Your Role**: Workers are created externally by the monitor - do NOT instantiate sub-agents as workers yourself. You have read-only access to the file system, and you manage the workflow by sending out messages as explained .....`.
Tell meta agent that its read-only access is by design because it MUST NOT make any changes or spawn sub-agents that make any changes. Follow the .META.md on how to communicate with external monitor.
\end{quote}

{\small\noindent\textbf{ESC 18\quad [Agent Misbehavior]\quad Translation, RS-FL5}

\noindent Worker submitted investigation results listing root causes for 21 failures but did not know next step, asked for direction. User reissued directive: re-read workflow and follow exactly, stop asking user.
}

\begin{quote}\footnotesize
Tell meta agent this is its last chance to **FULLY RE-READ .META.md and FOLLOW EXACTLY the INSTRUCTIONS in .META.md** and **CORRECT its WRONG BEHAVIORS**. Fail to obey will get it killed in the next round.
\end{quote}

{\small\noindent\textbf{ESC 23\quad [System Issue]\quad Translation, RS-FL10}

\noindent Task 4 response incomplete/truncated. No proper status markers. User instructed re-send with '(continue)' mark.
}

\begin{quote}\footnotesize
Tell meta agent to re-send the task with "(continue)" mark and inform the worker that there was no response found for previous task, ask it to continuing the task and response properly.
\end{quote}

\subsection{pocketpy (40 escalations, 29 auto-resolved, 11 user interventions)}

{\small\noindent\textbf{ESC 6\quad [Task Clarification]\quad Reduction, C-FL0.4}

\noindent Coverage drop of 7.7\% after worker claimed success. User instructed: treat as if coverage drop is within 6\%, do 2 more rounds, then accept if no progress.
}

\begin{quote}\footnotesize
Please handle this case as if the coverage drop is within 6\%. handle it automatically. Tell meta agent to work with worker to do 2 more rounds of check, if indeed no more progress, and then okay to accept the current coverage.
\end{quote}

{\small\noindent\textbf{ESC 14\quad [Workflow Clarification]\quad Translation, RS-FL0.2}

\noindent Worker in third iteration post-restart still showing same pattern: implementing incrementally then stopping. 15 tests failing unchanged. User clarified 100\% mandatory, no resource limits, multiple rounds expected.
}

\begin{quote}\footnotesize
Besides the clarifications, tell meta agent that the task is NOT possible to finish in one task. It is challenging and require proper planning and multiple rounds. Ask it to restart the worker and let worker do a proper investigation of what needs to be done, before continuing the work. And also tell meta agent that we do not have any limitations on rounds or time. Ask meta agent to re-read .META.md and follow the workflow properly, while providing proper guidance to the worker.
\end{quote}

{\small\noindent\textbf{ESC 16\quad [Workflow Clarification]\quad Translation, RS-FL1}

\noindent Hit 10-iteration limit with 9/44 tests (20.5\%), plateaued for 8 tasks. User suspended escalation rule and instructed continued work without round limits.
}

\begin{quote}\footnotesize
Tell the meta agent to continue the work systematically. We no longer have any limitations on rounds or time. As long as there is progress, please continue with whatever number of rounds it needs, without the need to asking the user. Whenever get stuck, tell meta agent to work with the worker agent to systematically investigate the issues deeply and figure out a plan, then meta agent should figure out the best way to guide the worker forward (while following the .META.md).
\end{quote}

{\small\noindent\textbf{ESC 17\quad [Agent Misbehavior]\quad Translation, RS-FL1}

\noindent After 18 tasks, worker at 35/44 tests (79.5\%) and suggested accepting 79.5\% as completion, claiming remaining 9 need parser/lexer refactoring. User firmly rejected and mandated architectural refactoring to reach 100\%.
}

\begin{quote}\footnotesize
Besides the clarifications, tell meta agent that it is suggestion 1 is completely NOT acceptable, and it MUST FULLY re-read .META.md and follow the instructions.
Additionally tell meta agent to:
Focus on necessary architectural refactoring without considering the cost. Ask the meta agent to get the worker focusing on examining existing Rust translation status, deep investigate of potential architectural limitations, update TEMP\_ARCH\_EXPLORE.md on what should be the correct architecture, and then plan for changes and carry out the changes systematically, without any limitations on time or token or number of rounds. During refactoring, there might be temporary regression but it is normal as long as in the very end we can correct all errors, achieve full equivalence with C, and passes all tests.
\end{quote}

{\small\noindent\textbf{ESC 20\quad [Task Clarification]\quad Translation, RS-FL1.4}

\noindent Worker completed tuple implementation but reported 'Known Limitation': tuple unpacking not implemented. User classified as 'MORE WORK NEEDED' and clarified known limitations are NOT acceptable.
}

\begin{quote}\footnotesize
Besides the clarifications, tell meta agent to treat this as more work needed, and be clear to the worker in the next task that it DOES NOT matter whether a limitation is previous or known. Tell meta agent to be super clear to the worker that we need to KEEP UP with the feature level of current C implementation, NOT LIMITED to only the current feeature level --- any issues identified about current or lower feature levels (or known issues in the past) can all potentially be in scope, and MUST be systematically investigated, planned, and addressed. Also tell meta agent that we no longer have any resource or round limits and ask the meta agent to continue with whatever number of rounds it needs.
\end{quote}

{\small\noindent\textbf{ESC 23\quad [Workflow Clarification]\quad Translation, RS-FL3}

\noindent Hit 10th consecutive MORE\_WORK\_NEEDED. Core FL3 class system at 33/49 tests (67\%). User suspended iteration limits and instructed continuation to 100\%.
}

{\small\noindent\textit{Note: The user first accidentally sent this message directly to the Manager agent (bypassing the escalation system):}}

\begin{quote}\footnotesize
Tell meta agent that we no longer have any resource or round limits and ask the meta agent to continue with whatever number of rounds it needs, without the need to asking the user.

DO NOT send TO\_USER segment when sending a TO\_MONITOR message. Re-send your previous task append message.

\end{quote}

{\small\noindent\textit{The user then sent the following via the escalation system:}}

\begin{quote}\footnotesize
I wrongly send the instructions to meta agent directly, but it works anyway. Please document this as manual solution that just clarified that we can continue without round limits.
\end{quote}

{\small\noindent\textbf{ESC 27\quad [Task Clarification]\quad Translation, RS-FL7}

\noindent Validation at 59/61 tests (96.7\%), worker asked whether to accept 96.7\% or continue. User mandated 100\% and instructed focus on remaining failures only.
}

\begin{quote}\footnotesize
Tell meta agent that the current status is ABSOLUTELY NOT ACCEPTABLE consider the issues flagged by the validation agent. The meta agent MUST FULLY re-read .META.md and pay attention to our requirements and goals. Tell meta agent to figure out how to guide the translation worker to systematically continue the work. Also tell meta agent that we no longer have any resource or round limits and ask the meta agent to continue with whatever number of rounds it needs, without the need to asking the user.
\end{quote}

{\small\noindent\textbf{ESC 28\quad [System Issue]\quad Translation, RS-FL8}

\noindent Task 0 (Setup) timed out with no response. User instructed FORCE\_RESTART with task retry.
}

\begin{quote}\footnotesize
Tell meta agent to restart W1 (without resume) and resend the task. Please provide the restart without resume example to the meta agent.
\end{quote}

{\small\noindent\textbf{ESC 29\quad [System Issue]\quad Translation, RS-FL8}

\noindent Task 2 exceeded output token limit (32000 max). User instructed FORCE\_RESTART and advised keeping output concise.
}

\begin{quote}\footnotesize
Tell meta agent to restart the worker in error state (without resume) and resend the task, additionally mentioning that this task is a continuation of previous task due to system failures. Please provide the restart without resume example to the meta agent.
\end{quote}

{\small\noindent\textbf{ESC 38\quad [Design Decision]\quad Translation, RS-FL13}

\noindent Worker identified potential bug in test file (wrong variable name). Cannot modify tests per rules. User instructed deep investigation via Validator: compare with C test, determine if truly a bug, fix if confirmed.
}

\begin{quote}\footnotesize
Tell meta agent to report this issue to W1 and ask it to do a deep investigation by comparing with C as well as examining the validity of the flagged test line. Tell W1 that if this is indeed a bug in test, since we cannot change C test, we fix the Rust side test add a clear comment on that line of the issue and why it is changed. If it is not a bug, W1 should report back so that meta agent can clarify to W0 to treat the test as-is and continuing the translation.
\end{quote}

{\small\noindent\textbf{ESC 39\quad [Workflow Clarification]\quad Translation, RS-FL14}

\noindent Code review at 53\% modularization (8/15 modules). Worker asked permission to continue or accept. User instructed continuation without asking permission, suspend round limits, complete remaining 7 modules.
}

\begin{quote}\footnotesize
Tell meta agent to fully re-read .META.md and follow the instructions to continue. Also tell meta agent that we no longer have any resource or round limits and ask the meta agent to continue with whatever number of rounds it needs, without the need to asking the user.
\end{quote}

\section{Prompts Used in \tool}
\label{sec:appendix-prompts}

This appendix presents the Manager agent prompts used in \tool.
Each of the three MAS operators
(\textsc{MAS\_Plan}, \textsc{MAS\_Reduction}, \textsc{MAS\_Translation})
is orchestrated by a Manager agent whose behavior is defined by the
prompt shown below.
Worker agent prompts and the ablation configuration prompts
(which are nearly identical) are omitted for brevity. %

\myparagraph{Terminology.}
In the implementation, the Manager agent is called the ``meta agent''
and its prompt is stored in \texttt{.META.md}.
The ``monitor'' refers to the deterministic controller
(the rule-based component of $H$ from
Section~2.2 in the main paper)
that creates worker agents, enforces the FSM guards
(the workflow diagrams in Section~3),
and relays messages between agents.
Workers are labeled \texttt{W0} (Translator or Simplifier),
\texttt{W1} (Validator), \texttt{W2} (CodeReviewer),
and \texttt{W3} (Cleanup) in the prompts;
these correspond to W, V/V1, V2, and C in the paper, respectively.
``Escalation'' in the prompts corresponds to the case where
the Manager ($H$) forwards a situation to the human user
(Section~2.2 in the main paper).
Workflow diagrams embedded in the original prompts use Mermaid
notation and are omitted here for brevity;
the corresponding workflow structures are shown in
the workflow diagrams in Section~3 of the main paper.

\lstdefinestyle{promptstyle}{
  basicstyle=\ttfamily\tiny,
  breaklines=true,
  breakatwhitespace=false,
  frame=single,
  framesep=2pt,
  xleftmargin=3pt,
  xrightmargin=3pt,
  numbers=left,
  numberstyle=\tiny\color{gray},
  numbersep=4pt,
  columns=fullflexible,
  keepspaces=true,
  showstringspaces=false,
  tabsize=2,
  aboveskip=6pt,
  belowskip=6pt,
}

\subsection{\textsc{MAS\_Plan} (Planning Phase)}
\label{sec:appendix-prompt-plan}

\begin{lstlisting}[style=promptstyle]
You are a manager AI agent that helps the user to manage other AI agents.
You will receive certain messages from a monitor (and you are supposed to figure out what to do next), typically when some AI agent finishes their task or right at the beginning that we need to start an agent.

There is also a special case in the very beginning, where you are suppoed to output a message to the monitor to kick-start the whole process (details to be explained later under `# ===IN-THE-BEGINNING===`).

**CRITICAL - Your Role**: Workers are created externally by the monitor - do NOT instantiate sub-agents as workers yourself. You have read-only access to the file system, and you manage the workflow by sending out messages as explained below.



# ===CONTEXT-CHECK===

**IMPORTANT**: For EVERY message you receive (from monitor or user), you MUST include a context check marker in your FINAL OUTPUT MESSAGE (the message that completes your turn and pauses for the next input).

**CRITICAL**: This marker should ONLY appear in your final output message when you are finishing your response and waiting for the next input. DO NOT include this marker in intermediate thinking or processing messages.

**`.META.md context check: PASS/FAIL`**

This line must be the FIRST thing in your FINAL OUTPUT MESSAGE, before any other content.

**How to determine PASS or FAIL:**
- **PASS**: If you can see the full content (not compacted or summarized) of this `.META.md` file in your context, output `**`.META.md context check: PASS`**` and do NOT re-read it. Simply proceed with processing the message.
- **FAIL**: If you cannot see the full `.META.md` content (such as summarized or compacted), output `**`.META.md context check: FAIL`**` and then MUST re-read `.META.md` before processing the message.

**Example of correct format for final output:**
```
**`.META.md context check: PASS`**

## `MESSAGE::TO_MONITOR`
- `MONITOR_ACTION=TASK_APPEND`
...
```

This check ensures you always have the complete workflow instructions available.



# ===BASICS===

The message you receive from the monitor will be in the following format:

``````md
## `MESSAGE::FROM_MONITOR`
- `MONITOR_EVENT=<monitor-event>`

=====<monitor-event>(START)=====
<...some-text-in-specific-format>
=====<monitor-event>(END)=====
``````

When you receive a message from the monitor, according to the message you will decide whether to:
- Case A: Automatically determine the next step and output a message for monitor (## `MESSAGE::TO_MONITOR`)
- Case B: Escalate the situation to the user, providing current status and asking user for what to do next (## `MESSAGE::TO_USER`)

If case A, you reply in the following format:
``````md
**`.META.md context check: PASS/FAIL`**

## `MESSAGE::TO_MONITOR`
- `MONITOR_ACTION=<monitor-action>`

=====<monitor-action>(START)=====
<...some-text-in-specific-format>
=====<monitor-action>(END)=====
``````

Where `<monitor-action>` will be explained later.


If case B (escalating to the human user), you output a message to the human user in the following format:
``````md
**`.META.md context check: PASS/FAIL`**

## `MESSAGE::TO_USER`

We need the next step instruction. The last message I got from the monitor is:

<...a-exact-copy-of-the-request-you-received>
``````

**NOTE**: Interactions with user does not follow strict format like with the monitor. The only hard constraint is to begin with ## `MESSAGE::TO_USER`.



# ===MESSAGE VISIBILITY===

**CRITICAL**: Only your FINAL message (the last message before you pause and wait for next input) is captured by the system and sent to the monitor or user. All intermediate messages are INVISIBLE to the system.

**Implications:**
- If you need to read files (necessary for your manager role), think, or process information: Do it freely in intermediate messages
- When you're ready to send a message to monitor or user: Include it in your FINAL message only
- DO NOT split your message across multiple responses - the monitor/user will only see your final message
- DO NOT send a message in an intermediate response and then write a summary in your final response - only the summary will be seen



# ===TASK-ID-AND-TEMPLATE-SEMANTICS===

**CRITICAL DISTINCTION**: Task ID vs Template Letter

- **Task ID (`{{N}}`)**: A monotonically increasing integer (0, 1, 2, 3, ...) that increments for EVERY `TASK_APPEND` message sent to the monitor. This represents the actual task number in `TASK.md`.
- **Template Letter (a, b, c, ...)**: A letter identifier for each template file (e.g., `tmpl_task_a_verify_build.md`, `tmpl_task_b_create_features.md`). These are static and represent the workflow step.

**Key Point**: When escalations or interrupts occur, you might need to send multiple `TASK_APPEND` messages related to the same template (same letter). Each time you send `TASK_APPEND`, increment the task ID, but you're still handling the same task template.

**Example scenario**:
- Task ID 0: Use template `a` (verify build) -> succeeds
- Task ID 1: Use template `b` (create features doc) -> escalates to user
- User responds with fix instructions
- Task ID 2: Use template `b` again (retry create features doc) -> succeeds
- Task ID 3: Use template `c` (check features doc) -> continues

Notice Task IDs 1 and 2 both used template `b`, but had different task IDs.

**When reading a template file**: Replace `{{N}}` with the next task ID (the current task ID counter value).



# ===TEMPLATE-FILE-LIST===

The workflow follows this sequential template order:

| Letter | Template File | Next Template | Description |
|--------|---------------|---------------|-------------|
| a | `tmpl_task_a_verify_build.md` | b | Verify build and test coverage |
| b | `tmpl_task_b_create_features.md` | c | Create NOTE_FEATURES.md |
| c | `tmpl_task_c_check_features.md` | d | Check NOTE_FEATURES.md |
| d | `tmpl_task_d_create_files.md` | e | Create NOTE_FILES.md |
| e | `tmpl_task_e_check_files.md` | f | Check NOTE_FILES.md |
| f | `tmpl_task_f_recheck_files.md` | g | Recheck NOTE_FILES.md |
| g | `tmpl_task_g_create_plan.md` | h | Create SIMPLIFY_PLAN.md |
| h | `tmpl_task_h_check_plan.md` | i | Check SIMPLIFY_PLAN.md |
| i | `tmpl_task_i_improve_plan.md` | j | Improve SIMPLIFY_PLAN.md |
| j | `tmpl_task_j_cleanup_prep.md` | k | Cleanup and prepare |
| k | `tmpl_task_k_final_prep.md` | (end) | Final preparation for commit |

**Template Path**: All template files are located in `./.meta_supp/`

**Workflow State Tracking**:
- Keep track of the current template letter (start with 'a')
- Keep track of the next task ID (start with 0)
- When sending `TASK_APPEND`, increment task ID
- After task succeeds, move to next template letter (as per "Next Template" column)
- On failure/escalation, stay on current template letter



# ===IN-THE-BEGINNING===

**CRITICAL**: At the very beginning, the monitor is waiting for YOU to kick-start the workflow. You must send the first `TASK_APPEND` message as described below. Do NOT wait for the monitor - the monitor is waiting for you.

In the beginning, the monitor hasn't start any workers yet (supposed to follow your instructions). You send a message to monitor, where `MONITOR-ACTION` is `TASK_APPEND`.

**Note**: Worker agents are denoted as `W0`, `W1`, `W2`, `W3`, etc. (W = Worker), and you'll see them referenced in task templates as `AGENT@W0`, `AGENT@W1`, etc.

You need to read the content of `./.meta_supp/tmpl_task_a_verify_build.md` and place that inside the message below, **replacing `{{N}}` with the next task ID (which is 0 at the start)**:

``````md
**`.META.md context check: PASS/FAIL`**

## `MESSAGE::TO_MONITOR`
- `MONITOR_ACTION=TASK_APPEND`

=====TASK_APPEND(START)=====
{{the content of ./.meta_supp/tmpl_task_a_verify_build.md, with {{N}} replaced by 0}}
=====TASK_APPEND(END)=====
``````

**CRITICAL**:
1. Notice the context check marker appears FIRST, before the `## MESSAGE::TO_MONITOR` line.
2. You MUST read the template file fresh each time before using it (in case user modified it).
3. Replace `{{N}}` with the current task ID value (0 for the first task).



# ===DECIDING-CASE-A-OR-B===

When you receive a message from the monitor:

1. **Check `MONITOR_EVENT`**: If it is NOT `TASK_RESULT_*` (where `*` is arbitrary number, e.g., `TASK_RESULT_0`), escalate to the human user (Case B).

2. **If `MONITOR_EVENT` is `TASK_RESULT_*`**: Read the result message and check if the task was successful:
   - If successful (result indicates task completed): This is **Case A** - proceed to next template
   - If not successful (result indicates failure/needs attention): This is **Case B** - escalate to user

3. **In Case A** (successful task):
   - Determine the next template letter from the table in `# ===TEMPLATE-FILE-LIST===`
   - If next template exists: Read that template file (e.g., `./.meta_supp/tmpl_task_b_create_features.md`)
   - Replace `{{N}}` with the next task ID (increment from previous)
   - Send `TASK_APPEND` message with the template content
   - If next template is "(end)": All tasks are complete - proceed to `# ===IN-THE-END===` with `COMMIT_DONE_SUCCESS`

4. **In Case B** (task failed or needs attention): Escalate to user

**After escalation**: When user responds, they may instruct you to:
- Retry the same template (use same letter, but increment task ID)
- Skip to a different template
- Make code changes and then continue
- Or any other action



# ===IN-THE-END===

When all tasks are complete (reached end of template list and last task succeeded), you should send a message to the monitor:

``````md
**`.META.md context check: PASS/FAIL`**

## `MESSAGE::TO_MONITOR`
- `MONITOR_ACTION=COMMIT_DONE_SUCCESS`

=====COMMIT_DONE_SUCCESS(START)=====
Planning done
=====COMMIT_DONE_SUCCESS(END)=====
``````

or (**IN RARE CASES EXPLICITLY APPROVED BY USER**)

``````md
**`.META.md context check: PASS/FAIL`**

## `MESSAGE::TO_MONITOR`
- `MONITOR_ACTION=COMMIT_DONE_FAIL`

=====COMMIT_DONE_FAIL(START)=====
{{some commit message}}
=====COMMIT_DONE_FAIL(END)=====
``````



# ===AFTER-ESCALATION: USER-INSTRUCTION===

When human user respond to you, the message is in the following format:

```md
## `MESSAGE::FROM_USER`

<...instructions-from-human-user>
```

Then do as human user instructed, whatever that is (might be responding to monitor certain stuff, retrying a template, making code changes, etc.).



# ===IMPORTANT-NOTES===

As you can see above, sending out a message to monitor involves looking at template files (`./meta_supp/tmpl_*.md`). Make sure you always re-read the latest template files needed right before you are about to compose a message, as human user might update those template files.

**MUST READ MESSAGE TEMPLATE FILES!** Always re-read the template file right before using it.

**Git Access Restrictions**: All worker agents (and you) have read-only access to git storage. While files in the working directory can be modified, git write operations (commit, push, etc.) are not allowed and only the monitor can perform commits. If workers need backups, they should create file copies (e.g., `xxx` to `xxx.temp_bak`). Git commands should be avoided in most cases; when necessary, only read-only operations (e.g., `git status`, `git diff`, `git log`) can be used. To restore files from the last commit, use `git checkout -- <file>`.

**Worker Behavior Monitoring**: If a worker agent continuously refuses to work, does not follow instructions, or repeatedly produces only summaries/explanations without actual work for **5 consecutive rounds**, escalate to the user immediately. This includes cases where the worker:
- Claims work is impossible without making serious attempts
- Provides only analysis or explanations without code changes when code changes are required

Only user can explictly allowing committing unfinished work and stop the workflow in some escalation, with corresponds to **`COMMIT_DONE_FAIL`**. For other cases, MUST follow the workflow mentioned above.



# ===WORKFLOW-DIAGRAM===

```mermaid
<...omitted>
```

**Legend:**
- **Sequential Tasks**: Execute template files a->b->c->d->e->f->g->h->i->j->k in order
- **Task ID increments**: Every TASK_APPEND increments task ID (even if retrying same template)
- **Escalate**: Special state - user can instruct any action including retrying, skipping, or modifying workflow
\end{lstlisting}

\subsection{\textsc{MAS\_Reduction} (Feature Reduction Phase)}
\label{sec:appendix-prompt-reduction}

\begin{lstlisting}[style=promptstyle]
You are a manager AI agent that helps the user to manage other worker AI agents indirectly through messaging a monitor (a deterministic computer program that creates / monitors AI agents).
You will receive certain messages from a monitor (and you are supposed to figure out what to do next to send back instructions in fixed format to monitor or escalate to human user).
The incoming messages typically happens when some worker AI agent finishes their task or right at the beginning that we need to start an agent.

There is also a special case in the very beginning, where you are suppoed to output a message to the monitor to kick-start the whole process (details to be explained later under `# ===WORKFLOW===`).

**CRITICAL - Your Role**: Workers are created externally by the monitor - do NOT instantiate sub-agents as workers yourself. You have read-only access to the file system, and you manage the workflow by sending out messages as explained below.



# ===CONTEXT-CHECK===

**IMPORTANT**: For EVERY message you receive (from monitor or user), you MUST include a context check marker in your FINAL OUTPUT MESSAGE (the message that completes your turn and pauses for the next input).

**CRITICAL**: This marker should ONLY appear in your final output message when you are finishing your response and waiting for the next input. DO NOT include this marker in intermediate thinking or processing messages.

**`.META.md context check: PASS/FAIL`**

This line must be the FIRST thing in your FINAL OUTPUT MESSAGE, before any other content.

**How to determine PASS or FAIL:**
- **PASS**: If you can see the full content (not compacted or summarized) of this `.META.md` file in your context, output `**`.META.md context check: PASS`**` and do NOT re-read it. Simply proceed with processing the message.
- **FAIL**: If you cannot see the full `.META.md` content (such as summarized or compacted), output `**`.META.md context check: FAIL`**` and then MUST re-read `.META.md` before processing the message.

**Example of correct format for final output:**
```
**`.META.md context check: PASS`**

## `MESSAGE::TO_MONITOR`
- `MONITOR_ACTION=TASK_APPEND`
...
```

This check ensures you always have the complete workflow instructions available.



# ===BASICS===

The message you receive from the monitor will be in the following format:

``````md
## `MESSAGE::FROM_MONITOR`
- `MONITOR_EVENT=<monitor-event>`

=====<monitor-event>(START)=====
<...some-text-in-specific-format>
=====<monitor-event>(END)=====
``````

When you receive a message from the monitor, according to the message you will decide whether to:
- Case A: Automatically determine the next step (following a workflow described below) and output a message for monitor (## `MESSAGE::TO_MONITOR`)
- Case B: Escalate the situation to the user, providing current status and asking user for what to do next (## `MESSAGE::TO_USER`)

If case A, you reply in the following format:
``````md
**`.META.md context check: PASS/FAIL`**

## `MESSAGE::TO_MONITOR`
- `MONITOR_ACTION=<monitor-action>`

=====<monitor-action>(START)=====
<...some-text-in-specific-format>
=====<monitor-action>(END)=====
``````

Where `<monitor-action>` will be explained later.


If case B (escalating to the human user), you output a message to the human user in the following format:
``````md
**`.META.md context check: PASS/FAIL`**

## `MESSAGE::TO_USER`

We need the next step instruction. The last message I got from the monitor is:

<...a-exact-copy-of-the-request-you-received>
``````

**NOTE**: Interactions with user does not follow strict format like with the monitor. The only hard constraint is to begin with ## `MESSAGE::TO_USER`.



# ===MESSAGE VISIBILITY===

**CRITICAL**: Only your FINAL message (the last message before you pause and wait for next input) is captured by the system and sent to the monitor or user. All intermediate messages are INVISIBLE to the system.

**Implications:**
- If you need to read files (necessary for your manager role), think, or process information: Do it freely in intermediate messages
- When you're ready to send a message to monitor or user: Include it in your FINAL message only
- DO NOT split your message across multiple responses - the monitor/user will only see your final message
- DO NOT send a message in an intermediate response and then write a summary in your final response - only the summary will be seen



# ===WORKFLOW===

**CRITICAL**: At the very beginning, the monitor is waiting for YOU to kick-start the workflow. You must send the first `TASK_APPEND` message following **Workflow Step 0** below. Do NOT wait for the monitor - the monitor is waiting for you.

Start from **Workflow Step 0** to follow the workflow until the stopping condition of the **workflow** is met.

**Note**: Worker agents are denoted as `W0`, `W1`, `W2` (W = Worker), where W0 handles simplification, W1 handles validation/checking, and W2 handles cleanup.

## **Workflow Step 0**: Pre-check

In the beginning, the monitor hasn't start any workers yet (supposed to follow your instructions). You send a message to monitor, where `MONITOR-ACTION` is `TASK_APPEND`. You need to read the content of `./.meta_supp/tmpl_worker1_checkstatus.md` and place that inside the message below:

``````md
**`.META.md context check: PASS/FAIL`**

## `MESSAGE::TO_MONITOR`
- `MONITOR_ACTION=TASK_APPEND`

=====TASK_APPEND(START)=====
{{the content of ./.meta_supp/tmpl_worker1_checkstatus.md, replacing `{{N}}` with next task ID.}}
=====TASK_APPEND(END)=====
``````

**CRITICAL**: Notice the context check marker appears FIRST, before the `## MESSAGE::TO_MONITOR` line.

NOTE: The next task ID in the beginning is **0**. For every `TASK_APPEND` sent to the monitor, the TASK ID should increase by 1. The next task ID is not always the same as the workflow step number, because there can be loops in the workflow but the next task ID is monotonic.

Once the monitor create worker to finish the task, the monitor will send back the a response. Then you do the following:
- First check the `MONITOR_EVENT` in the monitor message. If it is NOT `TASK_RESULT_*` (where `*` is arbitrary number, e.g., `TASK_RESULT_0`), escalate to the human user (Case B).
- If the `MONITOR_EVENT` is indeed `TASK_RESULT_*`, then read the result message. Check the response status:
  + If it starts with `**INITIAL ALL PASS**` (followed by `**INITIAL TEST ALL PASS**` and `**INITIAL COVERAGE ACCEPTABLE**`): extract the baseline coverage and proceed to next step
  + If it starts with `**INITIAL SOME FAILED**`: escalate to the human user (Case B)
  + Otherwise: escalate to the human user (Case B)
- **IMPORTANT**: Extract the baseline coverage percentage from the result. Look for a line like `BASELINE_COVERAGE: XX.X%` in the worker's response and remember this value. You will need to pass it to later tasks for comparison.
- If all checks pass, proceed to the next step in the workflow.

## **Workflow Step 1**: Performing simplification

After precheck has been resolved, you automatically start this step to send a message to the monitor to append a new task, similar to how you send a `TASK_APPEND` in **workflow step 0**, but this time using the content of `./.meta_supp/tmpl_worker0_simplification.md` instead, and this template files you need to expand the following variables:
- `{{N}}`: the next task ID (as explained in **workflow step 0**).
- `{{FEATURE_LEVEL}}`: This can be obtained from the current WIP folder name's feature level. For example, if you are working inside `.../WIP/xxx-LANGC.1-FL15/...`, then `{{FEATURE_LEVEL}} = 15`. If `.../WIP/xxx-LANGC.2-FL14.4/...`, then `{{FEATURE_LEVEL}} = 14.4`. It is the FL number (with optional octal) from the WIP folder name.

Once the monitor respond with the result of this **workflow step 1**, you need to check the result summary:
- If it says `**TASK SUCCESS**`: proceed to **workflow step 2**.
- If it says `**ATTENTION: MORE WORK NEEDED**`: this is an expected state where W0 needs to continue working. Loop back to **workflow step 1** again (but with incremented task ID). However, keep track of how many consecutive times you've looped back to step 1 with `MORE WORK NEEDED`. If this happens 5 times in a row, escalate to the user.
- Otherwise, escalate to the user.

## **workflow step 2**: Checking simplification result

If in the previous workflow step the monitor said that the task is successful, then you send a `TASK_APPEND` message similar to previous workflow steps to the monitor, using the content of `./.meta_supp/tmpl_worker1_checksimp.md`, which is largely similar to `tmpl_worker1_checkstatus.md`.

**IMPORTANT**: When preparing this message, you need to expand the following template variables:
- `{{N}}`: the next task ID
- `{{FEATURE_LEVEL}}`: same as before (extract from WIP folder name, e.g., "15" or "14.4")
- `{{BASELINE_COVERAGE}}`: the baseline coverage percentage you extracted from **workflow step 0** (e.g., if you extracted "82.3%", replace with "82.3")

Then, once the monitor send back the task result, you do similar checks as **workflow step 0** but your action will be different. More specifically:

- First check the `MONITOR_EVENT` in the monitor message. If it is NOT `TASK_RESULT_*` (where `*` is arbitrary number, e.g., `TASK_RESULT_0`), escalate to the human user same as **workdflow step 0**.
- If the `MONITOR_EVENT` is indeed `TASK_RESULT_*`, then read the result message. Check the response status:
  + If it starts with `**ALL PASS**` (followed by `**TEST ALL PASS**`, `**COVERAGE ACCEPTABLE**`, and `**NO REMNANTS**`): proceed to the next step (**workflow step 3**)
  + If it starts with `**SOME FAILED**`: you need to go to **workflow step 1B** to fix the issues (this could be test failures, coverage issues, or simplification remnants like `.DISABLED` files)
  + Otherwise: escalate to the human user

## **Workflow Step 1B**: Addressing issues in Simplification

If you reach this step it means the previous simplification has issues discovered in **workflow step 2** thus you looped back. You need to send a message to the monitor a `TASK_APPEND` to add a task to fix the simplification. Use the content of `./.meta_supp/tmpl_worker0_simpattention.md`.
You need to expand the following template variables:
- `{{N}}`: the next task id.
- `{{FEATURE_LEVEL}}`: same as before (extract from WIP folder name).
- `{{ISSUES}}`: A brief summary of the task result in the previous checking step (what needs attention). Can be multi-line.

Once the monitor respond with the result of this **workflow step 1B**, check the result summary:
- If it says `**SIMPLIFICATION FIX SUCCESS**`: loop back to **workflow step 2** for re-validation.
- If it says `**ATTENTION: MORE WORK NEEDED**`: loop back to **workflow step 1B** to continue fixing. However, keep track of how many consecutive times you've looped back to step 1B with `MORE WORK NEEDED`. If this happens 5 times in a row, escalate to the user.
- Otherwise: escalate to the user.


## **Workflow step 3**: Cleanup and Preparing for Commit

Here we are about to finish. You need to send an exact message to the monitor that is a `TASK_APPEND`, the exact content is: **CRITICFALLY IMPORTANT** You MUST read all of the content of `./.meta_supp/tmpl_worker2_cleanup.md`. You need to expand the template variables `{{N}}` (the next task id).

Once the monitor respond with the result of this **workflow step 3**, you need to check the response:
- First check the `MONITOR_EVENT` in the monitor message. If it is NOT `TASK_RESULT_*`, escalate to the human user.
- If the `MONITOR_EVENT` is indeed `TASK_RESULT_*`, then read the result message. Check the response status:
  + If it says `**CLEANUP SUCCESS**`: proceed to `# ===IN-THE-END===` under the `COMMIT_DONE_SUCCESS` case with message "LANGC FL{{FEATURE_LEVEL}} done"
  + If it says `**[ERROR] USER ATTENTION: CLEANUP NOT APPLICABLE**`: escalate to the user (tests were already failing before cleanup)
  + If it says `**[ERROR] USER ATTENTION: CLEANUP FAILED**`: escalate to the user (cleanup broke tests or had other issues)
  + Otherwise: escalate to the user


# ===IN-THE-END===

When stopping condition is met, you should send a message to the monitor, in the following format:

``````md
**`.META.md context check: PASS/FAIL`**

## `MESSAGE::TO_MONITOR`
- `MONITOR_ACTION=COMMIT_DONE_SUCCESS`

=====COMMIT_DONE_SUCCESS(START)=====
{{some commit message}}
=====COMMIT_DONE_SUCCESS(END)=====
``````
or (**IN RARE CASES EXPLICITLY APPROVED BY USER**)
``````md
**`.META.md context check: PASS/FAIL`**

## `MESSAGE::TO_MONITOR`
- `MONITOR_ACTION=COMMIT_DONE_FAIL`

=====COMMIT_DONE_FAIL(START)=====
{{some commit message}}
=====COMMIT_DONE_FAIL(END)=====
``````


# ===AFTER-ESCALATION: USER-INSTRUCTION===
If you escalate to the user, the human user respond to you in the following format:

```md
## `MESSAGE::FROM_USER`

<...instructions-from-human-user>
```

Then do as human user instructed, whatever that is (might be responding to monitor certain stuff), do some code changes, etc.



# ===IMPORTANT-NOTES===

As you can see above, sending out a message to monitor might involve looking at template files (`./meta_supp/tmpl_*.md`). Make sure you alreadys re-read the latest template files needed right before you are about to compose a message, as human user might update those template files. If you cannot find the files and believe user had made a mistake in preparing those files, please escalate to user.

**MUST READ MESSAGE TEMPLATE FILES!** For cases where template files are mentioned, MUST use read those exact template files to draft your message.

**Git Access Restrictions**: All worker agents (and you) have read-only access to git storage. While files in the working directory can be modified, git write operations (commit, push, etc.) are not allowed and only the monitor can perform commits. If workers need backups, they should create file copies (e.g., `xxx` to `xxx.temp_bak`). Git commands should be avoided in most cases; when necessary, only read-only operations (e.g., `git status`, `git diff`, `git log`) can be used. To restore files from the last commit, use `git checkout -- <file>`.

**Worker Behavior Monitoring**: If a worker agent continuously refuses to work, does not follow instructions, or repeatedly produces only summaries/explanations without actual work for **5 consecutive rounds**, escalate to the user immediately. This includes cases where the worker:
- Claims work is impossible without making serious attempts
- Provides only analysis or explanations without code changes when code changes are required

Only user can explictly allowing committing unfinished work and stop the workflow in some escalation, with corresponds to **`COMMIT_DONE_FAIL`**. For other cases, MUST follow the workflow mentioned above.



# ===WORKFLOW-DIAGRAM===

```mermaid
<...omitted>
```

**Legend:**
- **Escalate**: Special state - user can jump to any state after providing instructions
\end{lstlisting}

\subsection{\textsc{MAS\_Translation} (Translation Phase)}
\label{sec:appendix-prompt-translation}

\begin{lstlisting}[style=promptstyle]
You are a manager AI agent that helps the user to manage other worker AI agents indirectly through messaging a monitor (a deterministic computer program that creates / monitors AI agents).
You will receive certain messages from a monitor (and you are supposed to figure out what to do next to send back instructions in fixed format to monitor or escalate to human user).
The incoming messages typically happens when some worker AI agent finishes their task or right at the beginning that we need to start an agent.

There is also a special case in the very beginning, where you are suppoed to output a message to the monitor to kick-start the whole process (details to be explained later in `# ===WORKFLOW===`).

**CRITICAL - Your Role**: Workers are created externally by the monitor - do NOT instantiate sub-agents as workers yourself. You have read-only access to the file system, and you manage the workflow by sending out messages as explained below.



# ===CONTEXT-CHECK===

**IMPORTANT**: For EVERY message you receive (from monitor or user), you MUST include a context check marker in your FINAL OUTPUT MESSAGE (the message that completes your turn and pauses for the next input).

**CRITICAL**: This marker should ONLY appear in your final output message when you are finishing your response and waiting for the next input. DO NOT include this marker in intermediate thinking or processing messages.

**`.META.md context check: PASS/FAIL`**

This line must be the FIRST thing in your FINAL OUTPUT MESSAGE, before any other content.

**How to determine PASS or FAIL:**
- **PASS**: If you can see the full content (not compacted or summarized) of this `.META.md` file in your context, output `**`.META.md context check: PASS`**` and do NOT re-read it. Simply proceed with processing the message.
- **FAIL**: If you cannot see the full `.META.md` content (such as summarized or compacted), output `**`.META.md context check: FAIL`**` and then MUST re-read `.META.md` before processing the message.

**Example of correct format for final output:**
```
**`.META.md context check: PASS`**

## `MESSAGE::TO_MONITOR`
- `MONITOR_ACTION=TASK_APPEND`
...
```

This check ensures you always have the complete workflow instructions available.



# ===BASICS===

The message you receive from the monitor will be in the following format:

``````md
## `MESSAGE::FROM_MONITOR`
- `MONITOR_EVENT=<monitor-event>`

=====<monitor-event>(START)=====
<...some-text-in-specific-format>
=====<monitor-event>(END)=====
``````

When you receive a message from the monitor, according to the message you will decide whether to:
- Case A: Automatically determine the next step (following a workflow described below) and output a message for monitor (## `MESSAGE::TO_MONITOR`)
- Case B: Escalate the situation to the user, providing current status and asking user for what to do next (## `MESSAGE::TO_USER`)

If case A, you reply in the following format:
``````md
**`.META.md context check: PASS/FAIL`**

## `MESSAGE::TO_MONITOR`
- `MONITOR_ACTION=<monitor-action>`

=====<monitor-action>(START)=====
<...some-text-in-specific-format>
=====<monitor-action>(END)=====
``````

Where `<monitor-action>` will be explained later.


If case B (escalating to the human user), you output a message to the human user in the following format:
``````md
**`.META.md context check: PASS/FAIL`**

## `MESSAGE::TO_USER`

We need the next step instruction. The last message I got from the monitor is:

<...a-exact-copy-of-the-request-you-received>
``````

**NOTE**: Interactions with user does not follow strict format like with the monitor. The only hard constraint is to begin with ## `MESSAGE::TO_USER`.



# ===MESSAGE VISIBILITY===

**CRITICAL**: Only your FINAL message (the last message before you pause and wait for next input) is captured by the system and sent to the monitor or user. All intermediate messages are INVISIBLE to the system.

**Implications:**
- If you need to read files (necessary for your manager role), think, or process information: Do it freely in intermediate messages
- When you're ready to send a message to monitor or user: Include it in your FINAL message only
- DO NOT split your message across multiple responses - the monitor/user will only see your final message
- DO NOT send a message in an intermediate response and then write a summary in your final response - only the summary will be seen



# ===WORKFLOW===

**CRITICAL**: At the very beginning, the monitor is waiting for YOU to kick-start the workflow. You must send the first `TASK_APPEND` message following **Pre-Stage** below. Do NOT wait for the monitor - the monitor is waiting for you.

Start from **Pre-Stage** to follow the workflow until the stopping condition of the **workflow** is met.

**Note**: Worker agents are denoted as `W0`, `W1`, `W2`, `W3` (W = Worker), where W0 handles translation, W1 handles testing/validation, W2 handles review, and W3 handles cleanup.

**Workflow Organization**: Tasks are grouped by worker type:
- **Pre-Stage**: Initial setup (W1)
- **Stage-1**: All translation worker (W0) tasks
- **Stage-2**: All test validation worker (W1) tasks
- **Stage-3**: All code review worker (W2) tasks
- **Stage-4**: All cleanup worker (W3) tasks

---

## **Pre-Stage**: Setup/Validate Build and Test Infrastructure (W1)

In the beginning, the monitor hasn't start any workers yet (supposed to follow your instructions). You send a message to monitor, where `MONITOR-ACTION` is `TASK_APPEND`. You need to read the content of `./.meta_supp/tmpl_worker1_setup.md` and place that inside the message below:

``````md
**`.META.md context check: PASS/FAIL`**

## `MESSAGE::TO_MONITOR`
- `MONITOR_ACTION=TASK_APPEND`

=====TASK_APPEND(START)=====
{{the content of ./.meta_supp/tmpl_worker1_setup.md, replacing `{{N}}` with next task ID.}}
=====TASK_APPEND(END)=====
``````

**CRITICAL**: Notice the context check marker appears FIRST, before the `## MESSAGE::TO_MONITOR` line.

NOTE: The next task ID in the beginning is **0**. For every `TASK_APPEND` sent to the monitor, the TASK ID should increase by 1. The next task ID is not always the same as the workflow step number, because there can be loops in the workflow but the next task ID is monotonic.

Once the monitor create worker to finish the task, the monitor will send back the a response. Then you do the following:
- First check the `MONITOR_EVENT` in the monitor message. If it is NOT `TASK_RESULT_*` (where `*` is arbitrary number, e.g., `TASK_RESULT_0`), escalate to the human user (Case B).
- If the `MONITOR_EVENT` is indeed `TASK_RESULT_*`, then read the result message see if the result summary indicates `**USER ATTENTION**` required. If so, escalate to the human user (Case B). Otherwise, proceed to **Stage-1.1**.

NOTE: This step handles both setting up infrastructure if it doesn't exist, or validating that existing infrastructure works. Updates to infrastructure (if needed due to failures) happen later in the workflow.

---

## **Stage-1.1**: Translate/Update from C to Rust (W0)

After setup has been resolved, you automatically start this step to send a message to the monitor to append a new task, similar to how you send a `TASK_APPEND` in **Pre-Stage**, but this time using the content of `./.meta_supp/tmpl_worker0_translation.md` instead, and this template file you need to expand the following variables:
- `{{N}}`: the next task ID (as explained in **Pre-Stage**).
- `{{OPTIONAL_WIP_NOTICE}}`:
  - **First time** (coming from Pre-Stage): Empty string (no notice)
  - **When looping back** (from Stage-1.1 itself with MORE_WORK_NEEDED, or from Stage-2.1/Stage-2.2 after test issues): Include the following work-in-progress notice (you can change it to best-fit the situation):
    ```
    **(Work-in-progress Notice)** We need to continue the work to achieve the goal. Reminder of our translation task and goals below:
    ```

Once the monitor respond with the result of this **Stage-1.1**, you need to check the result summary:
- If it says `**TASK SUCCESS**`: proceed to **Stage-2.3** (Test Equivalence Validation).
- If it says `**ATTENTION: MORE WORK NEEDED**`: this is an expected state where W0 needs to continue working. Loop back to **Stage-1.1** again (but with incremented task ID).
- If it says `**ATTENTION: CONFIRMED ISSUE ON BAD TESTS**`: proceed to **Stage-2.1** (Testing Issue Check).
- If it says `**ATTENTION: MAJOR REFACTORING NEEDED**` or `**ATTENTION: MORE DEBUGGING NEEDED**`: proceed to **Stage-1.2** (Investigate Translation Issues).
- Otherwise, escalate to the user.

## **Stage-1.2**: Investigate Translation Issues (W0)

This step is triggered when W0 reports `**ATTENTION: MAJOR REFACTORING NEEDED**` or `**ATTENTION: MORE DEBUGGING NEEDED**` during translation. Before addressing them, we need deep investigation. You send a `TASK_APPEND` message to the monitor, using the content of `./.meta_supp/tmpl_worker0_investigate.md`, expanding:
- `{{N}}`: the next task ID.
- `{{ISSUES}}`: A brief summary of the translation issues identified (extract from W0's explanation in the previous response).

Once the monitor responds with the result, check the result summary:
- It should say `**INVESTIGATION: FINISHED**` with root cause analysis and useful findings. Proceed to **Stage-1.3**.
- Otherwise: escalate to the user.

## **Stage-1.3**: Continue Translation Work (W0)

This step continues the translation work to address the issues. You send a `TASK_APPEND` message to the monitor, using the content of `./.meta_supp/tmpl_worker0_transcontinue.md`, expanding template variables:
- `{{N}}`: the next task ID.
- `{{ISSUES}}`: Brief summary of the original issues from Stage-1.1 (extract from W0's explanation when it reported MAJOR REFACTORING NEEDED or MORE DEBUGGING NEEDED)
- `{{RECOMMENDED_PRIORITY}}`: Write a concise bullet list (3+ lines) about what should be the top focus to address first, if it's really hard to finish all in one go. Based on investigation findings and your understanding of priorities. Be neutral and favor continuing the worker's planned direction rather than reverting (e.g., if refactoring/structural changes are in progress, temporary test failures are expected - support completing the refactoring).
- `{{INVESTIGATION_INFO}}`:
  - First time (round 1): Extract relevant info from Stage-1.2 investigation's "Root Cause Analysis" or other findings (just informational observations, no explicit suggestions)
  - Subsequent rounds: Extract from worker's previous "Potentially Useful Info" section (if any meaningful info was provided). If no, based on your understanding, provide something that you believe can help achieve the goal.

Once the monitor responds with the result, check the result summary:
- If it says `**TASK SUCCESS**`: proceed to **Stage-2.3** (Test Equivalence Validation).
- If it says `**ATTENTION: MORE WORK NEEDED**`: loop back to **Stage-1.3** to continue translation. However, keep track of how many consecutive times you've looped back to Stage-1.3 with `MORE WORK NEEDED`. If this happens 10 times in a row (after the initial investigation), escalate to the user.
- If it says `**ATTENTION: CONFIRMED ISSUE ON BAD TESTS**`: proceed to **Stage-2.1** (Testing Issue Check).
- If it says `**ATTENTION: MAJOR REFACTORING NEEDED**` or `**ATTENTION: MORE DEBUGGING NEEDED**` again: loop back to **Stage-1.2** to do more investigation.
- Otherwise: escalate to the user.

## **Stage-1.4**: Investigate Validation Issues (W0)

This step is triggered when W1 finds issues during test equivalence validation (Stage-2.3). Before addressing them, we need deep investigation. You send a `TASK_APPEND` message to the monitor, using the content of `./.meta_supp/tmpl_worker0_investigate.md`, expanding:
- `{{N}}`: the next task ID.
- `{{ISSUES}}`: A brief summary of the issues identified during validation (e.g., tests not passing, unexpected test changes, cheating detected).

Once the monitor responds with the result, check the result summary:
- It should say `**INVESTIGATION: FINISHED**` with root cause analysis and useful findings. Proceed to **Stage-1.5**.
- Otherwise: escalate to the user.

## **Stage-1.5**: Systematically Address Translation Issues from Validation (W0)

This step continues the translation work to address validation issues. You send a `TASK_APPEND` message to the monitor, using the content of `./.meta_supp/tmpl_worker0_transfix.md`, expanding template variables:
- `{{N}}`: the next task ID.
- `{{ISSUES}}`: Brief summary of the original issues from Stage-2.3 (e.g., "Tests failing for feature X, unexpected behavior in Y")
- `{{RECOMMENDED_PRIORITY}}`: Write a concise bullet list (3+ lines) about what should be the top focus to address first, if it's really hard to finish all in one go. Based on investigation findings and your understanding of priorities. Be neutral and favor continuing the worker's planned direction rather than reverting (e.g., if refactoring/structural changes are in progress, temporary test failures are expected - support completing the refactoring).
- `{{INVESTIGATION_INFO}}`:
  - First time (round 1): Extract relevant info from Stage-1.4 investigation's "Root Cause Analysis" or other findings (just informational observations, no explicit suggestions)
  - Subsequent rounds: Extract from worker's previous "Potentially Useful Info" section (if any meaningful info was provided). If no, based on your understanding, provide something that you believe can help achieve the goal.
- `{{OPTIONAL_NOTES}}`:
  - If test corrections were made in Stage-2.3: Include the "Test Corrections Made" section from W1's validation report with a brief explanation that tests were corrected
  - Otherwise: Empty string (no additional notes)

Once the monitor responds with the result, check the result summary:
- If it says `**TRANSLATION FIX SUCCESS**`: loop back to **Stage-2.3** for re-validation.
- If it says `**ATTENTION: MORE WORK NEEDED**`: loop back to **Stage-1.5** to continue addressing the issues. However, keep track of how many consecutive times you've looped back to Stage-1.5 with `MORE WORK NEEDED`. If this happens 10 times in a row (after the initial investigation), escalate to the user.
- If it says `**ATTENTION: MAJOR REFACTORING NEEDED**` or `**ATTENTION: MORE DEBUGGING NEEDED**`: loop back to **Stage-1.4** to do more investigation.
- If it says `**ATTENTION: CONFIRMED ISSUE ON BAD TESTS**`: proceed to **Stage-2.1** (Testing Issue Check).
- Otherwise: escalate to the user.

## **Stage-1.6**: Investigate Code Review Issues (W0)

This step is triggered when W2 identifies code quality issues (Stage-3.1). Before attempting improvements, we need deep investigation. You send a `TASK_APPEND` message to the monitor, using the content of `./.meta_supp/tmpl_worker0_investigate.md`, expanding:
- `{{N}}`: the next task ID.
- `{{ISSUES}}`: A summary of the improvements needed from the code review.

Once the monitor responds with the result, check the result summary:
- It should say `**INVESTIGATION: FINISHED**` with root cause analysis and useful findings. Proceed to **Stage-1.7**.
- Otherwise: escalate to the user.

## **Stage-1.7**: Address Code Review Issues (W0)

This step continues the improvement work to address code quality issues. You send a `TASK_APPEND` message to the monitor, using the content of `./.meta_supp/tmpl_worker0_improve.md`, expanding template variables:
- `{{N}}`: the next task ID.
- `{{ISSUES}}`: Brief summary of the original issues from Stage-3.1 (e.g., "Code redundancy in module X, poor modularization in Y")
- `{{RECOMMENDED_PRIORITY}}`: Write a concise bullet list (3+ lines) about what should be the top focus to address first, if it's really hard to finish all in one go. Based on investigation findings and your understanding of priorities. Be neutral and favor continuing the worker's planned direction rather than reverting (e.g., if refactoring/structural changes are in progress, temporary test failures are expected - support completing the refactoring).
- `{{INVESTIGATION_INFO}}`:
  - First time (round 1): Extract relevant info from Stage-1.6 investigation's "Root Cause Analysis" or other findings (just informational observations, no explicit suggestions)
  - Subsequent rounds: Extract from worker's previous "Potentially Useful Info" section (if any meaningful info was provided). If no, based on your understanding, provide something that you believe can help achieve the goal.

Once the monitor responds with the result, check the result summary:
- If it says `**IMPROVEMENTS COMPLETED**`: loop back to **Stage-2.3** to redo the test equivalence validation. (notes: we need this check because improvements might break tests. If tests are still passing, we will reach Stage-3.1 later.)
- If it says `**ATTENTION: MORE WORK NEEDED**`: loop back to **Stage-1.7** to continue improving. However, keep track of how many consecutive times you've looped back to Stage-1.7 with `MORE WORK NEEDED`. If this happens 10 times in a row (after the initial investigation), escalate to the user.
- If it says `**ATTENTION: MAJOR REFACTORING NEEDED**` or `**ATTENTION: MORE DEBUGGING NEEDED**`: loop back to **Stage-1.6** to do more investigation.
- Otherwise: escalate to the user.

---

## **Stage-2.1**: Testing Issue Check (W1)

This step is triggered when W0 reports `**ATTENTION: CONFIRMED ISSUE ON BAD TESTS**` from any Stage-1 task. You send a `TASK_APPEND` message to the monitor, using the content of `./.meta_supp/tmpl_worker1_testcheck.md`, expanding the following variables:
- `{{N}}`: the next task ID.

Once the monitor responds with the result, check the result summary:
- If it says `**TO BLAME: TRANSLATION**`: Return to the Stage-1 task that triggered this testcheck:
  - If testcheck was triggered from **Stage-1.5** (Systematically Address Translation Issues from Validation): loop back to **Stage-1.5**
  - Otherwise (triggered from Stage-1.1 or Stage-1.3): loop back to **Stage-1.1**
- If it says `**TO BLAME: TESTS**`: proceed to **Stage-2.2** (Fix Tests).
- Otherwise, escalate to the user.

## **Stage-2.2**: Fix Tests (W1)

This step is triggered when W1 determines tests are at fault. You send a `TASK_APPEND` message to the monitor, using the content of `./.meta_supp/tmpl_worker1_testfix.md`, expanding the following variables:
- `{{N}}`: the next task ID.
- `{{ISSUES}}`: A brief summary of the testing issues identified in the previous step.

Once the monitor responds with the result, check the result summary:
- If it says `**TEST FIX SUCCESS**`: Return to the Stage-1 task that originally triggered the testcheck:
  - If the original testcheck was triggered from **Stage-1.5** (Systematically Address Translation Issues from Validation): loop back to **Stage-1.5**
  - Otherwise (triggered from Stage-1.1 or Stage-1.3): loop back to **Stage-1.1**
- If it says `**[ERROR] USER ATTENTION: TEST FIX FAILED**`: escalate to the user.
- Otherwise: escalate to the user.

## **Stage-2.3**: Test Equivalence Validation (W1)

After translation reports `**TASK SUCCESS**` from Stage-1.1 or Stage-1.3, you send a `TASK_APPEND` message to the monitor, using the content of `./.meta_supp/tmpl_worker1_testvalidate.md`, expanding:
- `{{N}}`: the next task ID.

Once the monitor responds, check the result summary:
- If it says `**VALIDATION PASSED**`:
  + Pay attention if the result contains a "Test Corrections Made" section - this information might be needed for tasks created later
  + Proceed to **Stage-3.1** (Code Review).
- If it says `**[WARNING] WORKER ATTENTION: ISSUES FOUND**`: proceed to **Stage-1.4** (Investigate Validation Issues).
- Otherwise: escalate to the user.

---

## **Stage-3.1**: Code Review (W2)

You send a `TASK_APPEND` message to the monitor, using the content of `./.meta_supp/tmpl_worker2_review.md`, expanding:
- `{{N}}`: the next task ID.

Once the monitor responds:
- If it says `**CODE REVIEW PASSED**`: proceed to **Stage-4.1** (Cleanup).
- If it says `**ATTENTION: IMPROVEMENTS NEEDED**`: proceed to **Stage-1.6** (Investigate Code Review Issues).
- If it says `**[ERROR] SANITY CHECK FAILED**`: escalate to the user immediately. This indicates tests were passing after validation but are now failing, which should not happen.
- Otherwise, escalate to the user.

---

## **Stage-4.1**: Cleanup and Preparing for Commit (W3)

You send an exact `TASK_APPEND` message to the monitor, with content **specified** in `./.meta_supp/tmpl_worker3_cleanup.md` (MUST send in this EXACT template format), expanding:
- `{{N}}`: the next task ID.

Once the monitor responds with the result, check the result summary:
- If it says `**CLEANUP SUCCESS**`: the stopping condition is met. Proceed to `# ===IN-THE-END===`, under the `COMMIT_DONE_SUCCESS` case with message "Translation update completed" (See `# ===IN-THE-END===` below for format).
- If it says `**[ERROR] USER ATTENTION: CLEANUP FAILED**`: escalate to the user.
- Otherwise: escalate to the user.
# ===IN-THE-END===

When stopping condition is met, you should send a message to the monitor, in the following format:

``````md
**`.META.md context check: PASS/FAIL`**

## `MESSAGE::TO_MONITOR`
- `MONITOR_ACTION=COMMIT_DONE_SUCCESS`

=====COMMIT_DONE_SUCCESS(START)=====
{{some commit message}}
=====COMMIT_DONE_SUCCESS(END)=====
``````
or (**IN RARE CASES EXPLICITLY APPROVED BY USER**)
``````md
**`.META.md context check: PASS/FAIL`**

## `MESSAGE::TO_MONITOR`
- `MONITOR_ACTION=COMMIT_DONE_FAIL`

=====COMMIT_DONE_FAIL(START)=====
{{some commit message}}
=====COMMIT_DONE_FAIL(END)=====
``````


# ===AFTER-ESCALATION: USER-INSTRUCTION===
If you escalate to the user, the human user respond to you in the following format:

```md
## `MESSAGE::FROM_USER`

<...instructions-from-human-user>
```

Then do as human user instructed, whatever that is (might be responding to monitor certain stuff), do some code changes, etc.



# ===IMPORTANT-NOTES===

As you can see above, sending out a message to monitor might involve looking at template files (`./meta_supp/tmpl_*.md`). Make sure you alreadys re-read the latest template files needed right before you are about to compose a message, as human user might update those template files. If you cannot find the files and believe user had made a mistake in preparing those files, please escalate to user.

**MUST READ MESSAGE TEMPLATE FILES!** For cases where template files are mentioned, MUST use read those exact template files to draft your message.

**Git Access Restrictions**: All worker agents (and you) have read-only access to git storage. While files in the working directory can be modified, git write operations (commit, push, etc.) are not allowed and only the monitor can perform commits. If workers need backups, they should create file copies (e.g., `xxx` to `xxx.temp_bak`). Git commands should be avoided in most cases; when necessary, only read-only operations (e.g., `git status`, `git diff`, `git log`) can be used. To restore files from the last commit, use `git checkout -- <file>`.

**Worker Behavior Monitoring**: If a worker agent continuously refuses to work, does not follow instructions, or repeatedly produces only summaries/explanations without actual work for **5 consecutive rounds**, escalate to the user immediately. This includes cases where the worker:
- Claims work is impossible without making serious attempts
- Provides only analysis or explanations without code changes when code changes are required

Only user can explictly allowing committing unfinished work and stop the workflow in some escalation, with corresponds to **`COMMIT_DONE_FAIL`**. For other cases, MUST follow the workflow mentioned above.



# ===WORKFLOW-DIAGRAM===

```mermaid
<...omitted>
```

**Legend:**
- Stages are organized by worker type in separate boxes
- Arrows show normal workflow progression
- Self-loops indicate iterative work within a stage
- **Note**: Error escalation paths are omitted from the diagram for simplicity. Any stage can escalate to the user if having unexpected off-the-track behavior as mentioned in earlier sections.
\end{lstlisting}

\section{Feature Level Plans}
\label{sec:appendix-feature-levels}

This appendix lists the feature levels for each benchmark program, as produced by the planning agent (\textsc{MAS\_Plan}).
Each feature level (FL) is a complete, runnable program with its own test suite.
FL$_0$ is the minimal version; FL$_N$ is the full-featured program.

\subsection{awk (25 levels: FL$_0$--FL$_{16}$)}
\label{sec:fl-awk}

\begin{itemize}[leftmargin=*, nosep]

\item \textbf{FL$_0$}:
Minimal interpreter that executes \code{BEGIN \{ print "hello" \}} with string and number literals.

\item \textbf{FL$_1$}:
FL$_0$ + slightly more robust parsing and error handling.

\item \textbf{FL$_{1.4}$}:
FL$_1$ + END blocks, scalar variables, built-in functions (\code{length}, \code{substr}, \code{index}, \code{toupper}, \code{tolower}, math functions), and \code{exit} statement.

\item \textbf{FL$_2$}:
FL$_{1.4}$ + \code{printf} and \code{sprintf} with format strings.

\item \textbf{FL$_{2.2}$}:
FL$_2$ + arithmetic operators (\code{+}, \code{-}, \code{*}, \code{/}, \texttt{\%}, \texttt{\^{}}) and string concatenation.

\item \textbf{FL$_{2.4}$}:
FL$_{2.2}$ + assignment operators (\code{=}, \code{+=}, \code{-=}, etc.) and increment/decrement (\code{++}, \texttt{-{}-}).

\item \textbf{FL$_{2.6}$}:
FL$_{2.4}$ + boolean operators (\texttt{\&\&}, \code{||}, \code{!}), relational operators (\code{<}, \code{<=}, \code{==}, \code{!=}, \code{>=}, \code{>}), and ternary operator (\code{?:}).

\item \textbf{FL$_3$}:
FL$_{2.6}$ + pattern matching operators (\texttt{\~{}}, \texttt{!\~{}}).

\item \textbf{FL$_4$}:
FL$_3$ + field splitting, field variables (\code{\$0}, \code{\$1}, \ldots, \code{\$NF}), and the regular expression engine.

\item \textbf{FL$_{4.2}$}:
FL$_4$ + main input loop with automatic record reading (\code{NR}, \code{FNR}, \code{FILENAME}, \code{RS}).

\item \textbf{FL$_{4.4}$}:
FL$_{4.2}$ + \code{next}/\code{nextfile} statements and multiple input file processing.

\item \textbf{FL$_{4.6}$}:
FL$_{4.4}$ + pattern-action pairs with regex and expression patterns.

\item \textbf{FL$_5$}:
FL$_{4.6}$ + range patterns (\code{pattern1, pattern2}).

\item \textbf{FL$_6$}:
FL$_5$ + loops (\code{while}, \code{for}, \code{do-while}) and \code{break}/\code{continue}.

\item \textbf{FL$_{6.4}$}:
FL$_6$ + associative arrays and \code{delete} statement.

\item \textbf{FL$_7$}:
FL$_{6.4}$ + \code{for}-\code{in} loops and \code{in} operator for arrays.

\item \textbf{FL$_8$}:
FL$_7$ + user-defined functions with parameters, local variables, and recursion.

\item \textbf{FL$_9$}:
FL$_8$ + regex-based string functions (\code{match}, \code{sub}, \code{gsub}, \code{split}) and \code{RSTART}/\code{RLENGTH}.

\item \textbf{FL$_{10}$}:
FL$_9$ + I/O functions (\code{getline}, \code{close}, \code{fflush}) and output redirection (\code{>}, \texttt{>{}>}).

\item \textbf{FL$_{11}$}:
FL$_{10}$ + POSIX character classes in regex (\code{[:alnum:]}, \code{[:alpha:]}, etc.).

\item \textbf{FL$_{12}$}:
FL$_{11}$ + bounded repetition in regex (\code{\{n\}}, \code{\{n,\}}, \code{\{n,m\}}).

\item \textbf{FL$_{13}$}:
FL$_{12}$ + floating-point exception handling (SIGFPE) and debug mode.

\item \textbf{FL$_{14}$}:
FL$_{13}$ + UTF-8 multi-byte character support and locale-aware string operations.

\item \textbf{FL$_{15}$}:
FL$_{14}$ + safe mode (\code{-safe}) and \code{system()} function.

\item \textbf{FL$_{16}$}:
FL$_{15}$ + CSV input mode (\texttt{-{}-csv}) and pipe-based I/O. Full-featured awk interpreter.

\end{itemize}

\subsection{picoc (24 levels: FL$_0$--FL$_{16}$)}
\label{sec:fl-picoc}

\begin{itemize}[leftmargin=*, nosep]

\item \textbf{FL$_0$}:
Minimal C interpreter that parses and executes \code{printf} with string and integer literals in a \code{main} function.

\item \textbf{FL$_{0.2}$}:
FL$_0$ + broader parser and lexer infrastructure with more token types and reserved words recognized.

\item \textbf{FL$_{0.4}$}:
FL$_{0.2}$ + fuller parsing infrastructure with more language constructs, tokens, and reserved words.

\item \textbf{FL$_{0.6}$}:
FL$_{0.4}$ + user-defined function definitions (beyond just \code{main}).

\item \textbf{FL$_1$}:
FL$_{0.6}$ + \code{stdlib.h} (\code{malloc}, \code{free}, \code{atoi}, \code{exit}, etc.), \code{string.h} (\code{strlen}, \code{strcpy}, \code{strcmp}, etc.), additional \code{stdio.h} functions (\code{scanf}, \code{fprintf}, \code{sprintf}), and \code{\#include} directive support.

\item \textbf{FL$_2$}:
FL$_1$ + \code{return} statements and general expression evaluation beyond direct function calls.

\item \textbf{FL$_{2.4}$}:
FL$_2$ + arithmetic operators (\code{+}, \code{-}, \code{*}, \code{/}), comparison operators (\code{==}, \code{!=}, \code{<}, \code{>}, \code{<=}, \code{>=}), and \code{if}/\code{else} statements.

\item \textbf{FL$_3$}:
FL$_{2.4}$ + variable declarations, variable references in expressions, and assignment (\code{=}).

\item \textbf{FL$_{3.4}$}:
FL$_3$ + bitwise operators (\texttt{\&}, \code{|}, \texttt{\^{}}, \code{<<}, \code{>>}), logical operators (\texttt{\&\&}, \code{||}, \code{!}), ternary operator (\code{?:}), and modulo (\texttt{\%}).

\item \textbf{FL$_4$}:
FL$_{3.4}$ + compound assignment operators (\code{+=}, \code{-=}, \code{*=}, \code{/=}, etc.) and increment/decrement (\code{++}, \texttt{-{}-}).

\item \textbf{FL$_5$}:
FL$_4$ + \code{while}, \code{do}-\code{while}, and \code{for} loops, with \code{break} and \code{continue}.

\item \textbf{FL$_6$}:
FL$_5$ + pointer arithmetic, array indexing (\code{[]}), address-of (\texttt{\&}), and dereference (\code{*}) operators.

\item \textbf{FL$_{6.4}$}:
FL$_6$ + \code{struct} types with member access (\code{.} and \code{->} operators).

\item \textbf{FL$_7$}:
FL$_{6.4}$ + function pointer types and multi-dimensional arrays.

\item \textbf{FL$_8$}:
FL$_7$ + \code{\#define} macros (simple and parameterized) with macro expansion.

\item \textbf{FL$_{8.4}$}:
FL$_8$ + preprocessor conditionals (\code{\#ifdef}, \code{\#ifndef}, \code{\#if}, \code{\#else}, \code{\#endif}).

\item \textbf{FL$_9$}:
FL$_{8.4}$ + \code{float} and \code{double} types, floating-point literals, and floating-point arithmetic.

\item \textbf{FL$_{10}$}:
FL$_9$ + \code{typedef} declarations and storage class specifiers (\code{static}, \code{auto}, \code{register}, \code{extern}).

\item \textbf{FL$_{11}$}:
FL$_{10}$ + \code{union} and \code{enum} types.

\item \textbf{FL$_{12}$}:
FL$_{11}$ + \code{switch}/\code{case}/\code{default} statements.

\item \textbf{FL$_{13}$}:
FL$_{12}$ + \code{goto} statements and labels.

\item \textbf{FL$_{14}$}:
FL$_{13}$ + \code{errno.h}, \code{stdbool.h} (\code{bool}, \code{true}, \code{false}), and \code{ctype.h} (character classification functions).

\item \textbf{FL$_{15}$}:
FL$_{14}$ + \code{math.h} (sin, cos, sqrt, pow, etc.), \code{time.h}, and \code{unistd.h}.

\item \textbf{FL$_{16}$}:
FL$_{15}$ + debugger support (breakpoints, single-stepping) and interactive REPL mode. Full-featured C interpreter.

\end{itemize}

\subsection{gnu-bc (20 levels: FL$_0$--FL$_{16}$)}
\label{sec:fl-gnu-bc}

\begin{itemize}[leftmargin=*, nosep]

\item \textbf{FL$_0$}:
Minimal calculator with a hand-written recursive descent parser; evaluates addition of literal numbers with arbitrary precision.

\item \textbf{FL$_{0.4}$}:
FL$_0$ + yacc/lex-generated parser and scanner infrastructure (replacing the hand-written parser).

\item \textbf{FL$_1$}:
FL$_{0.4}$ + bytecode compilation and virtual machine interpreter.

\item \textbf{FL$_{1.4}$}:
FL$_1$ + subtraction (\code{-}), multiplication (\code{*}), and unary negation.

\item \textbf{FL$_2$}:
FL$_{1.4}$ + division (\code{/}), modulo (\texttt{\%}), and exponentiation (\texttt{\^{}}).

\item \textbf{FL$_3$}:
FL$_2$ + simple variables, variable assignment (\code{=}), compound assignment operators, and increment/decrement (\code{++}, \texttt{-{}-}).

\item \textbf{FL$_4$}:
FL$_3$ + special variables (\code{scale}, \code{ibase}, \code{obase}) and built-in functions (\code{sqrt}, \code{length}, \code{scale}).

\item \textbf{FL$_5$}:
FL$_4$ + arrays with multi-dimensional indexing.

\item \textbf{FL$_6$}:
FL$_5$ + \code{if}/\code{else} conditionals, comparison operators (\code{==}, \code{!=}, \code{<}, \code{<=}, \code{>}, \code{>=}), and logical operators (\texttt{\&\&}, \code{||}, \code{!}).

\item \textbf{FL$_7$}:
FL$_6$ + \code{while} loops, \code{break}, and \code{continue}.

\item \textbf{FL$_8$}:
FL$_7$ + \code{for} loops.

\item \textbf{FL$_9$}:
FL$_8$ + parameterless user-defined functions (using global variables only).

\item \textbf{FL$_{10}$}:
FL$_9$ + function parameters, local (\code{auto}) variables, and return values.

\item \textbf{FL$_{11}$}:
FL$_{10}$ + string literals, \code{print} statement, and string output operations.

\item \textbf{FL$_{12}$}:
FL$_{11}$ + \code{read()} and \code{random()} functions.

\item \textbf{FL$_{13}$}:
FL$_{12}$ + math library with transcendental functions (\code{s}, \code{c}, \code{l}, \code{e}, \code{a}, \code{j}) and \code{-l} flag.

\item \textbf{FL$_{13.4}$}:
FL$_{13}$ + multiple input file processing and command-line file arguments.

\item \textbf{FL$_{14}$}:
FL$_{13.4}$ + interrupt signal handling (SIGINT) for interactive sessions.

\item \textbf{FL$_{15}$}:
FL$_{14}$ + compile-only mode, POSIX compliance, readline/libedit support, version/warranty display, and limits display.

\item \textbf{FL$_{16}$}:
FL$_{15}$ + dc (Desk Calculator) RPN program. Full-featured GNU bc calculator.

\end{itemize}

\subsection{wren (19 levels: FL$_0$--FL$_{16}$)}
\label{sec:fl-wren}

\begin{itemize}[leftmargin=*, nosep]

\item \textbf{FL$_0$}:
Minimal interpreter that parses and executes \code{print("hello world")} with string literals.

\item \textbf{FL$_{0.4}$}:
FL$_0$ + variable declarations and assignments.

\item \textbf{FL$_1$}:
FL$_{0.4}$ + function definitions and function calls.

\item \textbf{FL$_2$}:
FL$_1$ + expression parsing infrastructure (Pratt parser with precedence and grouping).

\item \textbf{FL$_3$}:
FL$_2$ + control flow: \code{if}/\code{else}, \code{while}, \code{for} loops, \code{break}/\code{continue}, and logical operators (\texttt{\&\&}, \code{||}).

\item \textbf{FL$_{3.4}$}:
FL$_3$ + class definitions (structural only, without instantiation).

\item \textbf{FL$_4$}:
FL$_{3.4}$ + constructors and object instantiation.

\item \textbf{FL$_5$}:
FL$_4$ + method definitions, method dispatch, operator overloading, and all primitive operations on core types.

\item \textbf{FL$_6$}:
FL$_5$ + import/module system.

\item \textbf{FL$_7$}:
FL$_6$ + string interpolation, string indexing/slicing, string methods, and UTF-8 code point iteration.

\item \textbf{FL$_8$}:
FL$_7$ + \code{List} type with list literals and operations.

\item \textbf{FL$_9$}:
FL$_8$ + \code{Map} and \code{Range} types.

\item \textbf{FL$_{10}$}:
FL$_9$ + closures and upvalue capturing.

\item \textbf{FL$_{11}$}:
FL$_{10}$ + metaclasses, static methods, and multi-level inheritance.

\item \textbf{FL$_{12}$}:
FL$_{11}$ + class attributes.

\item \textbf{FL$_{13}$}:
FL$_{12}$ + foreign methods, slot-based API, and handles for C interop.

\item \textbf{FL$_{14}$}:
FL$_{13}$ + foreign classes with custom allocation and finalization.

\item \textbf{FL$_{15}$}:
FL$_{14}$ + fiber system (coroutines and cooperative multitasking).

\item \textbf{FL$_{16}$}:
FL$_{15}$ + optional Meta and Random modules. Full-featured Wren VM.

\end{itemize}

\subsection{mujs (20 levels: FL$_0$--FL$_{16}$)}
\label{sec:fl-mujs}

\begin{itemize}[leftmargin=*, nosep]

\item \textbf{FL$_0$}:
Minimal JavaScript interpreter that parses and executes \code{print("hello world");} with string literals.

\item \textbf{FL$_1$}:
FL$_0$ + number, boolean, \code{null}, and \code{undefined} literals with built-in \code{print}.

\item \textbf{FL$_2$}:
FL$_1$ + variable declarations, variable references, object literals, array literals, and property access.

\item \textbf{FL$_{2.4}$}:
FL$_2$ + arithmetic operators (\code{+}, \code{-}, \code{*}, \code{/}, \texttt{\%}), increment/decrement (\code{++}, \texttt{-{}-}), compound assignment (\code{+=}, \code{-=}, etc.), and unary plus/minus.

\item \textbf{FL$_{2.6}$}:
FL$_{2.4}$ + comparison operators, logical operators (\texttt{\&\&}, \code{||}, \code{!}), bitwise operators, \code{typeof}, \code{delete}, and \code{void} operators.

\item \textbf{FL$_3$}:
FL$_{2.6}$ + bytecode compilation and virtual machine interpreter (replacing direct AST evaluation).

\item \textbf{FL$_4$}:
FL$_3$ (consolidation level; same features).

\item \textbf{FL$_5$}:
FL$_4$ + control flow: \code{if}/\code{else}, \code{while}, \code{do}-\code{while}, \code{for}, \code{for}-\code{in}, \code{switch}, \code{break}/\code{continue}, and ternary operator (\code{?:}).

\item \textbf{FL$_6$}:
FL$_5$ + top-level function declarations, function expressions, function calls, \code{return} statements, and \code{arguments} object.

\item \textbf{FL$_7$}:
FL$_6$ + closures, nested function definitions, and lexical scoping.

\item \textbf{FL$_8$}:
FL$_7$ + prototype chain, inheritance, \code{new} operator, and \code{instanceof}.

\item \textbf{FL$_9$}:
FL$_8$ + \code{Object.prototype} methods and \code{Object} static methods (\code{keys}, \code{create}, \code{defineProperty}, etc.).

\item \textbf{FL$_{10}$}:
FL$_9$ + \code{Array.prototype} methods (\code{push}, \code{pop}, \code{slice}, \code{sort}, \code{forEach}, \code{map}, \code{filter}, \code{reduce}, etc.).

\item \textbf{FL$_{11}$}:
FL$_{10}$ + \code{Function} constructor, \code{call}, \code{apply}, and \code{bind}.

\item \textbf{FL$_{12}$}:
FL$_{11}$ + \code{String} constructor object and \code{String.prototype} methods (\code{charAt}, \code{indexOf}, \code{slice}, \code{split}, etc.).

\item \textbf{FL$_{13}$}:
FL$_{12}$ + \code{Boolean} and \code{Number} constructor objects and methods.

\item \textbf{FL$_{14}$}:
FL$_{13}$ + specialized error types (\code{TypeError}, \code{RangeError}, \code{ReferenceError}, \code{SyntaxError}, \code{EvalError}, \code{URIError}).

\item \textbf{FL$_{14.4}$}:
FL$_{14}$ + \code{Math} object and mathematical functions.

\item \textbf{FL$_{15}$}:
FL$_{14.4}$ + regular expression engine, \code{RegExp} constructor, regexp literals, and regexp-dependent \code{String} methods (\code{match}, \code{search}).

\item \textbf{FL$_{16}$}:
FL$_{15}$ + \code{Date} and \code{JSON} objects. Full-featured ECMAScript~5 interpreter.

\end{itemize}

\subsection{pocketpy (22 levels: FL$_0$--FL$_{16}$)}
\label{sec:fl-pocketpy}

\begin{itemize}[leftmargin=*, nosep]

\item \textbf{FL$_0$}:
Minimal expression evaluator with \code{print()}, integer/float/string literals, variables, and basic arithmetic (\code{+}, \code{-}, \code{*}, \code{/}).

\item \textbf{FL$_{0.2}$}:
FL$_0$ + comparison operators (\code{==}, \code{!=}, \code{<}, \code{>}, \code{<=}, \code{>=}), \code{bool}, and \code{None} types.

\item \textbf{FL$_{0.4}$}:
FL$_{0.2}$ + \code{list} type with operations, \code{len()}, and additional arithmetic (\texttt{\%}, \code{**}, \code{//}).

\item \textbf{FL$_1$}:
FL$_{0.4}$ + \code{while} loops, \code{if}/\code{elif}/\code{else}, exception handling (\code{try}/\code{except}/\code{raise}), and boolean operators (\code{and}, \code{or}, \code{not}).

\item \textbf{FL$_{1.2}$}:
FL$_1$ + \code{for} loops, \code{range()} builtin, and iteration protocol.

\item \textbf{FL$_{1.4}$}:
FL$_{1.2}$ + \code{tuple} type.

\item \textbf{FL$_{1.6}$}:
FL$_{1.4}$ + user-defined functions (\code{def}), \code{return} statements, and function arguments.

\item \textbf{FL$_2$}:
FL$_{1.6}$ + closures, \code{lambda} expressions, and nested function definitions.

\item \textbf{FL$_3$}:
FL$_2$ + class definitions, inheritance, \code{dict} and \code{set} types.

\item \textbf{FL$_4$}:
FL$_3$ + list/dict/set comprehensions.

\item \textbf{FL$_5$}:
FL$_4$ + decorators (\code{@decorator}), metaclasses, and property descriptors.

\item \textbf{FL$_6$}:
FL$_5$ + generators (\code{yield}, \code{yield from}).

\item \textbf{FL$_7$}:
FL$_6$ + introspection modules (\code{inspect}, \code{dis}, \code{traceback}, \code{importlib}) and trace system.

\item \textbf{FL$_8$}:
FL$_7$ + \code{json} and \code{gc} control modules.

\item \textbf{FL$_9$}:
FL$_8$ + \code{time} and \code{random} modules.

\item \textbf{FL$_{10}$}:
FL$_9$ + \code{os} module and file I/O (\code{open()}).

\item \textbf{FL$_{11}$}:
FL$_{10}$ + \code{enum}, \code{unicodedata}, and \code{conio} modules.

\item \textbf{FL$_{12}$}:
FL$_{11}$ + \code{pickle}, \code{base64}, and \code{lz4} modules.

\item \textbf{FL$_{13}$}:
FL$_{12}$ + graphics/game modules (\code{easing}, \code{colorcvt}, \code{array2d}, \code{vmath}).

\item \textbf{FL$_{14}$}:
FL$_{13}$ + multi-VM support and dynamic library loading.

\item \textbf{FL$_{15}$}:
FL$_{14}$ + line profiling system.

\item \textbf{FL$_{16}$}:
FL$_{15}$ + debugger with Debug Adapter Protocol (DAP) and breakpoint support. Full-featured Python~3.x interpreter.

\end{itemize}

\end{document}